\documentclass{mn2e}

\usepackage{amsfonts}
\usepackage{amsmath}
\usepackage{graphicx}

\title[Evolution of early-type BCGs]
      {Evolution in the structural properties of 
       early-type Brightest Cluster Galaxies at small lookback time 
       and dependence on the environment}

\author[M. Bernardi]
{Mariangela Bernardi\thanks{E-mail: bernardm@physics.upenn.edu}\\
 Department of Physics \& Astronomy, University of Pennsylvania, 
      209 S. 33rd St., Philadelphia, PA 19104, USA}

\begin{document}
\pagerange{\pageref{firstpage}--\pageref{lastpage}}

\maketitle

\label{firstpage}

\begin{abstract}
At the present time, early-type brightest cluster galaxies 
in the SDSS MaxBCG and C4 catalogs have larger sizes than 
early-type galaxies of similar luminosity, 
whether these other objects are in the field, or are satellites in 
clusters.  BCG sizes are also stronger functions of luminosity
 $R_e\propto L$ 
than are the sizes of the bulk of the population; this remains true 
if one restricts attention to narrow bins in velocity dispersion.  
At fixed stellar mass {\em and} formation time, 
objects at lower redshift are larger and have smaller velocity 
dispersions -- i.e. the sizes increase and velocity dispersions 
decrease with age.   
In addition, at any given redshift, younger BCGs have slightly 
larger sizes than older BCGs of the same stellar mass; however, 
they have similar velocity dispersions.  
As a result, at redshifts $\sim 0.25$, corresponding to lookback 
times of order 3~Gyrs, BCGs are smaller than their lower redshift 
counterparts by as much as $\sim 70\%$ for the brightest BCGs:
the sizes evolve as $(1+z)^{0.85(M_r+21)}$.  
Qualitatively similar but weaker evolution in the sizes is also 
seen in the bulk of the early-type population: at $M_r<-22$ the sizes 
evolve as $(1+z)^{0.7(M_r+21)}$, while at $M_r>-22$ 
the evolution is approximately $(1+z)^{-0.7}$, independent of $M_r$. 
The velocity dispersion-luminosity correlation also evolves:\, $(1+z)^{-0.2(M_r+21)}$ at $M_r < -22$ (as for the BCGs) and $(1+z)^{0.2}$ 
for fainter galaxies.  
The size-- and velocity dispersion--stellar mass correlations 
yield consistent results, although, in this case, accounting for 
selection effects is less straightforward.  
These trends, in particular the fact that the velocity dispersions 
at fixed stellar mass decrease with age, are most easily understood 
if early-type BCGs grew from many dry minor mergers rather than 
a few major mergers.  Only in such a scenario can BCGs be the 
descendents of the super-dense galaxies seen at $z\sim 2$; major 
dry mergers, which increase the size in proportion to the mass, 
cannot bring these galaxies onto the BCG $R_e-M_*$ relation at 
$z\sim 0$.  

We also compared the ages and sizes of our early-type BCGs with 
other cluster galaxies (satellites).  BCGs are larger than satellites 
of similar luminosity or stellar mass at the same redshift.  
Although both satellites and BCGs trace the same weak age$-L$ or 
age$-M_*$ relation, this can be understood by noting that BCGs are 
typically about 1~Gyr older than the satellites in their group, and 
they are about 0.5~mags more luminous. Finally, we find that the mean 
satellite luminosity is approximately independent of BCG luminosity, 
in agreement with recent predictions based on the luminosity-dependence 
of clustering.
\end{abstract}


\section{Introduction}
There has been recent interest in the sizes of galaxies: at fixed 
stellar mass, galaxies appear to be more than three times
smaller at $z\sim 2$ than at $z\sim 0$ (Trujillo et al. 2006;
Cimatti et al. 2008; Van Dokkum et al. 2008; 
Younger et al. 2008; Buitrago et al. 2008; Tacconi et al. 2008; 
Chapman et al. 2008; Franx et al. 2008; Damen et al. 2008; 
Damjanov et al. 2009; Saracco et al. 2008).  
Similar evolution in the size-luminosity relation of radio 
galaxies was seen almost a decade ago \cite{Roche98}.    
This evolution is difficult to arrange in models where the galaxies 
form from a simple monolithic collapse.  

On the other hand, qualitatively similar behaviour for dark matter 
halos has been expected for some thirty years:  a virialized halo 
at a given epoch is approximately 200 times denser than the critical 
density at that epoch, whatever its mass \cite{gg72}.  Thus, at fixed 
mass, the virial radius scales approximately as $(1+z)^{-1}$.  
Mass-independent densities appear to be a good description of 
cluster-mass halos locally \cite{as07,maxbcgLensing,chandra08} 
and at higher redshifts (e.g., Meneux et al. 2008).  
These halos are expected to have formed from essentially dissipationless 
mergers, so, for systems dominated by dissipationless merging, we expect 
that, at fixed mass, the radii are larger at late times. 

One arrives at the same conclusion if one considers mergers along 
parabolic orbits \cite{oh77,Ciotti08}.  For example, the velocity 
dispersion of the product of a parabolic merger of two equal mass 
galaxies is the same as that of its progenitors (this assumes mass 
and energy conservation).  The virial theorem requires that if 
the mass has doubled with no change to the velocity dispersion, 
then the size must also have doubled, making the density smaller 
by a factor of four.  This is the most extreme case: if the 
progenitor masses were unequal, then the density of the product is 
less than that of the more massive progenitor, but by a factor of 
less than four.  
Again, dissipationless mergers act to decrease the density.

Galaxy formation was not dissipationless \cite{fe80,bh91}:  
gas dissipation has played an important role, although this is 
expected to have been more true at high redshift \cite{cores08}.
Subsequent major mergers between disk galaxies are thought to 
be the main way in which elliptical galaxies form \cite{toomre72}, 
possibly followed by dry dissipationless mergers in which 
both the size and the stellar mass of the final object increase
\cite{mk85,cfa06,berkeley06,Khochfar06,cores08}.  
So, to answer the question of where, today, are the super-dense 
objects seen at high-$z$, it has been suggested that they must 
have undergone dissipationless mergers since then, so as to have 
gone unnoticed today.  (Cimatti et al. 2008 note that there is a 
class of local objects which may be direct descendents of the 
high-$z$ samples for which the merger hypothesis is unnecessary -- 
these are the fast rotators in the sample of Bernardi et al. 2008 
which have large $\sigma$ but small sizes.  But these objects are 
too rare to be the typical descendents. Recently, Trujillo et al. 2009
also found a very low number density of old superdense massive galaxies 
in the present Universe).

However, Fan et al. (2008) have shown a model which may be able 
to reproduce the observed evolution in the size$-M_*$ relation.  
They assume that the AGN feedback which expels gas from the central 
regions produces a sudden reduction of mass in the core, as a result 
of which the stellar distribution puffs-up.  In this model, the sizes 
increase, but the stellar masses do not.  Since the peak of AGN 
activity was around $z\sim 2$, most of the size evolution in this 
model is over by $z\sim 1$. In contrast, dry mergers and evolution 
in hierarchical models is expected to continue to the present day  
\cite{munich06,durham07}.  

Most observational studies have concentrated on objects at $z>1$.  
van der Wel et al. (2008) measured the size of a sample of massive 
early-type galaxies at $z\sim 1$ and found that they are about a 
factor of two smaller than at $z\sim 0$.
One of the main goals of this paper is to investigate the 
possibility that the sizes are evolving even at small redshift 
($z<0.3$), paying particular attention to those objects for which 
dry dissipationless merging is thought to have been most common -- 
early-type BCGs \cite{mk81,mk85,oh91,Lauer07,Bernardi07,Tran08}.  
Late-type BCGs, and, indeed, late-type galaxies, are not studied in this 
paper, so we will often not bother to specify the qualifier `early-type' 
when we discuss BCGs (and similarly, when we discuss non-central galaxies).

Early-type BCGs at $0.4<z<1$ have been compared with more local 
samples recently.  
One study concludes that the stellar mass appears to not have grown 
significantly since $z\sim 1$ \cite{EDisCSbcg}, but does not include 
a study of the BCG sizes.  Another suggests that the sizes may have 
been smaller at high redshift, but some of this apparent evolution was 
almost certainly a consequence of not looking at fixed restframe 
wavelength or stellar mass \cite{nszdg02}.  

Section~\ref{sample} describes our early-type BCG sample in which we 
have size, velocity dispersion, stellar mass and age estimates out 
to $z\sim 0.3$.  
In Section~\ref{evolve} we show that, at fixed (evolution corrected) 
luminosity BCGs were larger in the past, and they had smaller 
velocity dispersions.  This remains true if we replace luminosity 
with stellar mass, although in this case the measurement is 
complicated by selection effects (as we illustrate in Appendix A,
where we also discuss how correlated errors in stellar mass and 
age can compromise the observed correlations).  
We also show that the objects which formed earlier 
are smaller, but their velocity dispersions are not larger -- 
whereas the former is expected in monolithic collapse 
models the latter is harder to arrange.

In Section~\ref{censat} we compare the sizes and ages of BCGs to 
those of other (early-type) cluster galaxies.  This complements 
recent work indicating that non-central or satellite luminosities 
should be approximately independent of the mass of a group or the 
luminosity of the central BCG \cite{Skibba06,Skibba07}.  

A final section summarizes our results.  Appendix~B discusses how 
our luminosity-size correlation for BCGs compares with other recent 
determinations.  

Complementary studies of the sizes and velocity dispersion of the 
bulk of the local early type population are presented in 
Shankar \& Bernardi (2009) and Shankar et al. (2009).  
These other studies show that the size-luminosity relation of 
galaxies with $L_r < 10^{11}$~$L_{\odot}$ at $z\sim 0$ is independent 
of the age of the stellar population, implying that the size-stellar 
mass relation does depend on age:  older galaxies are smaller than 
younger galaxies of the same stellar mass.
However, they find a weak dependence of the velocity dispersions 
(at fixed mass) on galaxy age -- this is difficult to accomodate 
in a monolithic-based collapse model.
The amplitude of these trends for massive galaxies 
($L_r > 10^{11}$~$L_{\odot}$) is even smaller. 

As discussed below, we show that early-type BCGs/massive galaxies 
are different, and have evolved differently, at least at small lookback 
times, from the bulk of early-type galaxies; their properties are more 
consistent with formation from predominantly dissipationless mergers 
than from a monolithic collapse.  

\section{The sample}\label{sample}
In what follows, we will study the luminosities, sizes, velocity 
dispersions and stellar masses of early-type BCGs identified in 
the SDSS.  Spectra are available for objects with $m_r<17.7$, 
whereas the photometry allows galaxies to be reliably identified 
down to about 4 magnitudes deeper.  
The photometric quantities (magnitudes, half-light radii) we 
use in what follows are not exactly the same as those output by 
the SDSS DR6 database.  
Rather, we use the prescriptions (Equations~1 -- 4) in 
Hyde \& Bernardi (2009) to correct the SDSS parameters for 
known sky subtraction problems 
with the photometry of bright objects in crowded fields.  Typically, 
the corrected magnitudes are slightly brighter and the sizes 
correspondingly larger (there is a well-known correlation for 
correlated errors in early-type galaxy photometry which has 
$r_e/I_e^{0.8}\approx$constant).  The 
velocity dispersions are measured through a fiber of radius 
1.5~arcsec; they are then corrected to $r_e/8$ (as is standard 
practice).  The size and velocity dispersion can be combined to 
estimate a dynamical mass; we do this by setting
 $M_{\rm dyn} = 5R_e\sigma^2/G$.
For a subset of objects, stellar masses, and luminosity-weighted age 
estimates are available from Gallazzi et al. (2005).  
Throughout, angular diameter and luminosity distances were 
computed from the measured redshifts assuming a Hubble constant 
of $H_0=70$~km~s$^{-1}$~Mpc$^{-1}$ 
in a geometrically flat background model dominated by a 
cosmological constant at the present time:
  $(\Omega_0,\Lambda_0) = (0.3,0.7)$.  

We would like to study early-type BCGs over a relatively large range 
in redshift.  Since we are most interested in evolution, we would 
like the sample to be homogeneously selected over the entire redshift 
range.  This motivates the use of the {\tt MaxBCG} catalog which was 
assembled by Koester et al. (2007) from the SDSS.  It spans the range 
$0.07 < z < 0.3$; the 13823 groups and clusters in it each contain 
at least 10 galaxies brighter than $0.4L_*$ in the $r$-band (i.e., 
brighter than about $-19.5$~mags in $r$).  

\begin{figure}
 \centering
 \includegraphics[width=0.95\hsize]{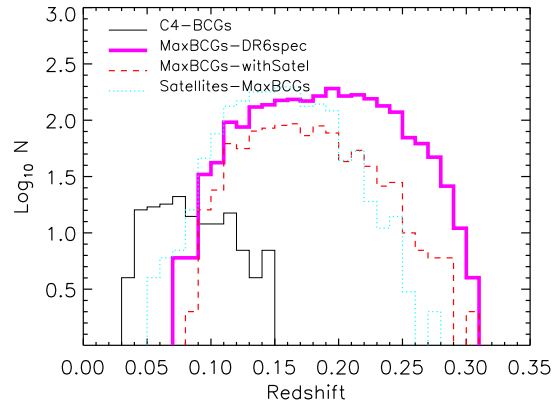}
 \caption{Redshift distribution of BCGs from the C4 (thin solid) 
          and MaxBCG (thick solid) catalogs.  Dot-dashed histogram shows 
          the redshift distribution of the objects we identify as 
          satellites, and dashed histogram shows the redshift 
          distribution of the subset of BCGs which host these 
          satellites.}
 \label{BCGz}
\end{figure}

\begin{figure}
 \centering
 \includegraphics[width=0.95\hsize]{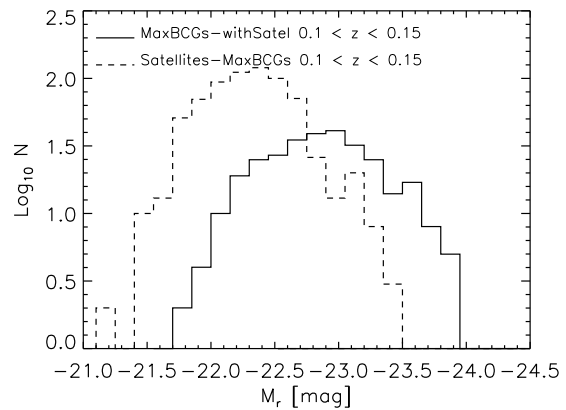}
 \caption{Luminosity distribution of early-type MaxBCGs with 
          early-type satellites (solid), and satellites 
          (dashed) over the range $0.1<z<0.15$.  }
 \label{BCGl}
\end{figure}

At the very least, we would like to study BCG luminosities and sizes 
over this redshift range, but, if possible, we would like to study 
velocity dispersions and stellar masses as well.  Of 13823 clusters 
in the MaxBCG catalog, 5413 are reported to have BCGs with spectroscopic 
redshifts.  However, upon matching with the SDSS-DR6 
(Adelman-McCarthy et al. 2008), we find 7832 objects with spectroscopic 
redshifts.  
A subset of 4912 has deVaucouleur apparent magnitude between 14.5 
and 17.5.  
Requiring {\tt fracDev>0.8} in both the $g$- and $r$-bands is a 
good way to select early-types; this reduces the sample size to 4350.  
Of these, 2634 have stellar mass estimates from Gallazzi et al. (2005), 
and only 2012 of these have estimated velocity dispersions.  For the 
objects with stellar mass estimates, Gallazzi et al. (2005) also 
provide (luminosity weighted) age estimates.  
The results which follow that are based on $L$ rather than $M_*$ 
do not depend on whether or not we used 4350 or 2634 galaxies.  
(Requiring that velocity dispersions were estimated but not caring 
about $M_*$ makes the sample size 3277.)

We will also be interested in how BCGs compare to non-central or 
satellite galaxies.  We do this by searching an appropriately chosen 
volume around each MaxBCG in our sample as follows.  The catalog 
contains an estimate of the number of galaxies, $N_{\rm gal}$ in the 
group associated with each BCG.  We define a velocity dispersion  
 $\sigma_{group} \equiv 100\, \sqrt{1.33\,N_{\rm gal}}$~km~s$^{-1}$
and a radius 
 $r_{group} \equiv \sigma_{group}/7/70$~Mpc.
The relation between group radius and velocity dispersion is 
chosen to approximately match the expected scaling for dark matter 
halos.  The resulting scaling between group velocity dispersion and 
$N_{gal}$ is close to that reported by Becker et al. (2007) in their 
analysis of this catalog.  
We identify as satellites all objects which are within 
$r_{group}$ across and $2\sigma_{\rm group}$ along the 
line of sight to a BCG in the catalog.  
Having identified the satellites, we would now like to exclude 
those that are of later type.  We do so by applying the same 
selection cuts as we did to identify early-type BCGs.  This 
leaves a sample of 1734 objects -- less than half the full 
sample of satellites.  (If the BCG does not sit at the center 
of its cluster, then we will be making errors of inclusion and 
exclusion at the cluster boundaries; but since we are already 
making approximations about which objects along the line of sight 
are true members, which lead to qualitatively similar errors, 
we do not attempt to correct for this effect.)

Note that our satellite sample is smaller than the BCG sample; 
this might seem in conflict with our previous statement that each 
maxBCG cluster has at least 10 satellites brighter than $0.4L_*$.  
This is because we are requiring that satellites have measured 
spectra; because of the SDSS magnitude limit, spectra of $0.4L_*$ 
objects are not available beyond $z\sim 0.05$.  For example, 
of the 1734 satellites in our sample, 1555 have $M_r < -22$.
Thus, the vast majority of these objects are more luminous than 
$3L_*$ -- they represent the bright-end of the satellite luminosity 
function.  In addition, if an SDSS fiber was placed on a BCG, then 
objects within 55 arcsecs of it will not have an SDSS spectrum.  
These fiber-collisions affect a larger physical scale for the higher 
redshift clusters, further reducing the number of satellites with 
measured spectra.  Note that this means we tend to miss those satellites 
which are closest to their BCGs -- dynamical friction arguments suggest 
that these may be amongst the most massive satellites in each cluster.

Johnston et al. (2007) note that about 20\% of the BCGs are 
misidentified, and that this fraction increases at lower $N_{gal}$.  
In such cases where a satellite has been classified as a BCG, fiber 
collisions almost certainly prevent the BCG from having a spectrum 
and so entering our sample of satellites.  This will serve to make 
the BCG sample more like the satellites, but not vice versa -- a 
point we return to later.

At lower redshifts, we use the C4 catalog of Miller et al. (2005).  
Although this is a rather different catalog, we show below that if 
we extrapolate the trends we see in the MaxBCG catalog to smaller 
$z$, then they are in good agreement with those in the C4 catalog.  
(The most important difference is that the MaxBCG algorithm searches 
for a red sequence in the photometric sample, whereas the C4 method 
simply searches for objects with similar colors in the spectroscopic 
sample, so it does not select against groups of blue galaxies.  
Since we only use the early-type BCGs from this catalog anyway, the 
real difference is that, because the C4 method is based on the 
spectroscopic sample, it may miss the true BCG because of fiber 
collisions.  This is a known problem, for which the catalog provides 
a flag.  We only use the subset of C4 clusters for which the 
photometric and spectroscopic BCG are the same.)  

Furthermore, there is previous work on the C4 BCG catalog; 
studying this catalog allows us to tie our measurements to those 
in the literature (see Appendix~B).
This is important because, although all the objects we analyze are 
drawn from the SDSS database, the photometric quantities (magnitudes, 
half-light radii) are not exactly the same as those output by the 
survey.  Rather, we use the prescriptions in Hyde \& Bernardi (2009) 
to correct the SDSS parameters for known sky subtraction problems 
with the photometry of bright objects in crowded fields.  

Figure~\ref{BCGz} shows the distribution of early-type BCG redshifts 
in the C4 (solid) and MaxBCG (MaxBCG-DR6spec, solid) catalogs 
(the C4 BCGs probe lower redshifts), the redshift distribution of 
MaxBCG early-type satellites (dotted), and the redshift distribution 
of early-type MaxBCGs with early-type satellites (MaxBCG-withSatel, dashed).  
The satellite distribution does not extend to as high redshifts, 
because satellites are fainter than BCGs (by definition).  
The dashed histogram shows the redshift distribution of those BCGs 
which have satellites with spectroscopic information.  
Most of the high redshift BCGs have no satellites, due to the 
combined effects of the magnitude limit and fiber-collisions.
However, over the redshift range $0.1<z<0.15$ there are approximately 
three satellites per BCG; we will only use the objects in this range 
when we compare BCG and satellite properties.  
Figure~\ref{BCGl} shows the distribution of (evolution corrected) 
luminosities over this range:  the mean BCG luminosity is about 
half a magnitude brighter.

\begin{figure}
 \centering 
 \vspace{-2cm}
 \includegraphics[width=0.95\hsize]{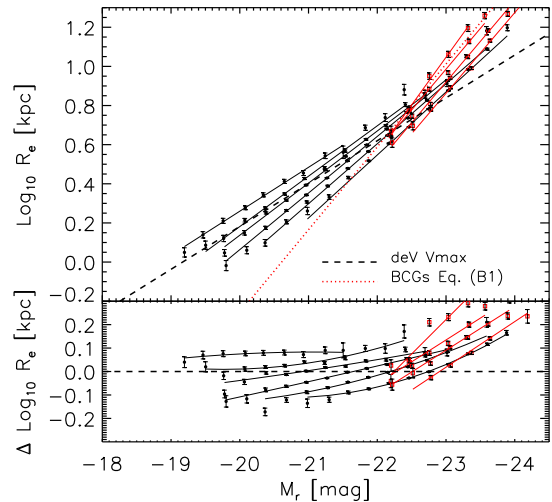}
 \caption{Size-luminosity relation for the bulk of the population 
          (dashed) and BCGs (dot-dashed).  Filled symbols show this 
          relation when the sample is restricted to a narrow range 
          in $\sigma$:  results for 6 bins of width 0.1~dex, 
          starting from $\log_{10}\sigma=1.9$, are shown.  
          Open circles show a similar analysis of the BCGs:  4 bins 
          of width $0.05$~dex, starting from $\log_{10}\sigma=2.3$.}
 \label{LRsigbin}
\end{figure}

\begin{figure*}
 \centering
 \includegraphics[width=0.475\hsize]{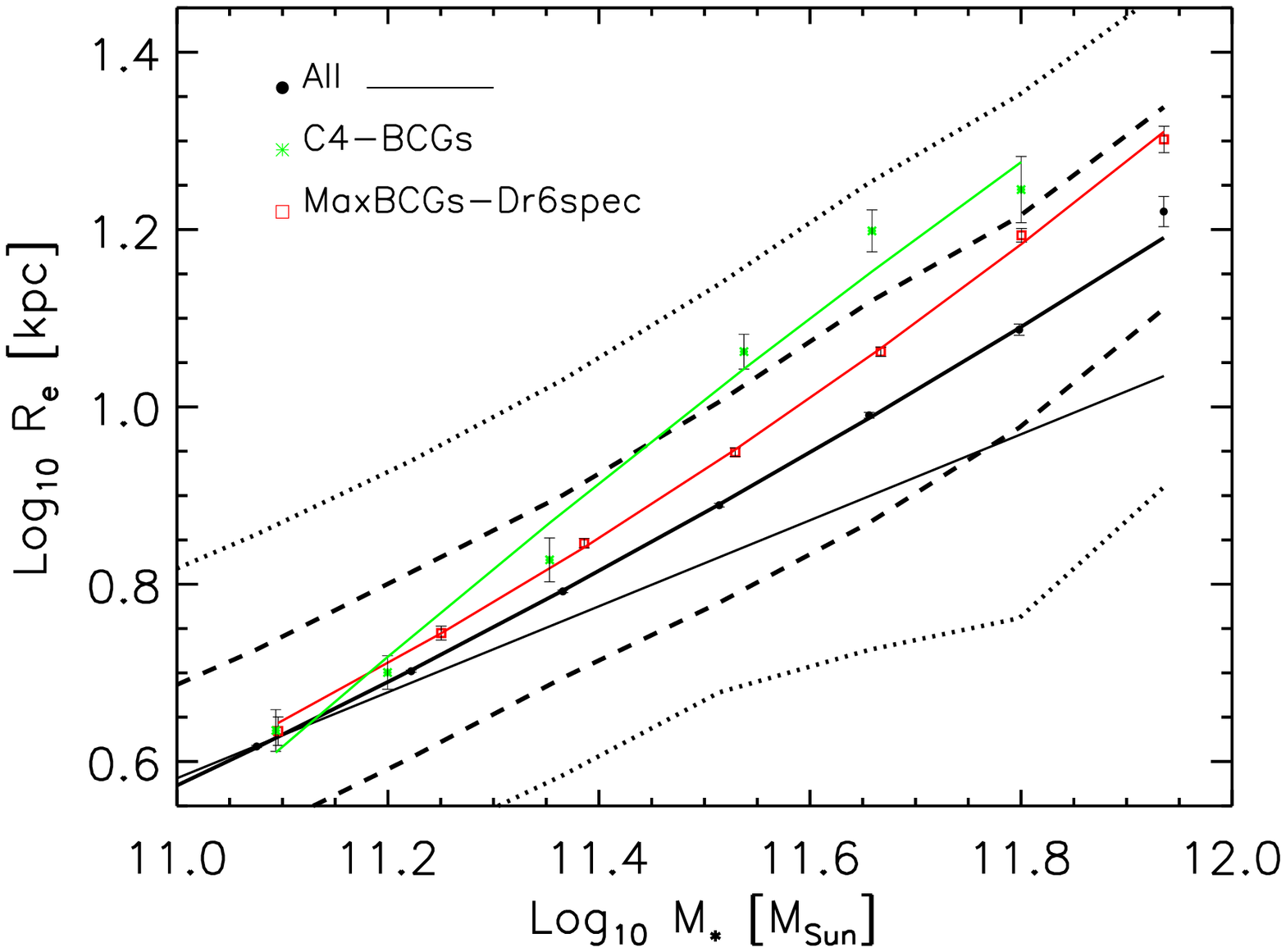}
 \includegraphics[width=0.475\hsize]{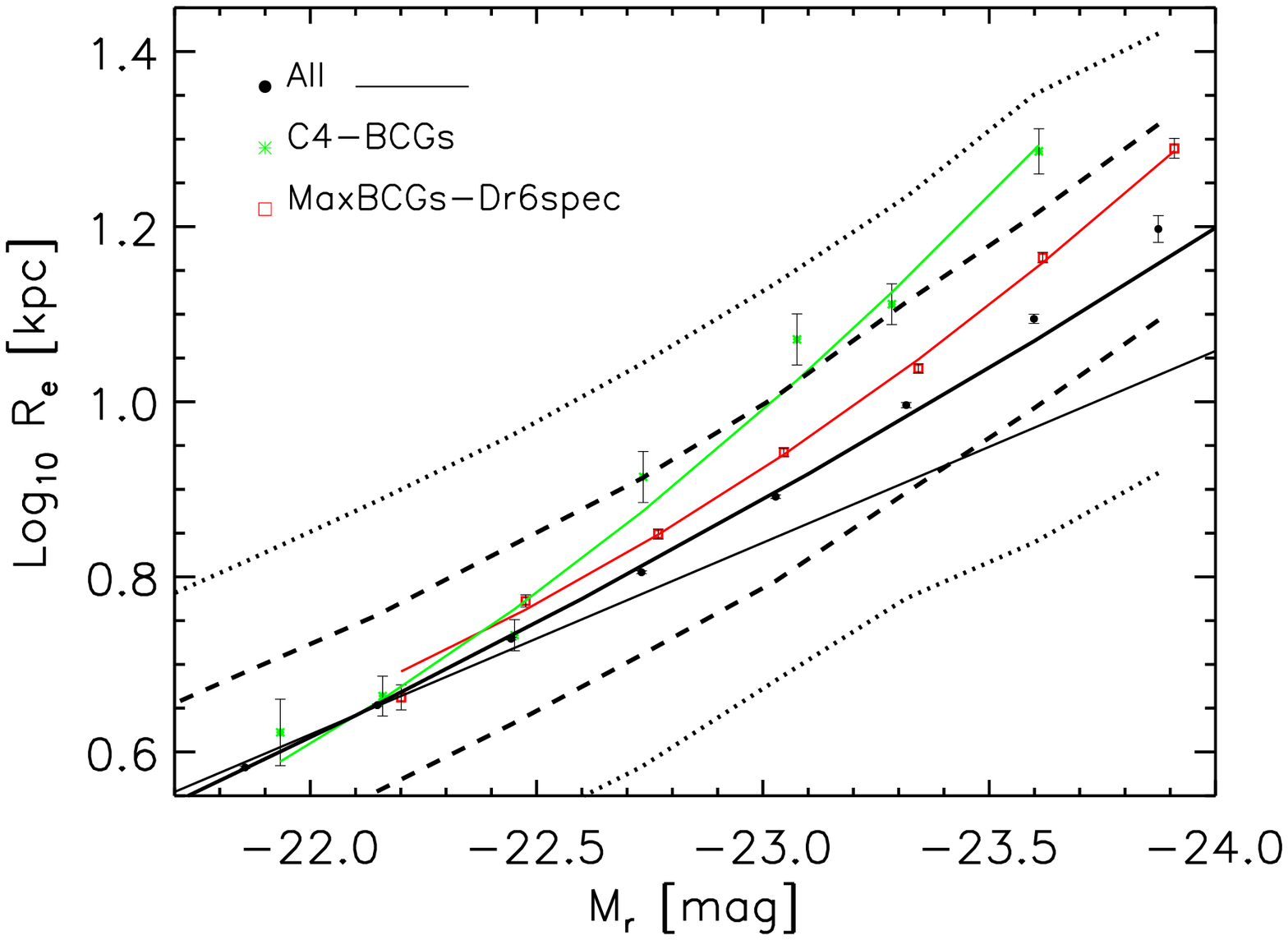}
 \includegraphics[width=0.475\hsize]{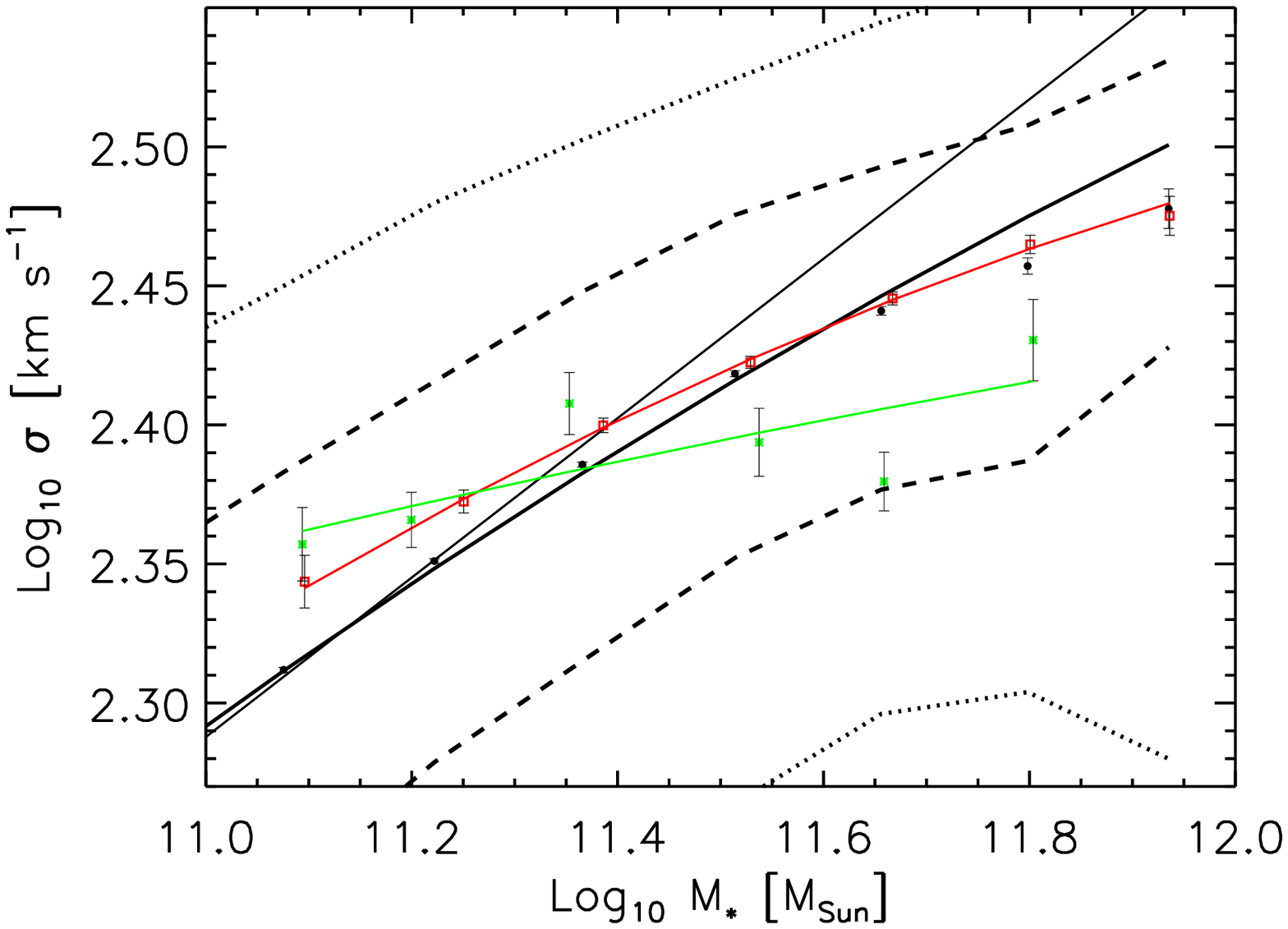}
 \includegraphics[width=0.475\hsize]{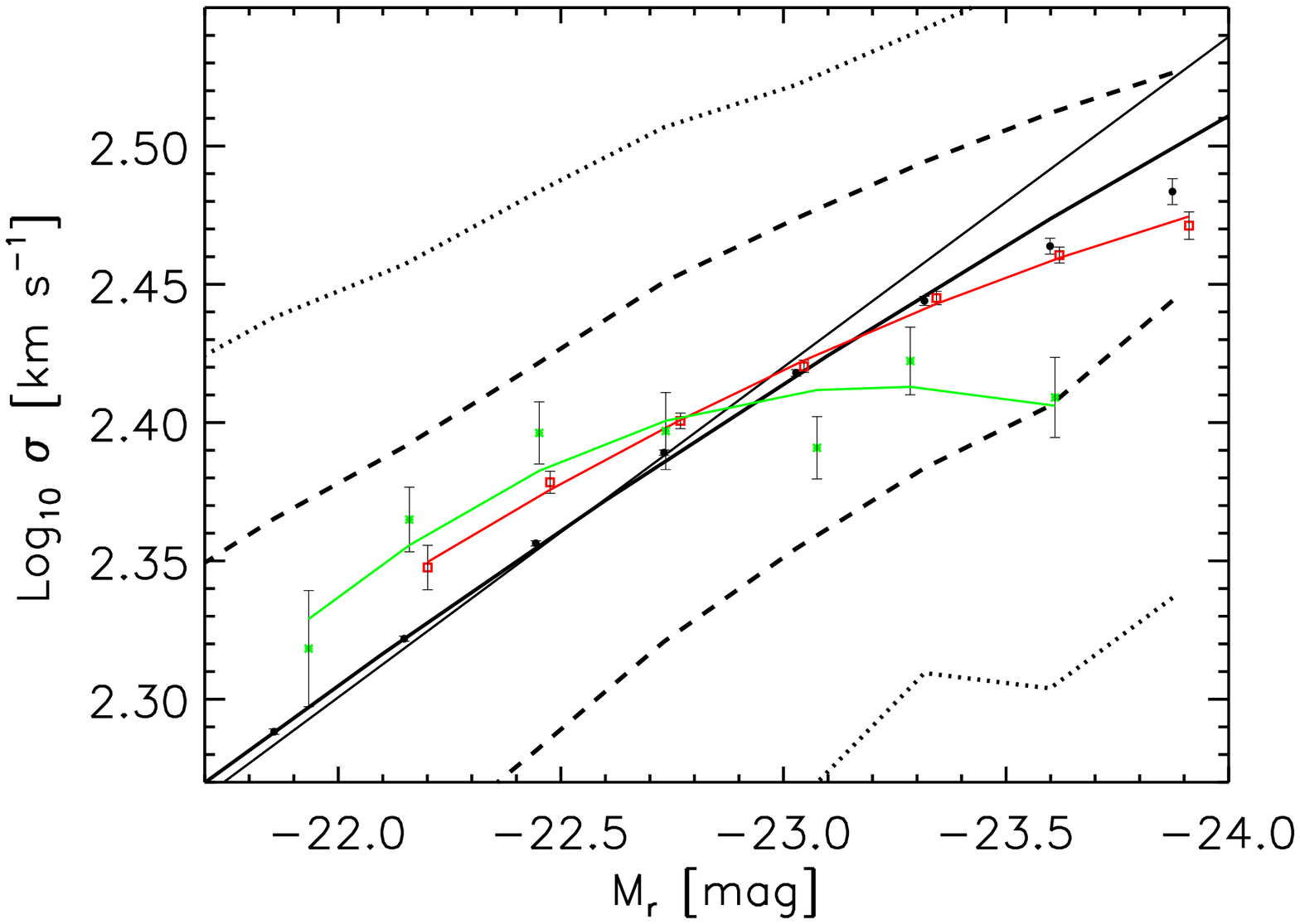}
 \caption{Size-luminosity and stellar mass relations for BCGs 
        in the C4 (stars) and MaxBCG (open squares) samples (top), 
        and similarly when size is replaced by velocity dispersion (bottom). 
        Filled circles 
        show the median value and its uncertainty for the bulk of 
        the early-type population; thin and thick black solid lines 
        show the linear and quadratic fits from Table~1 of 
        Hyde \& Bernardi (2009), respectively. Dashed and dotted curves show 
        the regions which enclose 68\% and 95\% of the objects. 
        At fixed $L$ or $M_*$, the C4 BCGs are larger than MaxBCGs, 
        and both are larger than the mean relation traced by the 
        bulk of the population.  The objects with largest sizes 
        have the smallest velocity dispersions.}
 \label{c4maxBCGbulk}
\end{figure*}

\begin{figure*}
 \centering
 \includegraphics[width=0.475\hsize]{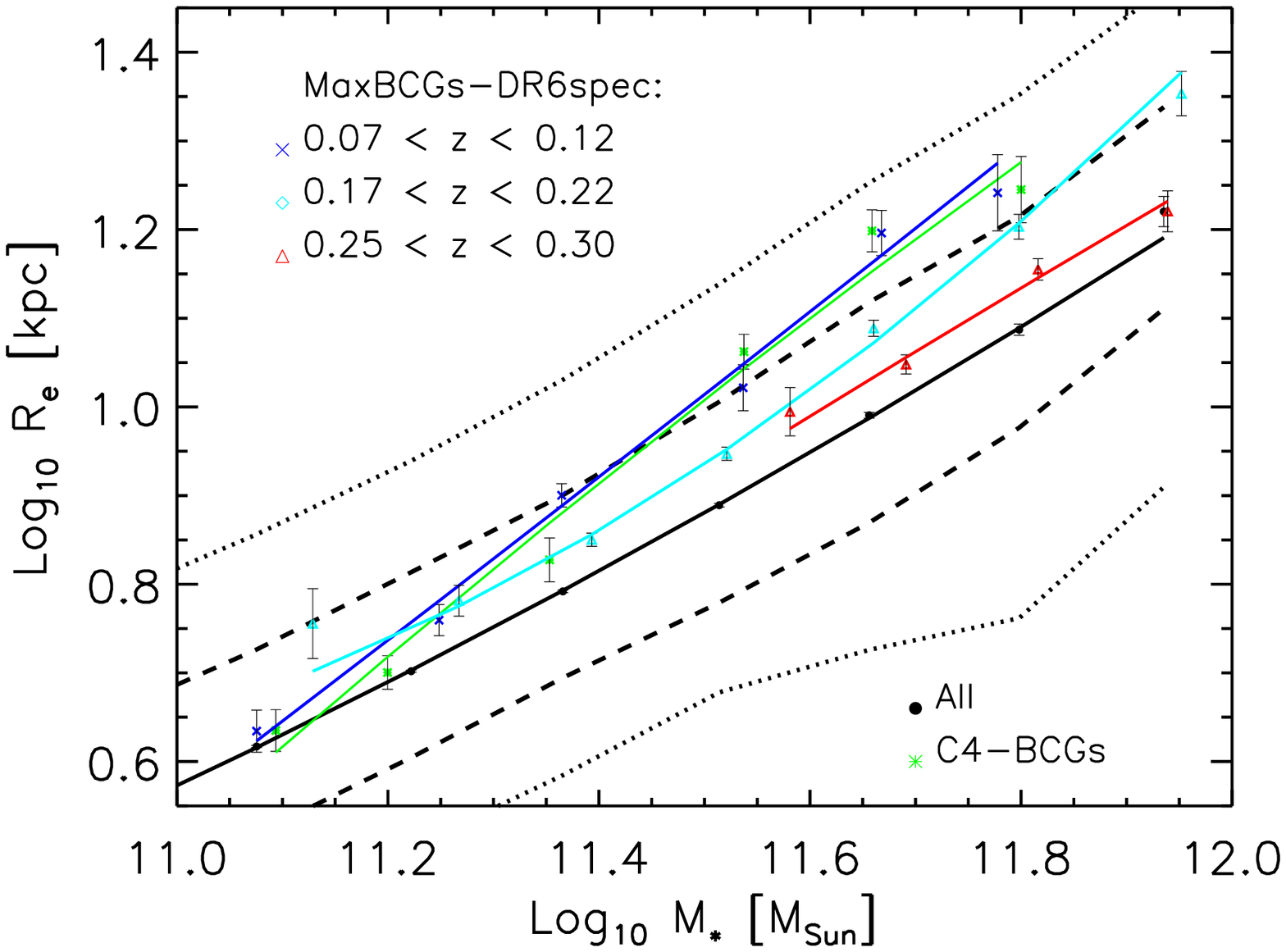}
 \includegraphics[width=0.475\hsize]{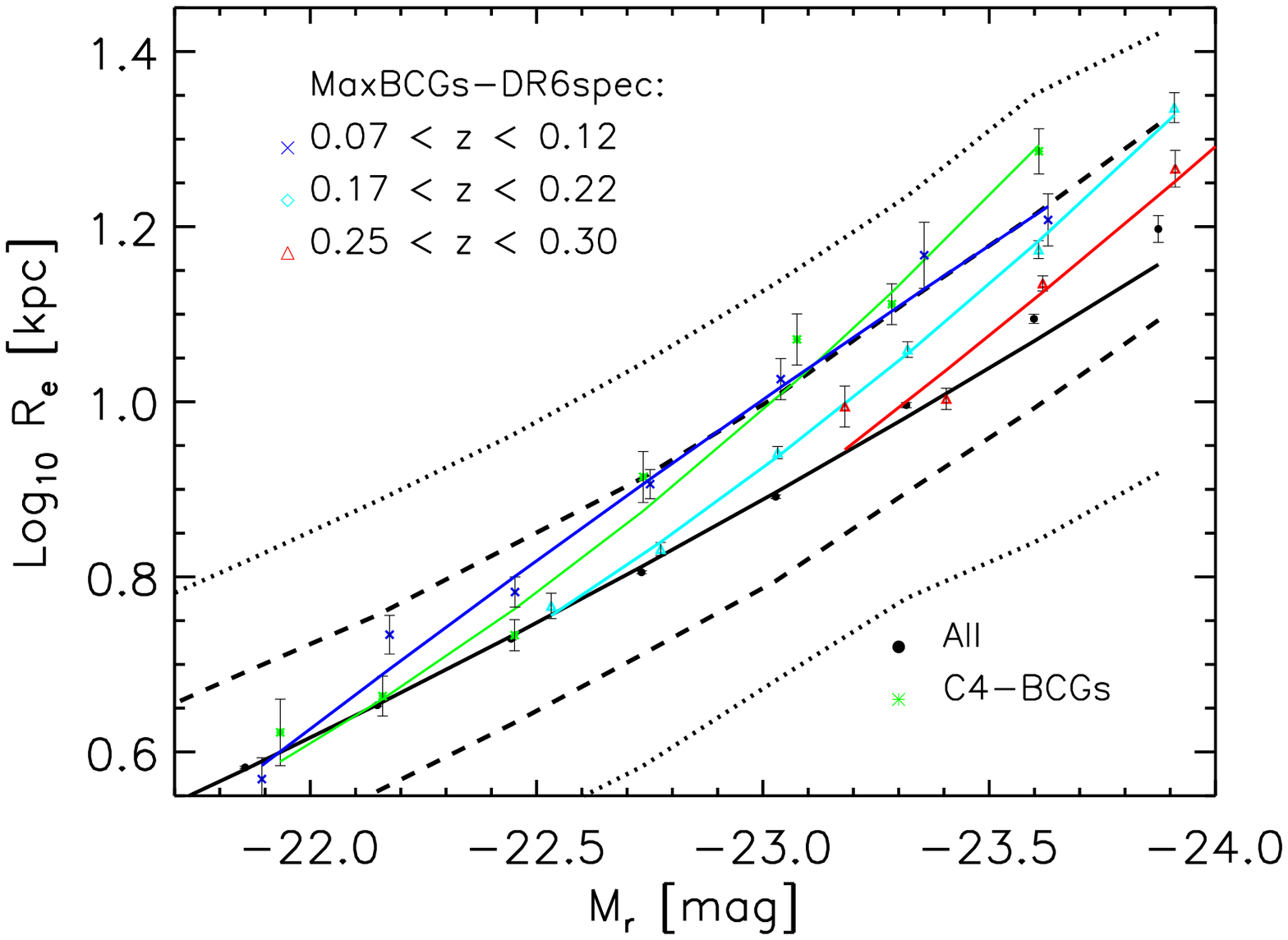}
 \includegraphics[width=0.475\hsize]{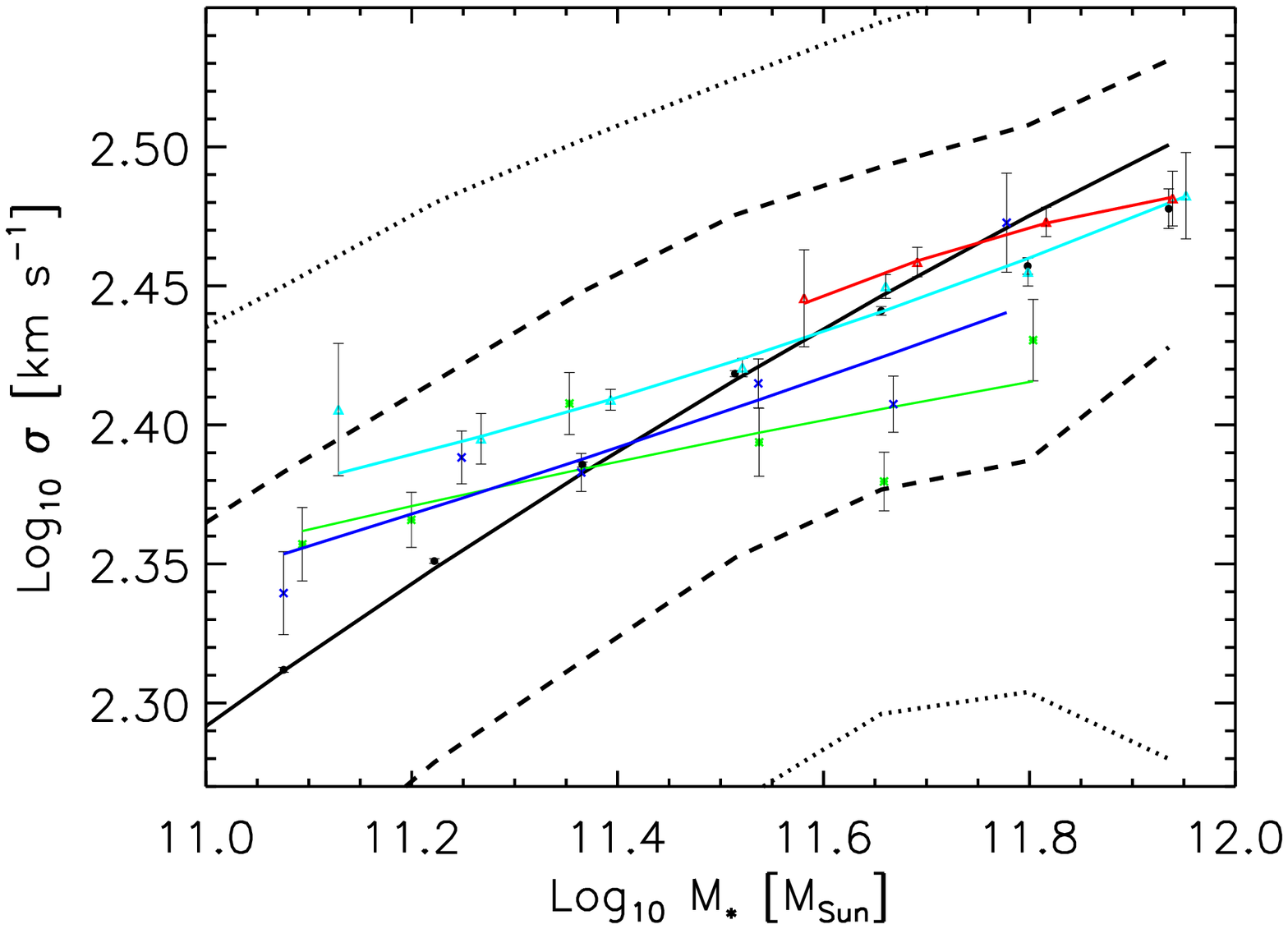}
 \includegraphics[width=0.475\hsize]{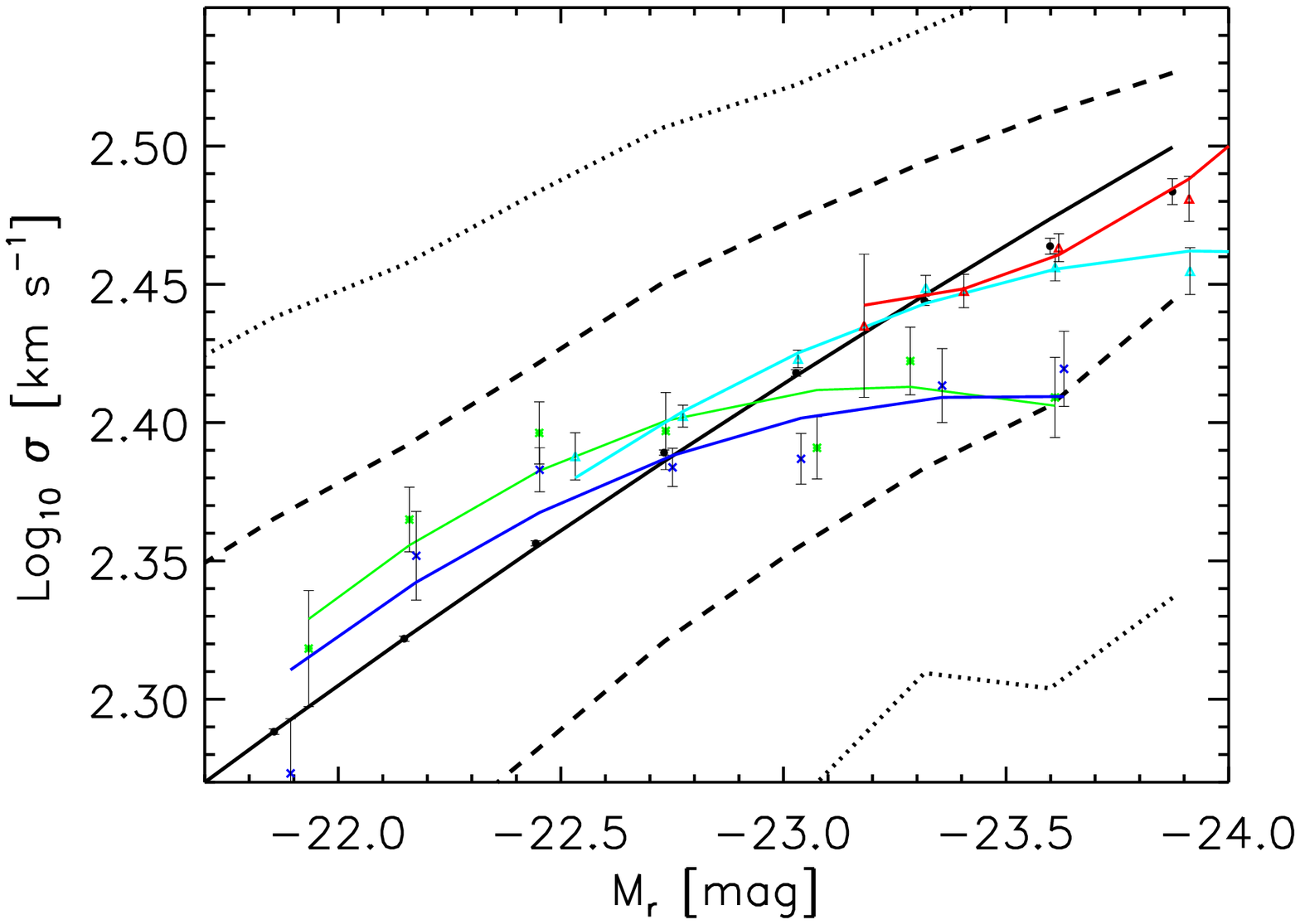}
 \caption{Same as previous figure, but now for objects from the 
        MaxBCG sample only, subdivided into redshift bins as 
        indicated.  
        At fixed luminosity or stellar mass, the BCGs in the lowest 
        redshift bin are larger and have smaller velocity dispersions.}
 \label{maxBCGonly}
\end{figure*}

\section{The size-luminosity and size-stellar mass relations}\label{evolve}
The main goal of this section is to present measurements of the 
correlation between (restframe) size and (evolution corrected) 
luminosity in a few redshift bins.  (The measured luminosities are 
corrected for known problems associated with the SDSS sky subtraction 
algorithm following Equation~2 in Hyde \& Bernardi 2009.  They are 
then corrected for evolution, by adding $0.9z$ to all absolute 
magnitudes.  The sizes are corrected for the known sky subtraction problems
and for the fact that early-type sizes depend on wavelength 
following Equations~3 and 6 in Hyde \& Bernardi 2009, respectively.)  
Inferences about evolution in this correlation depend upon how much 
one believes that the effects of luminosity evolution have been 
removed.  For this reason, one might have thought it preferable to 
study the correlation between size and stellar mass.  In what follows, 
we will show these relations side-by-side, but emphasize that because 
the SDSS is magnitude limited, evolution in correlations with $M_*$ 
should be interpretted carefully.  Appendix~A discusses why.  
Appendix~B compares the BCG size-luminosity relation we find here 
with other determinations in the recent literature.  

Note that although our $M_*$ estimates are from Gallazzi et al. (2005), 
we correct them slightly to account for the fact that they actually 
come from multiplying an estimate of $M_*/L$ by the observed 
luminosity.  The $L$ used by Gallazzi et al. did not account for 
the SDSS sky subtraction problem, so we divide their $M_*$ 
estimates by the same correction factor we used for the magnitudes 
(Equation~4 in Hyde \& Bernardi 2009).

\subsection{Abnormally large sizes}

The sizes and luminosities of early-type galaxies are correlated:  
 the mean size increases with luminosity as $R\propto L^{0.6}$ 
(e.g. Hyde \& Bernardi 2009).  
However, BCGs follow a steeper relation:  $R\propto L$ (see Appendix B), 
and it has long been argued that this is evidence for formation histories 
that are dominated by dry mergers.  The argument is not so 
straightforward, however.  This is because objects with small values 
of $\sigma$ are not BCGs, so BCGs have a narrower distribution in 
$\sigma$ than the bulk of the population.  
However, for the bulk of the population, the $R-L$ correlation 
at fixed $\sigma$ is considerably steeper, $R\propto L^{0.9}$, 
than when averaged over all $\sigma$ (Bernardi et al. 2003; 
Bernardi et al. 2008).  So one may ask if the steeper relation 
for BCGs can be attributed to the fact that they are biased to 
larger $\sigma$.  

Figure~\ref{LRsigbin} presents a direct comparison:
the dashed and dot-dashed lines in the top panel show the $R_e-L$ 
relation for the bulk and for the early-type BCGs, and the various 
solid lines show this relation for narrow bins in $\sigma$:  
the lines are offset to smaller $R_e$ as $\sigma$ increases.  
The bottom panel shows these relations after subtracting-off 
the dashed line.  This shows clearly that, except for the smallest 
bin in $\sigma$, the other relations are all steeper.  
However, the BCGs are steeper still:  the $R_e-L$ relation of 
BCGs is steeper than that of the bulk, even at fixed $\sigma$.  
In the next subsections, we study other evidence that BCGs 
are a different population.

\subsection{Evidence for evolution}

The top panels in Figure~\ref{c4maxBCGbulk} compare the 
size-luminosity and stellar mass relations for the BCGs in the 
C4 and MaxBCG catalogs, with the relations traced out by the bulk 
of the early-type galaxy population.  At fixed $L$, the C4 BCGs 
are larger than MaxBCGs, and both are larger than the mean relation 
traced by the bulk of the population.  These differences are most 
pronounced for the most luminous objects; there is essentially 
no effect at $M_r>-22.5$ or $\log_{10} (M_*/M_\odot)<11.2$.  

\begin{figure}
 \centering
 \includegraphics[width=0.95\hsize]{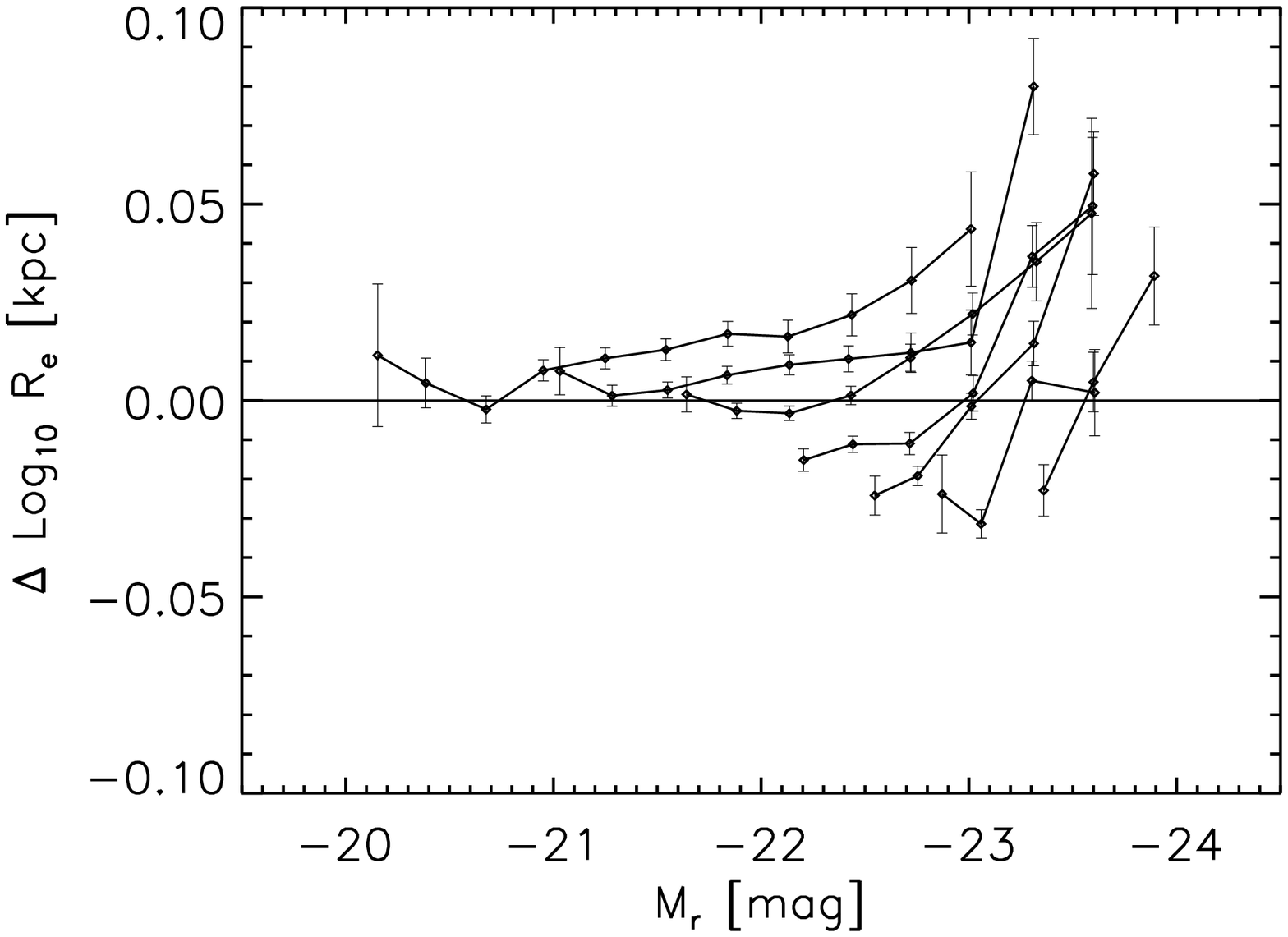}
 \includegraphics[width=0.95\hsize]{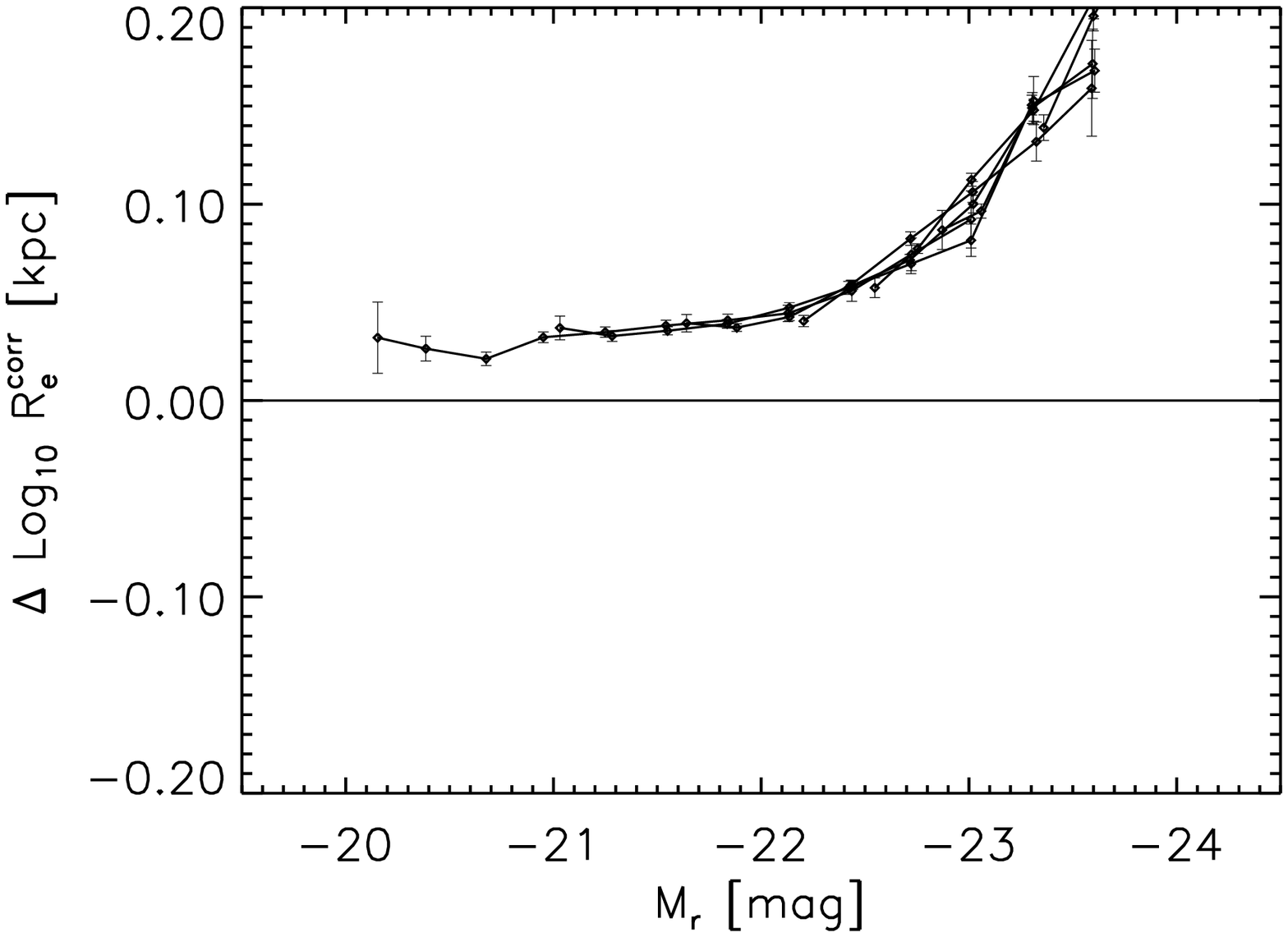}
 \caption{Residuals from the size-luminosity relation 
      $\Delta \log_{10} R_e \equiv \log_{10}(R_e/{\rm kpc}) 
                                  - (4.72 + 0.63\,M_r + 0.02\,M_r^2)$, 
      for the bulk of the early-type galaxy population in different
      redshift bins:  
      $0.07< z\le 0.1$, $0.1< z\le 0.13$, $0.13< z\le 0.16$, 
      $0.16< z\le 0.19$, $0.19< z\le 0.22$, $0.22< z\le 0.25$ and 
      $z > 0.25$.  At $M_r > -22$, there is a tendency for the objects 
      at higher redshift to have slightly smaller $\sigma$. 
      At $M_r < -22$, the difference in size between two different 
      redshift bins increases for brighter galaxies, in agreement 
      with the evolution seen for BCGs in Figure~\ref{maxBCGonly}.  
      Bottom panel shows the result of applying the corrections in 
      Equations~1 and 3.  }
 \label{bulkRz}
\end{figure}

\begin{figure}
 \centering
 \includegraphics[width=0.95\hsize]{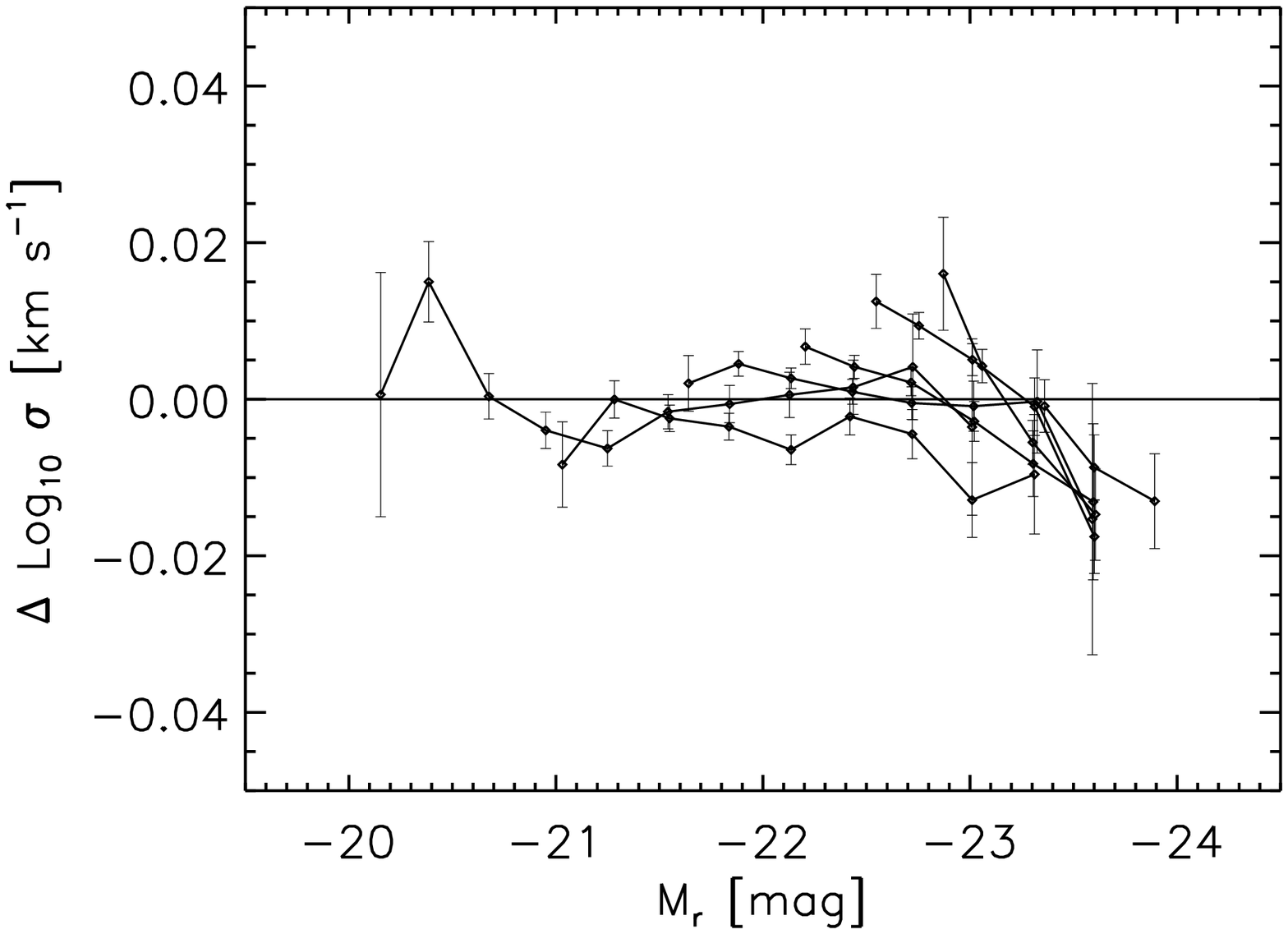}
 \includegraphics[width=0.95\hsize]{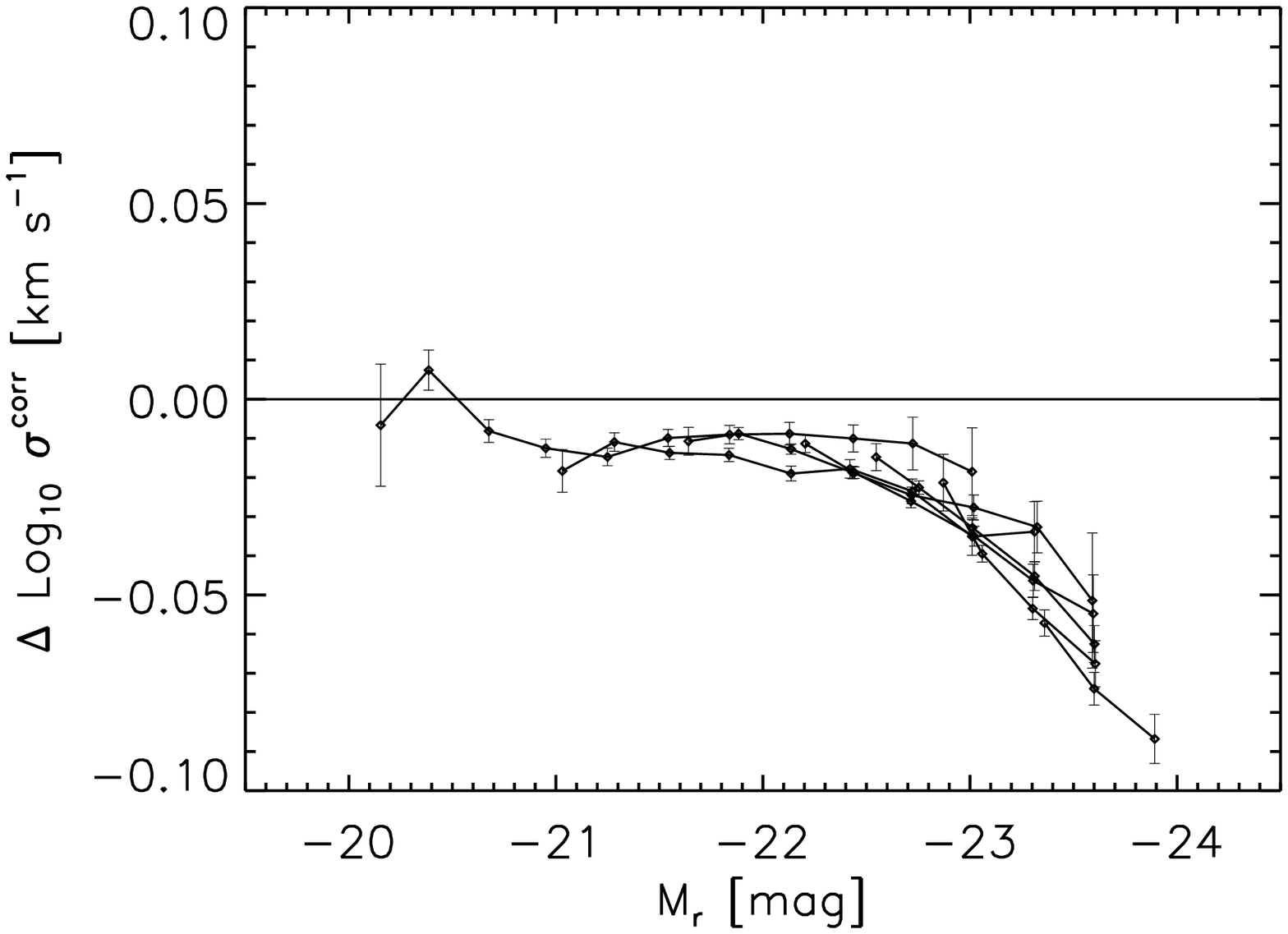}
 \caption{Same as the previous figure, but now for the 
      $\sigma$-luminosity relation:  
      $\Delta \log_{10} \sigma \equiv \log_{10}(\sigma/{\rm km s^{-1}}) - (-2.97 - 0.37\,M_r - 0.006\,M_r^2)$, 
      for the bulk of the early-type galaxy population in different
      redshift bins. Bottom panel shows the result of applying the 
      corrections in Equations~2 and 4.  
 }
 \label{bulkVz}
\end{figure}

The bottom panels show that complementary differences are seen 
when size is replaced with velocity dispersion:  at the brightest 
luminosities ($M_r<-23$) C4 BCGs have the smallest velocity 
dispersions.  While the trends at the bright end are the ones of 
most interest in the present context, we note that, at fainter 
luminosities, BCGs tend to have larger $\sigma$ for their $L$ than 
the bulk of the population.  

That BCGs have larger sizes and smaller velocity dispersions than 
the bulk is no surprise -- what {\em is} surprising is the significant 
difference between the two BCG samples.  Although it is possible that 
this is related to the fact that the two samples span different 
redshift ranges, it is also possible that systematic differences 
between how the catalogs were assembled are to blame.

To eliminate the second possibility, Figure~\ref{maxBCGonly} 
shows a similar analysis, but now restricted to MaxBCG objects 
only.  Since this sample is relatively large, we divided it 
into subsamples in redshift:  $0.07<z<0.12$, $0.17<z<0.22$ and
$0.25<z<0.30$.  This shows clearly that, even within the MaxBCG 
catalog itself, the lower redshift BCGs tend to have larger sizes 
and smaller velocity dispersions than their higher redshift 
counterparts of similar luminosity or stellar mass.  
Moreover, the MaxBCGs in the lowest redshift bin tend to 
follow similar scaling relations to those defined by the C4 BCGs.  

Finally, Figure~\ref{bulkRz} shows the $R_e-L$ relation for the 
bulk of the early-type galaxy population. (Recall that the measured 
luminosities have been corrected for evolution by adding $0.9z$ 
to all absolute magnitudes, and the sizes are corrected for the 
fact that early-type sizes depend on wavelength.) 
Each set of symbols shows data from a number of redshift bins:  
$0.07< z\le 0.1$, $0.1< z\le 0.13$, $0.13< z\le 0.16$, $0.16< z\le 0.19$,
$0.19< z\le 0.22$, $0.22< z\le 0.25$ and $z > 0.25$.  
To reduce the range of sizes, in the top panel we have subtracted 
out a fiducial relation to better see if there is any evolution: 
we actually show
 $\Delta \log_{10}R_e \equiv 
         \log_{10}(R_e/{\rm kpc}) - (4.72 + 0.63\,M_r + 0.02\,M_r^2)$ 
(from Table~1 of Hyde \& Bernardi 2009).  
The top panel in Figure~\ref{bulkVz} shows a similar analysis 
of the velocity dispersions, for which 
 $\Delta \log_{10}\sigma \equiv 
         \log_{10}(\sigma/{\rm km s^{-1}}) - (-2.97 - 0.37\,M_r - 0.006\,M_r^2)$ 
(from Table~1 of Hyde \& Bernardi 2009).

There is a hint that the higher redshift objects have smaller sizes.
At $M_r < -22$, the difference in size between two different redshift 
bins increases for brighter galaxies, in agreement with the evolution 
seen for BCGs (Figure~\ref{maxBCGonly}).
Thus, at the bright end ($M_r<-22$), we find that the evolution depends 
on the luminosity of the galaxy:   
the sizes evolve as $(1+z)^{0.7(M_r+21)}$ and the velocity dispersions 
as $(1+z)^{-0.2(M_r+21)}$.  At fainter luminosities ($M_r>-22$), the 
evolution is weaker; we approximate it as $(1+z)^{-0.7}$ and $(1+z)^{0.2}$.  
Hence, to correct the sizes and velocity dispersions to $z=0$ one 
could use: 
\begin{eqnarray}
 \label{correctMbright}
 \log_{10} \left(\frac{R_e^{\rm corr}}{\rm kpc}\right) 
  &=& \log_{10} \left(\frac{R_e}{\rm kpc}\right) - 0.7(M_r+21)\log(1+z)\\
 \log_{10} \left(\frac{\sigma^{\rm corr}}{{\rm kms}^{-1}}\right) 
  &=& \log_{10} \left(\frac{\sigma}{{\rm kms}^{-1}}\right) + 0.2(M_r+21)\log(1+z)
 \label{correctMbrightV}
\end{eqnarray}
if $M_r < -22$ and by 
\begin{eqnarray}
 \log_{10} \left(\frac{R_e^{\rm corr}}{\rm kpc}\right) 
  &=& \log_{10} \left(\frac{R_e}{\rm kpc}\right) + 0.7\log(1+z)\\
 \log_{10} \left(\frac{\sigma^{\rm corr}}{{\rm kms}^{-1}}\right) 
  &=& \log_{10} \left(\frac{\sigma}{{\rm kms}^{-1}}\right) - 0.2\log(1+z)
 \label{correctMfaint}
\end{eqnarray}
if $M_r > -22$.
The bottom panels in the two figures show the result of applying 
these corrections.  We show below that the scaling at the bright end 
is slightly smaller than the luminosity dependent 
evolution in the sizes of our BCG sample which goes as $(1+z)^{0.85(M_r+21)}$.

\begin{figure*}
 \centering 
\includegraphics[width=0.475\hsize]{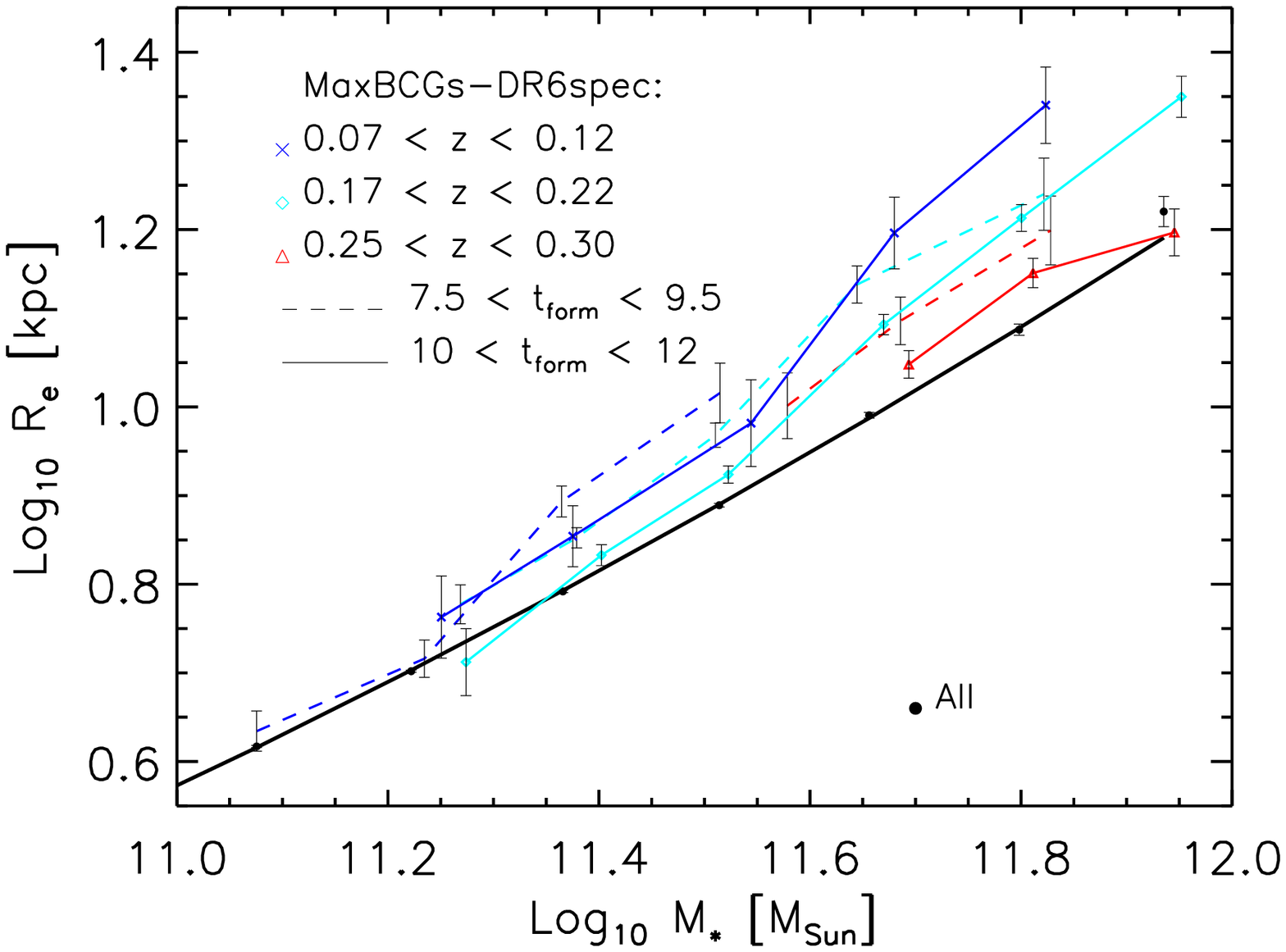}
\includegraphics[width=0.475\hsize]{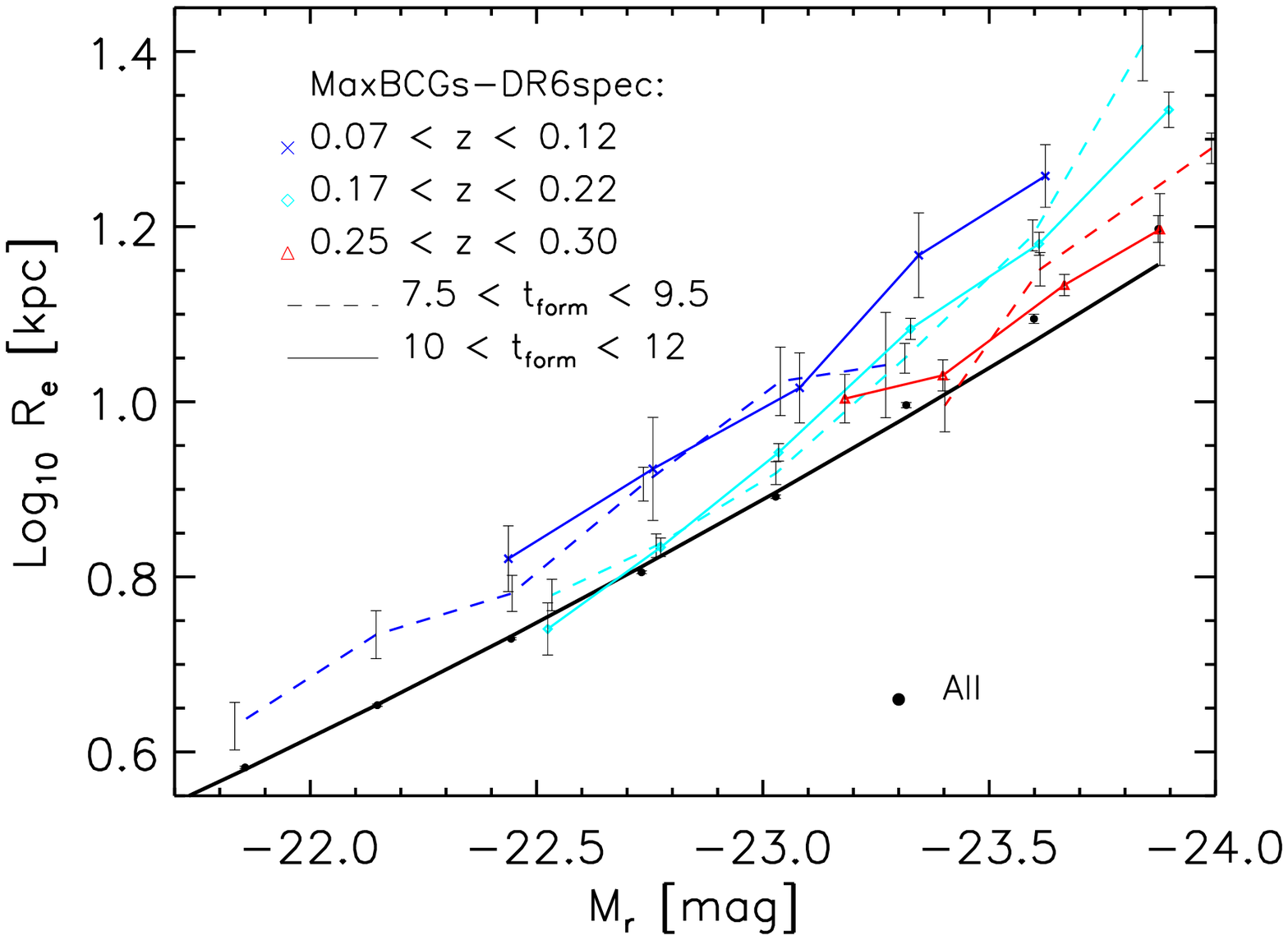}
\includegraphics[width=0.475\hsize]{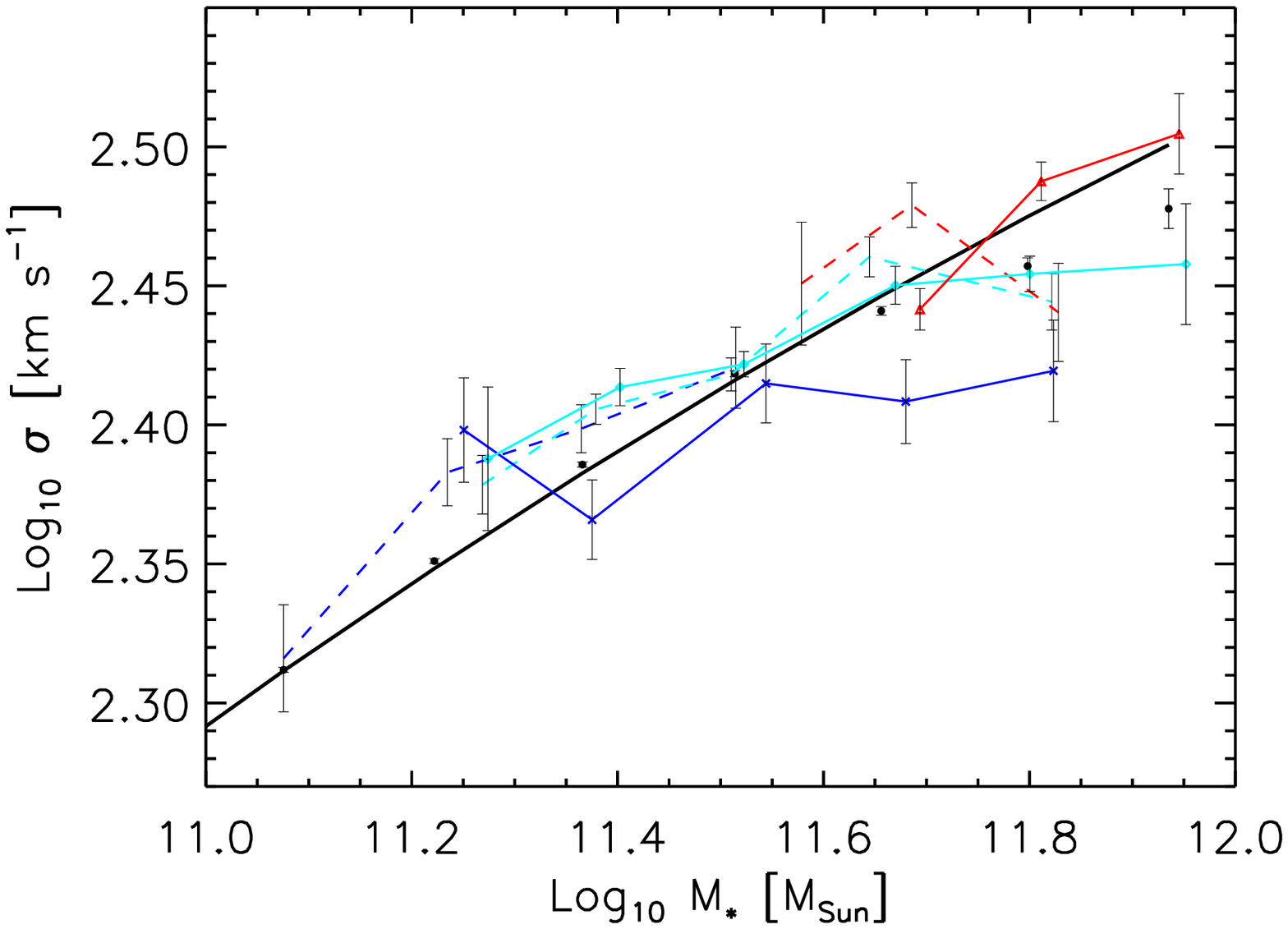}
\includegraphics[width=0.475\hsize]{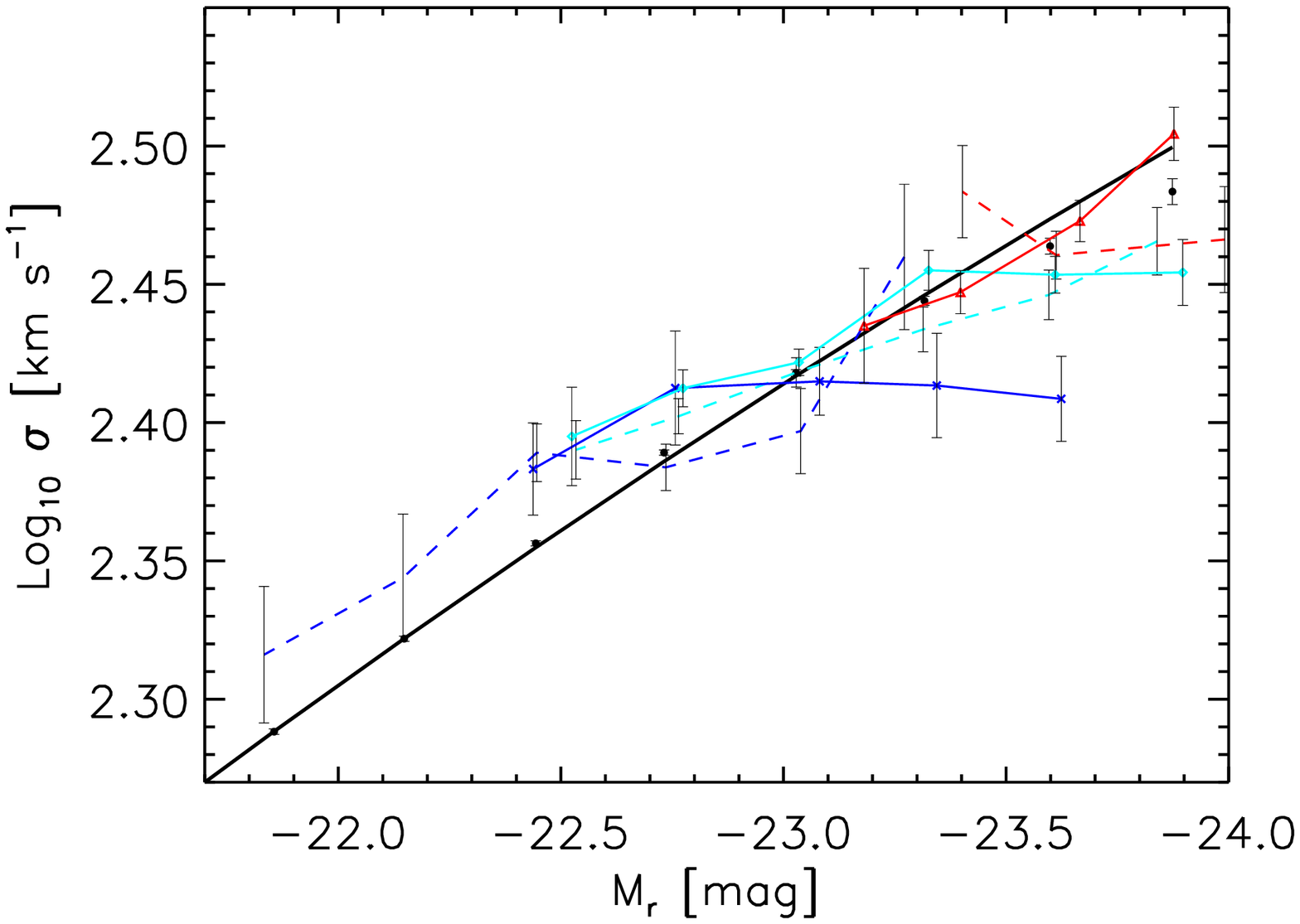}
 \caption{Same as Figure~\ref{maxBCGonly}, but now the objects in each 
      redshift bin are subdivided by the lookback time to when their 
      stars formed.  At fixed formation time, objects at low redshift 
      have larger sizes and smaller velocity dispersions than their 
      higher redshift counterparts of similar $M_*$ (left) or 
      $L$ (right).}
 \label{age2}
\end{figure*}

\subsection{Dependence on age and formation time}
If we are seeing evolution, then it is interesting to ask if this 
depends on the age or formation time of the stellar population.  
E.g., the simplest monolithic collapse models predict that, for 
fixed formation time, there should otherwise be no dependence on 
age.  To address this, we use age estimates of the stellar populations 
in these galaxies (from Gallazzi et al. 2005); these, with the 
observed redshifts, yield the lookback time to when the stars 
formed.  (Because the age and $M_*$ estimates have significant 
uncertainties, and they are correlated, it is important to use ages 
that are output from the same models which estimate $M_*$.)

Figure~\ref{age2} is similar to Figure~\ref{maxBCGonly}, but now 
the objects in each redshift bin have been divided into two subgroups, 
based on the formation time of the stars.  These two bins correspond 
approximately to $z_{\rm form}\sim 2.5$ (solid lines) and $1.25$ 
(dashed lines).  

We begin with a comparison of the solid lines in the panels on 
the left.  These show that, for fixed formation time, older objects 
have larger sizes (top left) and smaller velocity dispersions 
(bottom left) than younger objects of the same stellar mass.  
The same is true of the dashed lines in these panels:  
size increases and velocity dispersion decreases as the galaxy 
population ages.  The increase in size is more easily accomodated 
in dry merger models, and the decrease in velocity dispersion 
suggests that these mergers were minor.

\begin{figure}
 \centering 
\includegraphics[width=0.95\hsize]{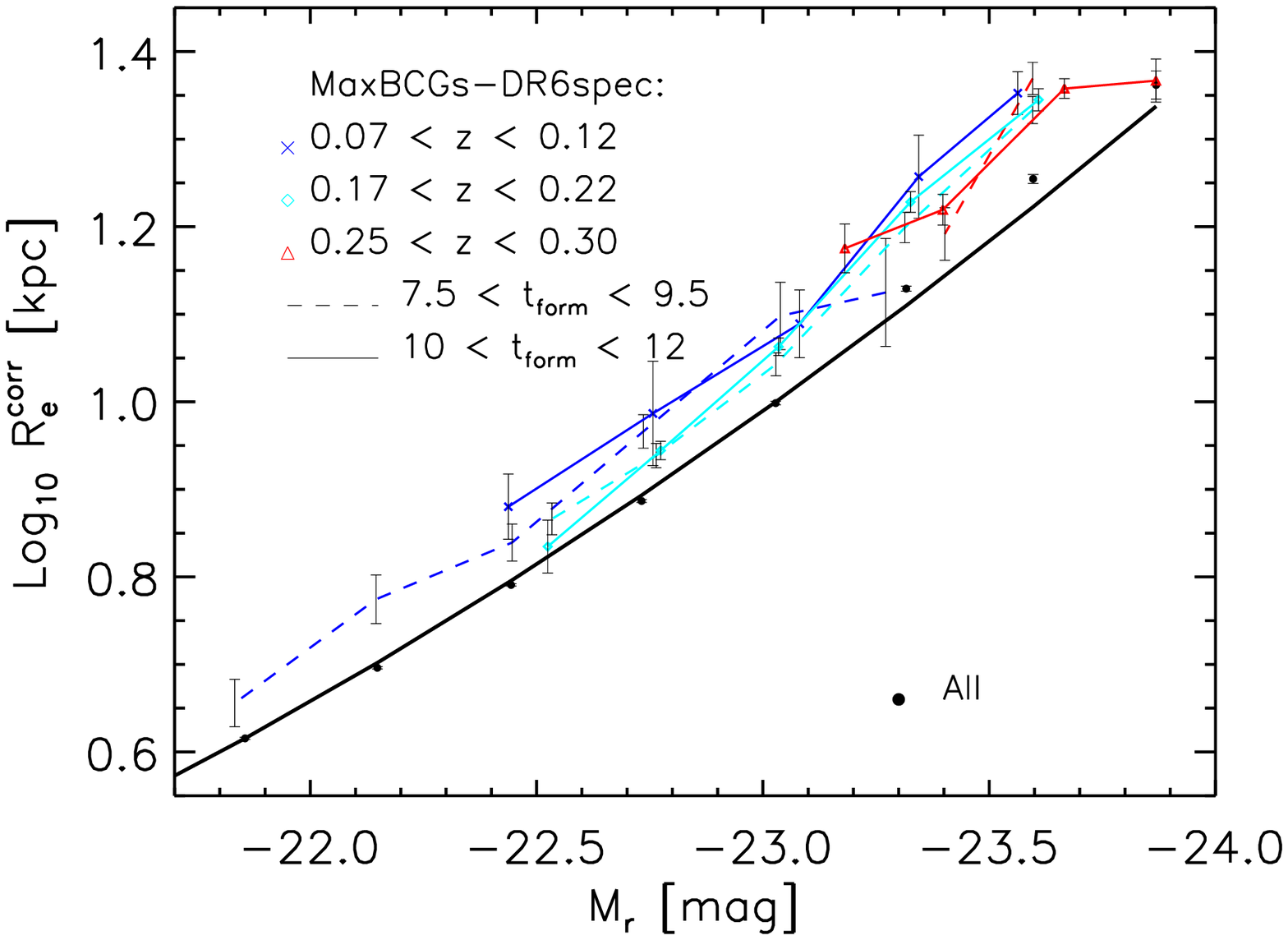}
\includegraphics[width=0.95\hsize]{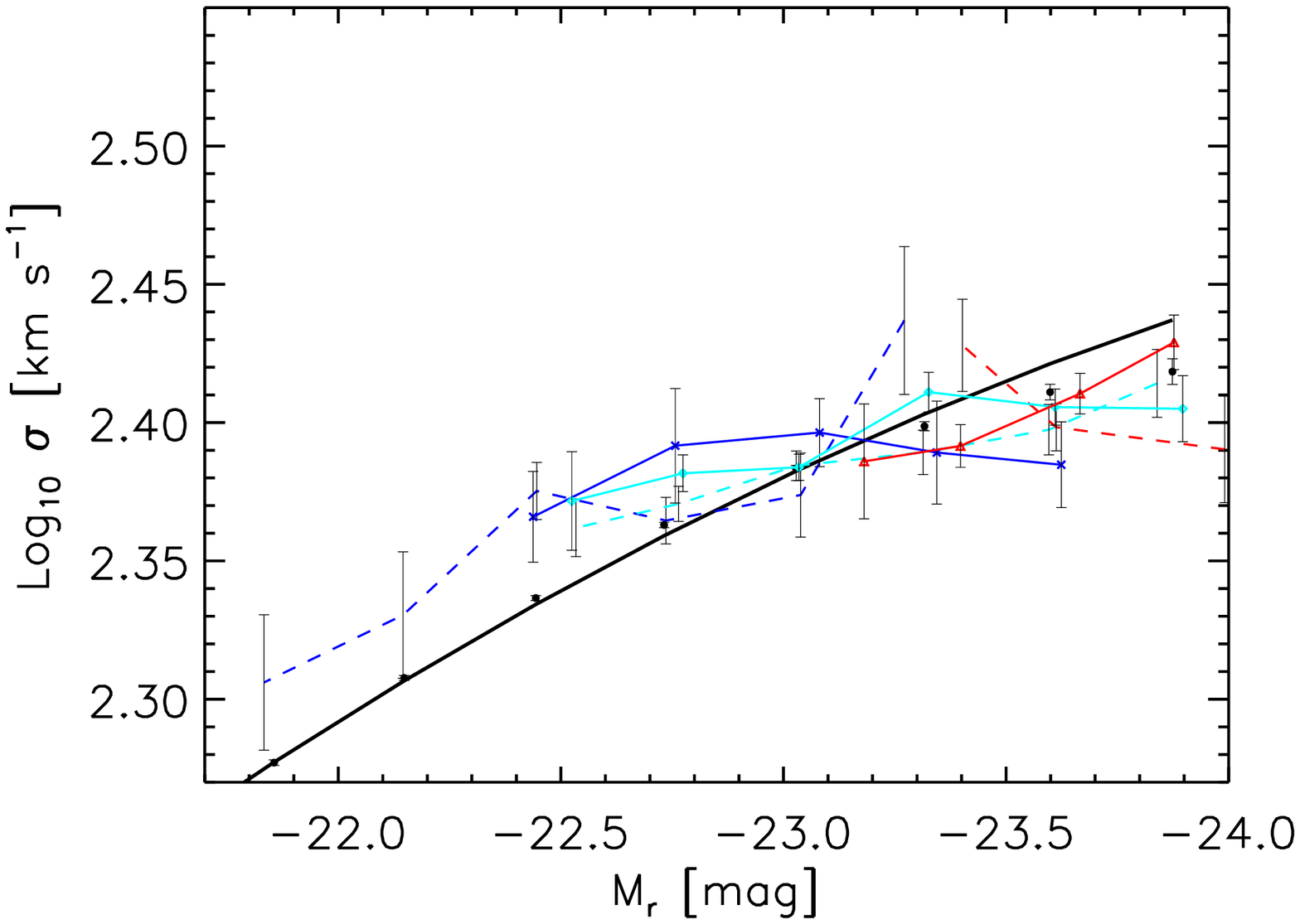}
 \caption{Same as right-hand panels of previous figure, but now after 
  correcting the sizes and velocity dispersions for evolution using 
  Equations~1 and 2.}
 \label{RVcorrected}
\end{figure}

Comparison of the dashed and solid curves for a given redshift 
of observation shows that, for a given stellar mass, the objects 
which formed more recently have larger sizes.  While this is 
qualitatively consistent with having formed when the Universe 
was less dense, the difference is much less than the factor of 
$\log_{10}(3.5/2.25)=0.2$~dex one might naively have expected.  
Similarly, the velocity dispersions of the objects which formed 
more recently are not much smaller than when the formation redshift 
was higher.  Presumably, this is because the sizes of the older 
objects have increased from their initial values, and the velocity 
dispersions have decreased (as suggested by comparing the solid 
lines with one another, and the dashed lines with one another).  

There are important qualitative differences when one uses 
luminosity rather than $M_*$, meaning that care must be taken when 
translating trends seen in plots with $L$ into trends with $M_*$.  
The top right panel shows that, at a given redshift of observation, 
the size-luminosity correlation does not depend on formation time 
(solid and dashed lines overlap), and the $\sigma-L$ relation 
does not either (bottom right panel).  This may be understood as follows.  
The panel on the right is obtained by shifting each galaxy 
in the panel on the left by its $(M_*/L)^{-1}$.  
In a model where the stars age passively (whatever the assembly 
history), the older population has a larger $M_*/L$:  
the expected difference in $M_*/L$ between the two age bins is 
about 0.1~dex ($M_*/L \propto t^{0.75}$ 
or so, where $t$ is the age of the population).  
So, if we start from the $R_e-M_*$ relation, then the solid 
curves in each redshift bin should shift towards the left, 
bringing them closer to the dashed ones.  

Finally, Figure~\ref{RVcorrected} shows the effect of correcting 
the sizes and velocity dispersions of our early-type BCGs using 
an equation similar to Equation~(\ref{correctMbright}) 
(since the size evolution of BCGs is better described by 
$(1+z)^{0.85(M_r+21)}$ we replace 0.7 with 0.85) and 
Equation~(\ref{correctMbrightV}).  
These corrections bring the curves associated with different 
redshifts into better agreement, suggesting that they have 
captured most of the evolution.  

We conclude that, at fixed luminosity or stellar mass, the high 
redshift BCGs are denser, with the effect being more pronounced 
for objects with the largest luminosities or stellar masses.  
The most straightforward interpretation of this observation is 
that {\it the sizes and velocity dispersions of luminous BCGs 
are evolving in a manner which is qualitatively consistent with 
assembly histories that are dominated by dissipationless mergers 
which are still happening at low redshift}; the large size of our 
sample has allowed a detection of this even though it spans only 
a small lookback time.

Before we conclude this section, we emphasize again that one must 
be cautious when replacing luminosity with stellar mass, since 
selection effects can complicate the measurement (as we illustrate 
in Appendix A).  Working with age and $M_*$ presents additional 
complications because errors in age and $M_*/L$ are correlated 
(see Appendix~\ref{ageM*} for further discussion).

\begin{figure*}
 \centering
 \includegraphics[width=0.475\hsize]{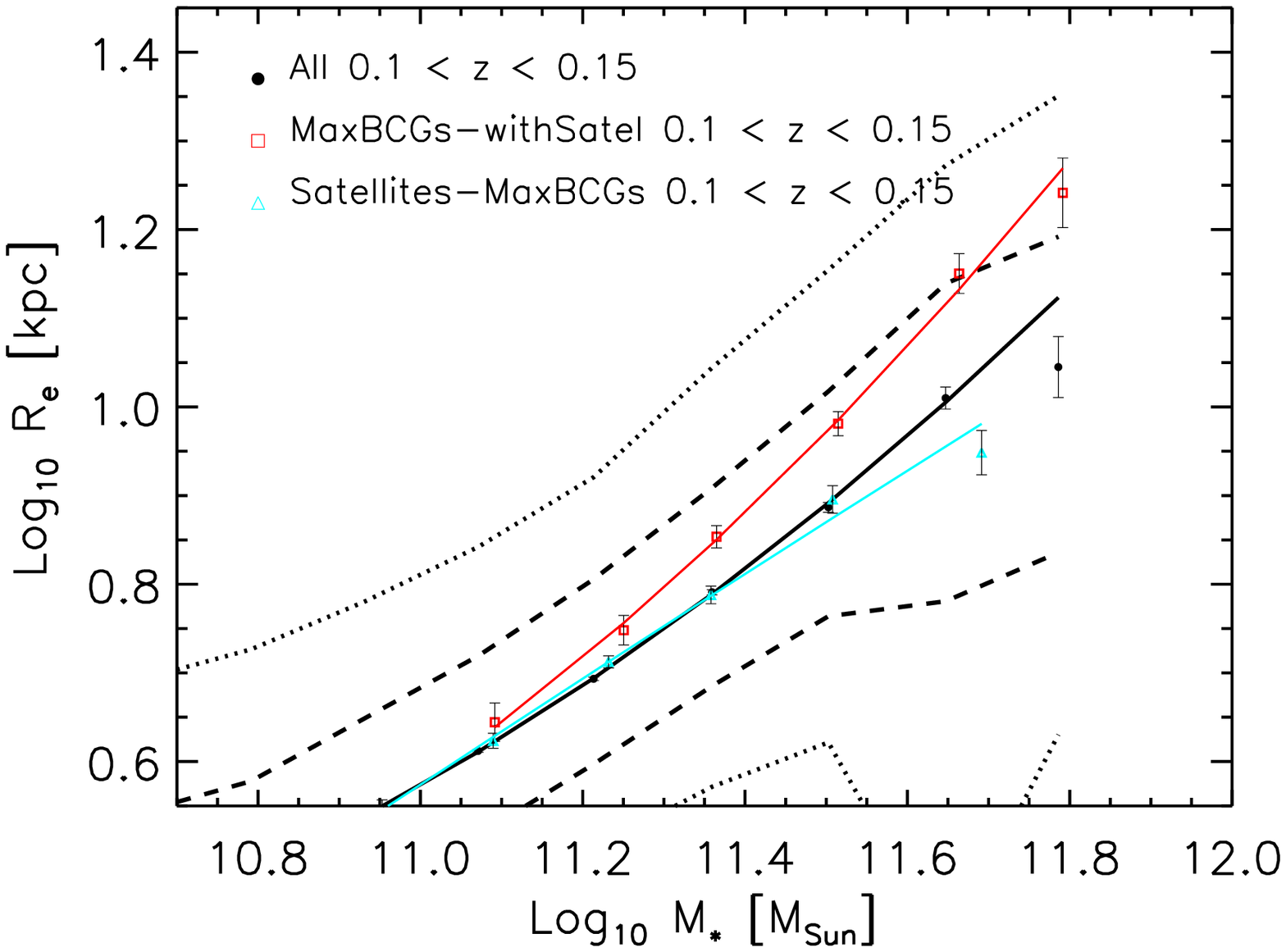}
 \includegraphics[width=0.475\hsize]{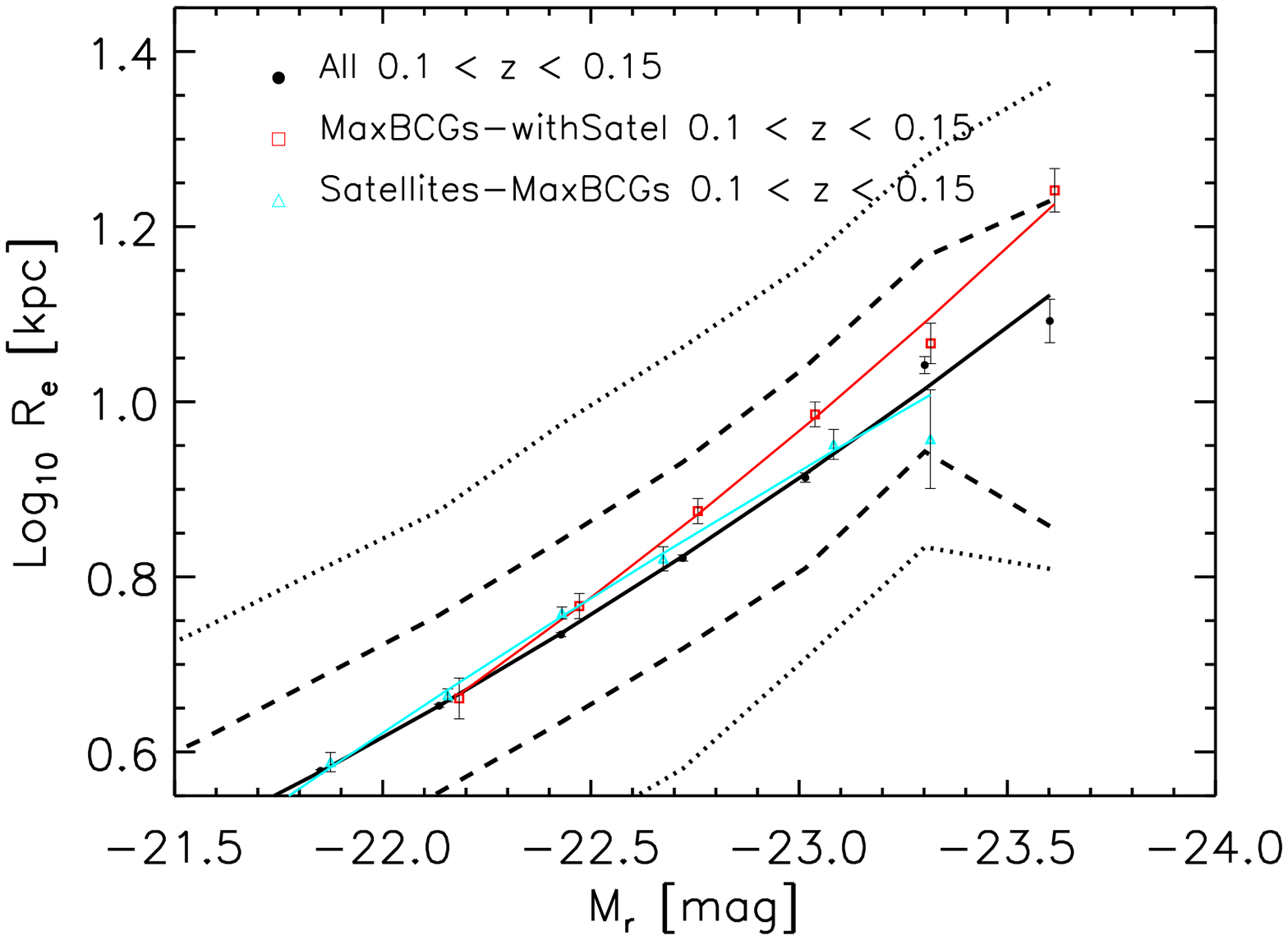}
 \includegraphics[width=0.475\hsize]{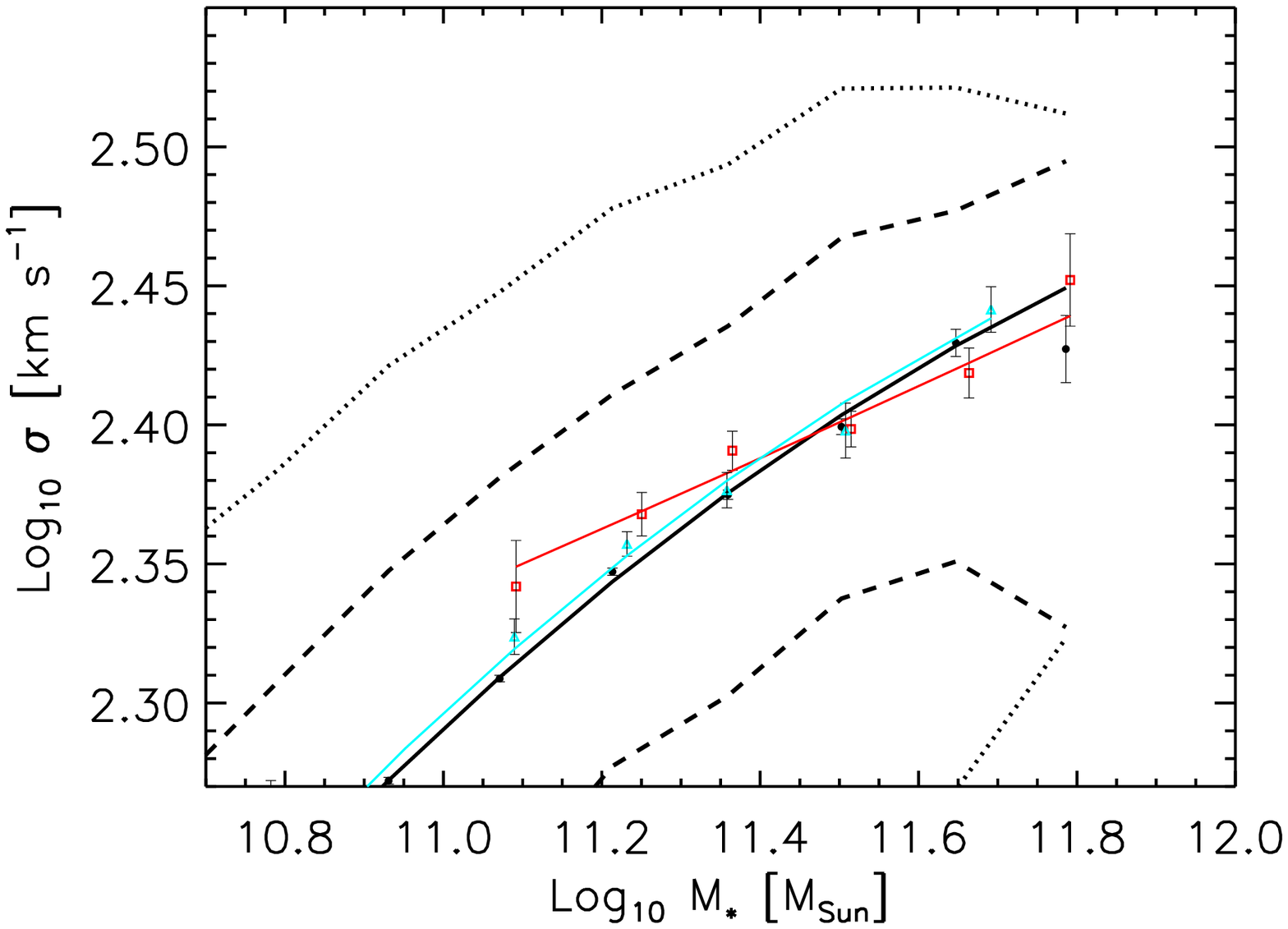}
 \includegraphics[width=0.475\hsize]{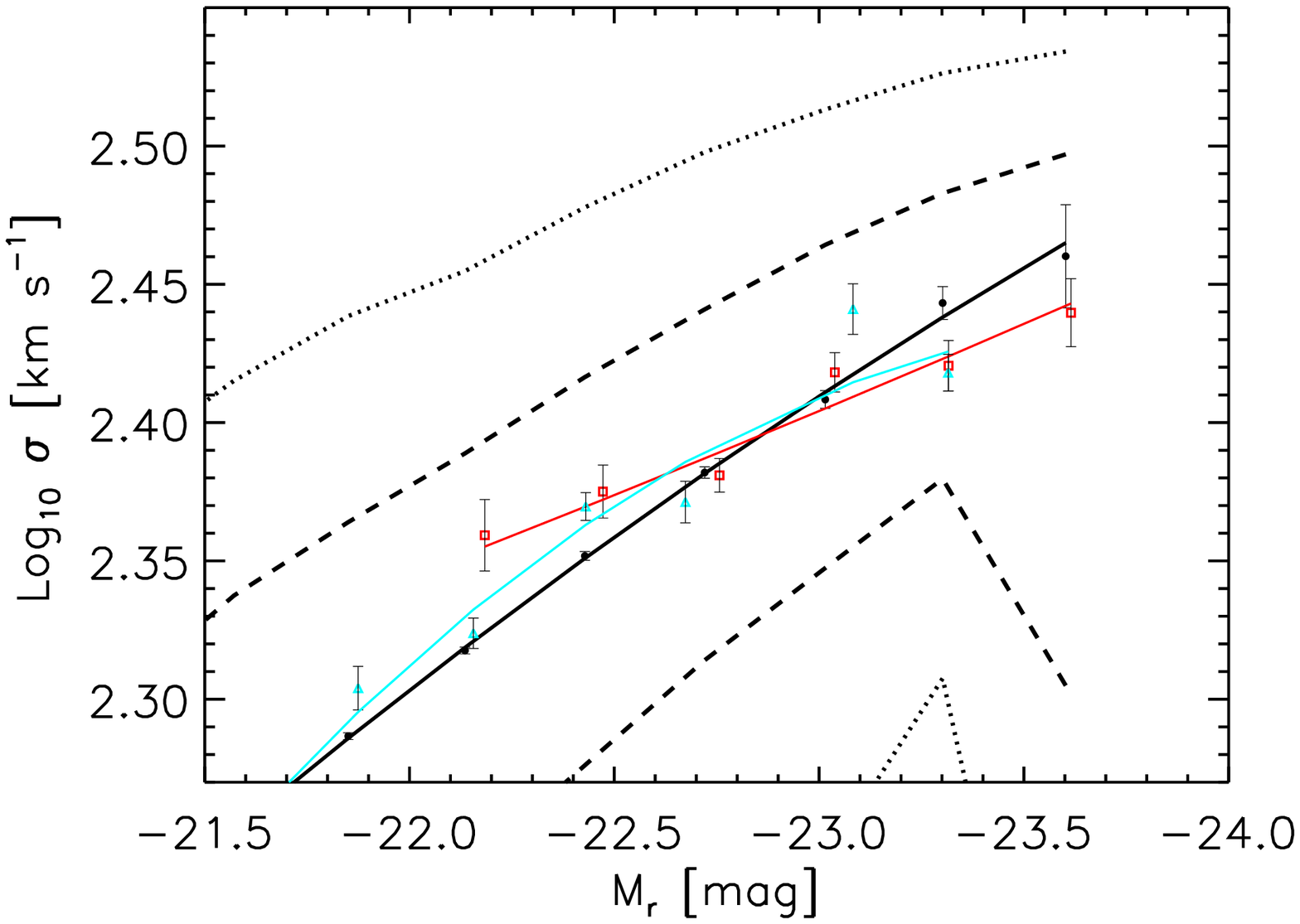}
 \caption{Comparison of sizes and velocity dispersions of BCGs 
          and non-central/satellite cluster galaxies of similar 
          luminosity and stellar mass, in a narrow redshift bin.}
 \label{cs}
\end{figure*}

\section{Centrals and satellites}\label{censat}
Early-type BCGs tend to have larger sizes than the bulk of the 
early-type galaxy population.  
So one might wonder if the large sizes of BCGs are something that 
is characteristic of the group/cluster environment, or if this is 
specific to BCGs.  

To address this, Figure~\ref{cs} compares the scaling relations of 
the objects we identified as early-type satellites, with those for 
early-type BCGs and for the bulk of the early-type population.  
This comparison indicates that the non-central/satellite early-types 
tend to be very similar to the bulk of the early-type population;
it is the BCGs which are different.  At $\log_{10}(M_*/M_\odot)>11.4$ 
they have unusually large sizes with the effect increasing at large 
$M_*$; at lower masses ($\log_{10}(M_*/M_\odot)<11.4$), there is a 
hint that BCG velocity dispersions are larger than those of satellites
and of the bulk of the early-type population (a hint also seen in 
Figure~\ref{maxBCGonly}).  

Our findings appear to contradict those of Weinmann et al. (2008)  
who report that the sizes of early-type central galaxies are {\em not} 
larger than the sizes of early-type satellites of the same stellar mass 
(see their Fig.~4). We suspected that some discrepancy arised because 
they used Petrosian-based quantities which are ill-suited for this sort 
of analysis \cite{hb09}. However, more recently Guo et al. (2009) have fit
Sersic profiles to a subset of the same sample of galaxies and found
a similar result: no difference in the size of central and satellite 
early-type galaxies of the same stellar mass. The difference between
our results and theirs may be explained by the fact that they 
studied groups which are less massive than ours and so they do not
have satellites at the high mass end where we find the differences to be 
most significant.

\begin{figure}
 \centering
 \includegraphics[width=\hsize]{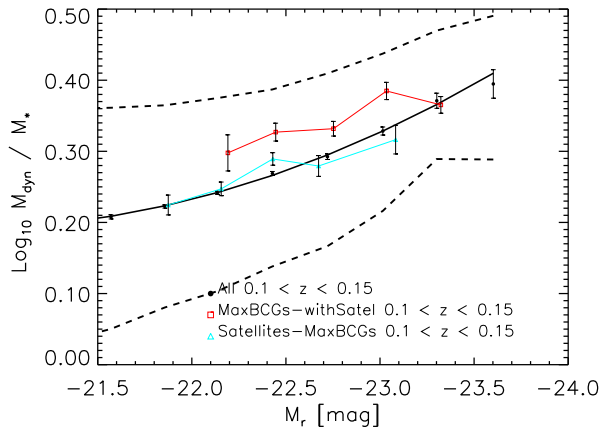}
 \caption{The dynamical to stellar mass ratio versus absolute magnitude
          of BCGs and non-central/satellite cluster galaxies of similar 
          luminosity and stellar mass, in a narrow redshift bin.}
 \label{mdmscs}
\end{figure}

The ratio of dynamical to stellar mass is another quantity of recent 
interest \cite{hb09}; it increases at large masses, suggesting that 
star formation is inefficient at large mass.  
Figure~\ref{mdmscs} shows that this ratio is about 0.05~dex larger 
for BCGs than it is for the bulk of the population of the same 
luminosity, whereas the satellites are similar to 
the bulk of the population.  At the bright end ($M_r<-22.8$), 
the difference is primarly due to the differences in sizes -- the 
velocity dispersions of BCGs are similar to those of satellites of 
the same luminosity.  
In a model where BCGs formed from dissipationless mergers this is 
easily understood:  the offset to large $M_{\rm dyn}/M_*$ is associated 
not with lower star formation efficiency, but with the subsequent 
assembly of the stars which has increased the sizes more than the 
velocity dispersions.  At the faint end ($M_r>-22.5$), the 
difference is probably due to the velocity dispersions rather than to the 
stellar masses.  

\begin{figure*}
 \centering 
 \includegraphics[width=0.475\hsize]{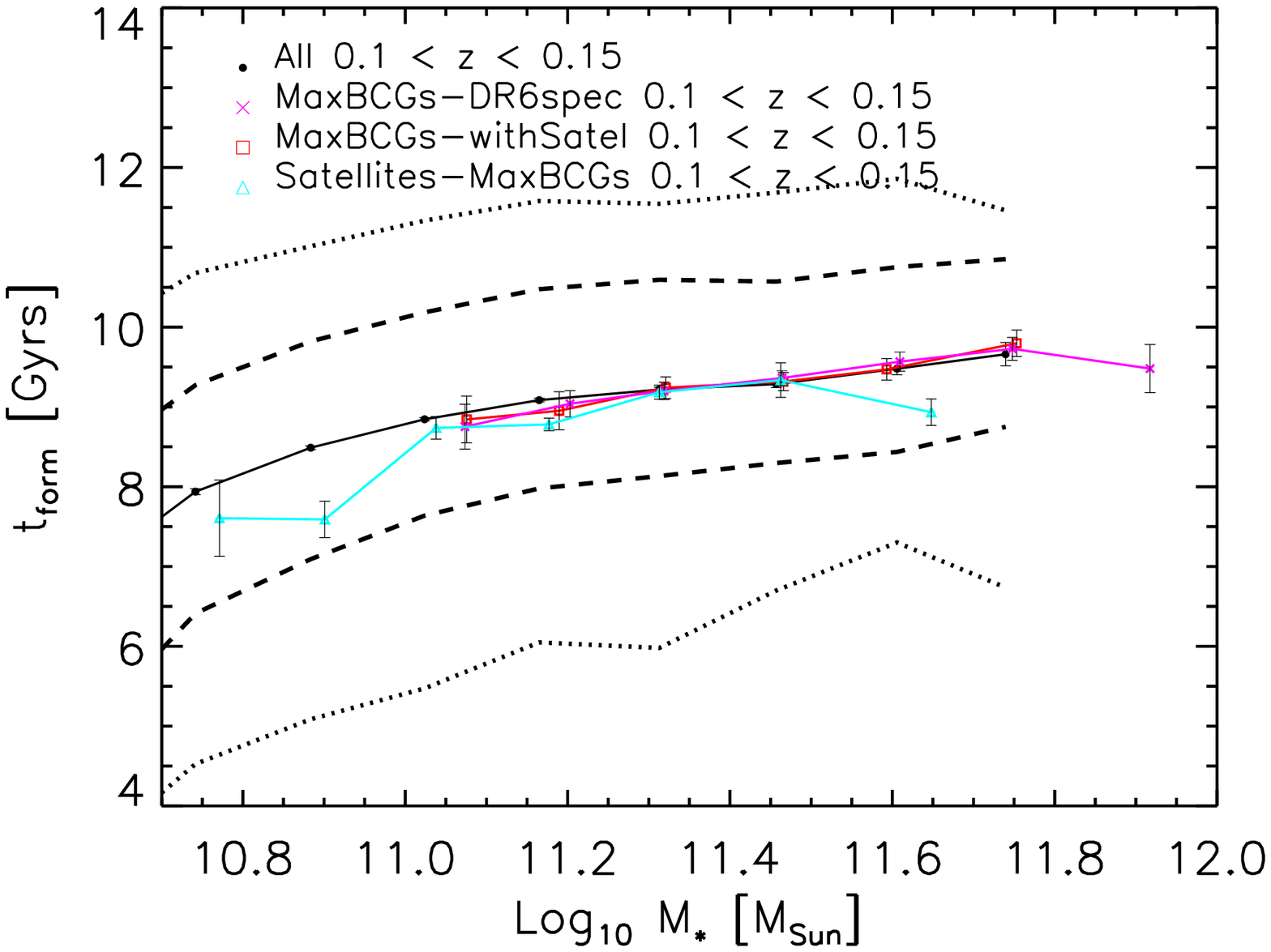}
 \includegraphics[width=0.475\hsize]{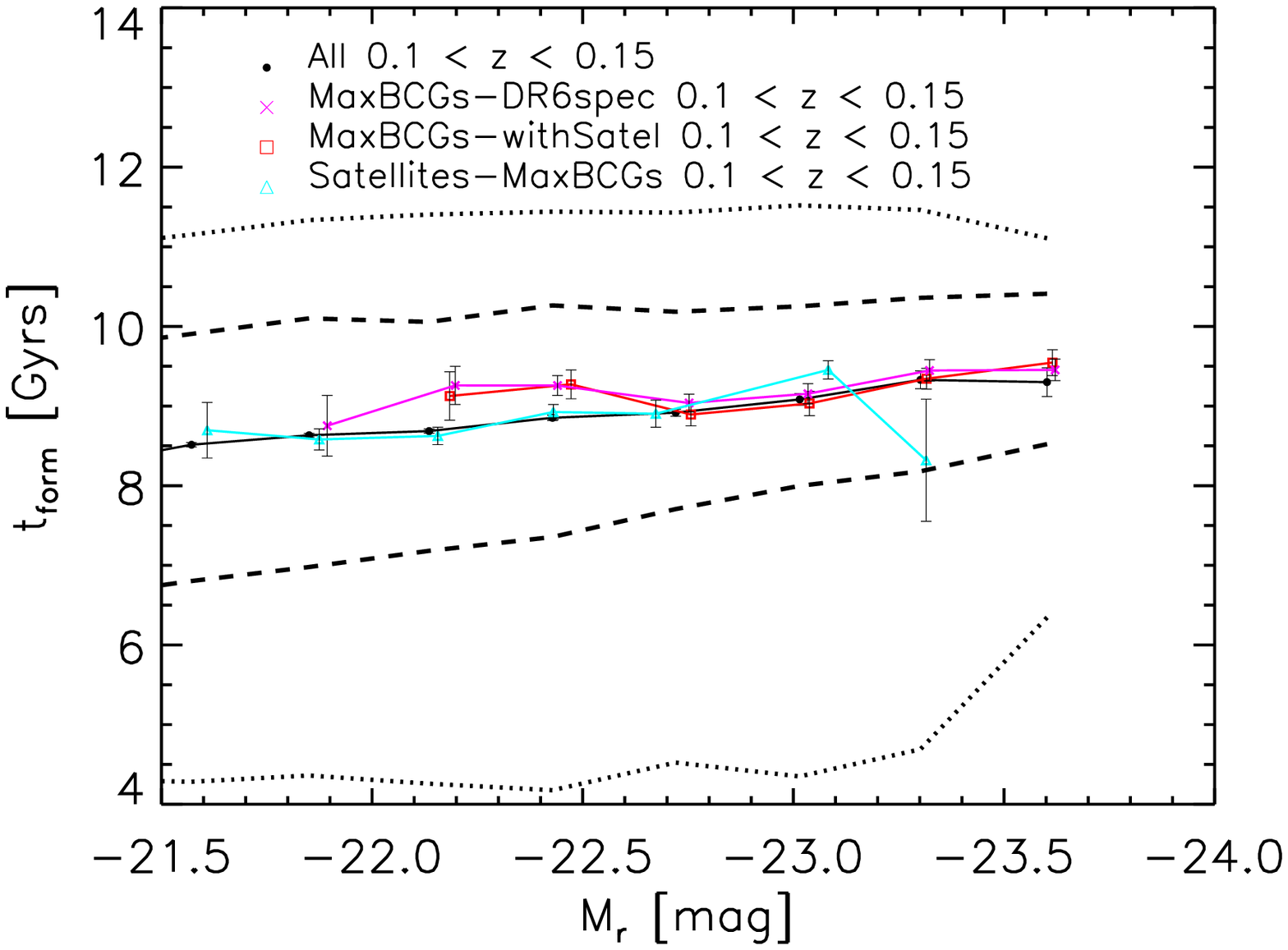}
 \caption{Correlation between lookback time to formation and stellar 
     mass (left) or luminosity (right) for BCGs (squares) and 
     satellites (triangles), over the redshift range where the 
     number counts differ by a factor of about 3.  
     Filled circles show the locus traced by the bulk of the 
     early-type population; dashed and dotted lines show the 
     region containing 68\% and 95\% of the objects.}
 \label{mstf_zoom}
\end{figure*}

We have also studied the correlation between the ages and 
luminosities or stellar masses of BCGs and satellites.  
(Appendix~A discusses why, at smaller $M_*$ than we show here, the 
trends with stellar mass may be strongly affected by selection 
effects -- see Figure~\ref{mstf}.)  
Figure~\ref{mstf_zoom} shows that at luminosities of about 
$L_*$ (i.e., $\sim -21.2$ mag in the $r-$band) and 
larger, the bulk of the population defines an age$-M_*$ or age$-L$ 
relation:  massive or more luminous galaxies tend to be slightly older.  
There is a hint that, about a 
magnitude brighter than $L_*$, BCGs are slightly older ($\sim 0.5$ Gyr) 
than other objects of the same luminosity; but there is no difference at 
brighter or fainter $L$, and there is no difference when compared with 
objects of the same $M_*$.  
Satellites and BCGs (whether or not they have satellites) follow the 
same age$-M_*$ relation as the bulk of the population.  
This is remarkable, given that we see this trend over a redshift 
range where the counts of centrals and satellites are approximately 
proportional to one another (Figure~\ref{BCGz}).  
Given that the BCGs in the same volume are more luminous 
(Figure~\ref{BCGl}), and have larger stellar masses on average, 
one might have expected the BCGs to also be older.  
Our results indicate that, if they have the same stellar mass, 
satellites and BCGs have the same age.

To understand why this happens, we now compare the luminosity-weighted 
age difference between satellites and their BCG, as a function of 
BCG luminosity.  
Figure~\ref{tbcg-tsat} shows that BCGs tend to be older than their 
satellites, by about $0.5 - 1$~Gyrs, over about two magnitudes in BCG 
luminosity -- a range over which the mean BCG age changes by about 
1~Gyr.  (Our results are unchanged if we use the number- rather than 
luminosity-weighted age difference.)  The younger ages with the 
smaller luminosities conspire to keep the satellites and BCGs along 
the same age-$L$ or age-$M_*$ relations. 

\section{Discussion}\label{discuss}
Early-type BCGs have larger sizes than other early-type galaxies 
of similar luminosity or stellar mass.  If restricted to a narrow 
bin in velocity dispersion, the size-$L$ relation of the bulk of 
the population is steeper, suggesting that perhaps it is the 
fact that BCGs are biased towards larger velocity dispersions that 
is the origin of this difference.  However, at fixed $\sigma$, the 
BCG $R_e-L$ scaling relation is steeper still (Figure~\ref{LRsigbin}).

Moreover, the sizes of BCGs appear to be evolving:  
higher redshift BCGs had smaller sizes than their local counterparts 
of the same (evolution corrected) luminosity or stellar mass 
(Figure~\ref{maxBCGonly}).  
The evolution in the $R_e-L$ relation of the early-type BCG population 
(and in general of galaxies with $\sim M_r < -22$) is more evident 
than that of the bulk of the early-type population at fainter luminosity: 
the BCG sizes evolve as $(1+z)^{0.85(M_r+21)}$ (Figures~\ref{RVcorrected}).  
The evolution in the $\sigma-L$ relation over this same period 
suggests that velocity dispersions evolve as $(1+z)^{-0.2(M_r+21)}$.  

For the bulk of the early-type population, the sizes of bright 
objects ($M_r<-22$) evolve as $(1+z)^{0.7(M_r+21)}$ and the velocity 
dispersions evolve similarly to BCGs (Figures~\ref{bulkRz} 
and~\ref{bulkVz}).  
The scaling for the bulk of the early-type population that we see 
from studying small lookback times is consistent with that reported by 
van der Wel et al. (2008) from a comparison of $z=0$ with $z=1$ 
objects:  they find that the sizes evolve as $(1+z)^{-0.98\pm0.11}$ for 
objects brighter than $M_r\approx -22$ (they did not study luminosity 
or mass-dependent trends).  

\begin{figure}
 \centering 
 \includegraphics[width=0.95\hsize]{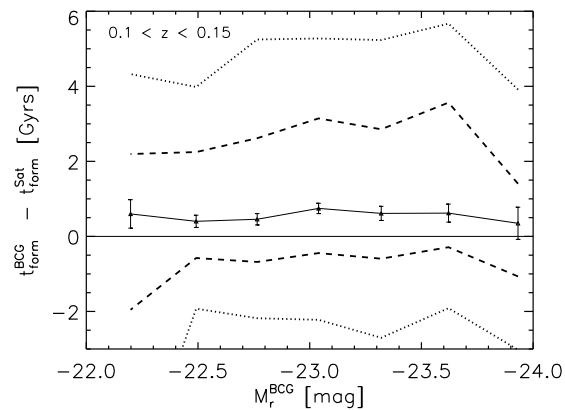}
 \caption{Difference between the mean luminosity-weighted age of the 
          satellites in a group, and that of the BCG, as a function of 
          BCG luminosity.}
 \label{tbcg-tsat}
\end{figure}

At fainter luminosities ($M_r>-22$), the evolution is weaker; it 
goes as $(1+z)^{-0.7}$ and $(1+z)^{0.2}$.
See Shankar \& Bernardi (2009) for a more detailed analysis of 
the bulk of the population, who conclude that the more massive 
galaxies ($L_r > 10^{11}$~$L_{\odot}$) show stronger evidence of 
the effects of dissipationless mergers.  

The cleanest tests of the evolution we see come from restricting 
the BCG sample to narrow bins in formation time.  
At fixed formation time and stellar mass, the objects observed at 
lower redshift are larger, and their velocity dispersions are 
smaller (compare solid curves in Figure~\ref{age2}).  
The evolution, which we detect over lookback times as small as 1~Gyr, 
is difficult to reconcile with the simplest monolithic collapse 
models, and is most pronounced for the most luminous objects.  
The recent growth in BCG sizes is in qualitative agreement with 
hierarchical galaxy formation models in which the assembly of BCGs 
continues to the present day \cite{munich06,durham07}.  

Figure~\ref{age2} also shows that the objects which formed earlier 
are smaller, but their velocity dispersions are not larger 
(compare dashed with solid curves in Figure~\ref{age2}).  
Whereas the former is expected in the simplest monolithic collapse 
models -- the universe was denser at high redshift, so one expects 
the younger objects of a given mass to be smaller and have larger 
velocity dispersions -- the latter is harder to arrange.  Thus, both 
trends in Figure~\ref{age2} -- the evolution of the sizes and velocity 
dispersions of objects (of fixed stellar mass) that formed at the 
same time, and the dependence of the sizes and velocity dispersions 
on formation time -- are difficult to accomodate in the simplest 
monolithic collapse models.  

Recently, Fan et al. (2008) have suggested a {\it puffing-up} scenario
for the evolution in the size$-M_*$ relation --- it postulates that it 
is the sizes which evolve, not the stellar masses.  
This model exploits the fact that the 
super-dense galaxies mentioned in the Introduction are observed at 
about the epoch at which QSOs are most active; feedback from the 
AGN activity at $z\sim 2$ or 3 is assumed to expel gas from the 
central regions.  The sudden reduction of mass in the core makes 
the surrounding stellar distribution puff up, after which the 
objects settle down to new (larger) sizes.  {\it This is expected to 
have been completed by $z\sim 1$, whereas our observations of 
evolving sizes are at low redshift, so it seems unlikely that 
this mechanism can explain our measurements.}  Also, there is 
little evidence for recently outflowing gas in our BCG sample.  
Nevertheless, we note that if the stellar masses have not changed, 
then Figure~\ref{age2} suggests that at 
$\log_{10}(M_*/M_\odot) > 11.5$, the sizes have 
increased by a factor of 1.5, and the velocity dispersions have 
decreased by a factor of 1.15, between $z=0.27$ and $z=0.09$.  
In addition, the dependence of the size and (especially) velocity 
dispersion of BCGs on formation time (at a given $M_*$) 
predicted by Fan et al. ($\Delta \log_{10} R_e > 0.2$ and 
$\Delta \log_{10} \sigma > 0.1$) is significantly larger than what we 
observe in Figure~\ref{age2} (compare dashed with solid curves for a 
given redshift).


These results (i.e. the evolution in size and velocity dispersion 
at small lookback times, and the weak dependence on formation time 
of the $\sigma-L$ relation) are more consistent with models which 
assume that galaxies formed from predominantly dissipationless 
mergers \cite{mk85,cfa06,berkeley06,cores08}. 
In such models, the stars formed in gas rich mergers at high redshift 
(explaining the small sizes at early times), but were assembled into 
BCGs at later times by gas-poor, dissipationless mergers. In these 
models, the stellar masses, the sizes and the velocity dispersions can 
evolve, and {\it the question arises as to whether the mergers were major} 
(approximately equal size pieces) {\it or minor.  
Our results suggest that the recent mergers for BCGs were minor.}  
This is because equal mass mergers have the growth in stellar mass 
approximately the same as the growth in size, with little change in 
velocity dispersion \cite{Ciotti08}.  However, our 
Figure~\ref{age2} shows that the velocity dispersion decreases.
 
If the mergers are minor, and the mass increases by a factor of 
$(1+f)$ in each merger, where $f\ll 1$, then the size increases 
by $(1+2f)$ and $\sigma^2$ decreases by $(1-f)$ in each merger.  
In this case, a given change in size implies a smaller change in 
mass than if the changes were caused by a major merger.  In this 
case, we can estimate a required change in mass and size by sliding 
the $R_e-M_*$ relation at $z=0.27$ upwards and to the right until 
it sits above the $z\sim 0.1$ relation:  this suggests that 
$f\approx 0.4$.  The predicted evolution in the $\sigma-M_*$ 
relation can now be compared with that observed; while it is in 
the right sense, it is a little too strong.  
The problem can be alleviated slightly if we account for the fact 
that the stellar mass estimates at low redshift are slightly smaller 
than they should be, because of mass losses associated with stellar 
evolution. If so, the implied growth in mass agrees well with that 
expected in the hierachical models \cite{munich06,durham07}.  
However, Almeida et al. (2007) predict that the velocity dispersions 
of luminous red galaxies (most luminous early-type BCGs are LRGs) 
were smaller, not larger, in the past.  

Note that a 0.4:1 merger is not what we would call minor; we 
are supposing that the mass increase of 40\% was due to a sequence 
of minor mergers (e.g. four mergers each adding $\sim$10\% to the 
mass and increasing the size by $\sim$20\%).  
In this context, it is also interesting that motion along an 
$R_e\propto M_*$ line (major mergers) cannot bring the 
superdense galaxies recently seen at $z\sim 2$ onto the $z\sim 0$ 
$R_e-M_*$ relation.  However, minor mergers (motion along a line 
of slope 2 in the $\log(R)-\log(M_*)$ plane) may bring them onto  
the local relation traced by BCGs.  The objects at $z\sim 2$ have 
large $M_*$ even by $z\sim 0$ standards, so it is not implausible 
that they are the progenitors of today's BCGs.  This requires 
mass growth factors of order 4 or 5 (0.6~dex), coupled with an 
increase in size by a factor of order 10. This is consistent with 
our estimate of the observed BCG size evolution $(1+z)^{0.85(M_r+21)}$: 
setting $M_r \sim -23.5$ and $z=2$ we get an evolution of a factor
of $\sim 10$. 

Recent results support our conclusion that minor mergers are important.  
Hopkins et al. (2008) suggest that if mergers are major, then the 
fraction of superdense massive galaxies which survived intact since 
their formation at $z >2$ could reach $1-10$\%.  However, 
Trujillo et al. (2009) show that the actual number density of 
superdense galaxies at $z\sim 0$ is much smaller (the few which 
do exist appear to be young, so they are unlikely to be 
descendents of the $z\sim 2$ objects).  Minor mergers must account 
for the difference.  

Minor mergers are also prefered because constraints from the bright 
end of the luminosity function suggest little evolution since 
$z\sim 1$ (e.g. Wake et al. 2006; Brown et al. 2007; Cool et al. 2008), 
even though cluster masses are expected to have grown substantially, 
through mergers, during this time (e.g. Sheth \& Tormen 1999).  
Of course, as a result of such mergers, the fractional mass growth 
of the BCG need not be the same as that of its cluster, since 
some of the added stellar mass must make the intercluster light 
(e.g. Skibba et al. 2007).  
And indeed, comparison of the clustering of the most luminous galaxies 
at $z\sim 0.7$ with that more locally suggests that cluster merging 
has resulted in some stellar mass growth of the BCGs 
(White et al. 2007; Wake et al. 2008); while massive halos double 
in mass over the last 7 Gyr, the stellar masses of their BCGs are 
expected to have changed by about 30\% (Brown et al. 2008).  

If this estimate of the mass change is accurate, then major mergers 
cannot account for the factor of 2 change in size which van der 
Wel et al. (2008) report is typical for massive early-types over this 
time -- and which our measurements suggest is even more dramatic for BCGs.  
On the other hand, minor mergers are better able to reconcile the 
observations of dramatic size evolution with little mass evolution.  
Indeed, Bournaud et al. (2007) have recently highlighted the fact 
that multiple minor mergers may be the dominant channel for early-type 
galaxy formation.  Simulations show that, for such a merger, the 
fractional increase in size can be larger than that of the mass 
(e.g., Halo A in Naab et al. 2007), consistent with our simple 
analytic estimate above.  

\begin{figure}
 \centering 
 \includegraphics[width=0.95\hsize]{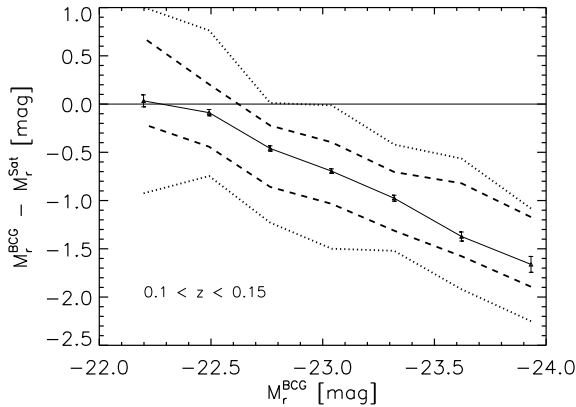}
 \caption{Difference between the luminosity of a satellite galaxy 
          and that of its BCG, as a function of BCG luminosity.}
 \label{Lbcg-Lsat}
\end{figure}

We also compared the ages and sizes of our early-type BCGs with 
other cluster galaxies (satellites).  BCGs are larger than 
early-type cluster galaxies of similar luminosity or stellar mass 
at the same redshift (Figure~\ref{cs}).  Although satellites and 
BCGs trace the same weak age-$L$ or age$-M_*$ relation 
(Figure~\ref{mstf_zoom}), this can be understood by noting that 
BCGs are typically about 1~Gyr older than the satellites in their 
group (Figure~\ref{tbcg-tsat}), and they are about 0.5~mags 
more luminous (Figure~\ref{BCGl}).

The mean satellite luminosity is approximately independent of BCG 
luminosity -- a prediction \cite{Skibba06} which has recently been 
confirmed from measurements in other group catalogs 
\cite{Skibba07,Hansen07}.  
Figure~\ref{Lbcg-Lsat} shows that this remains true if both BCGs 
and satellites are required to be early-types.  
Skibba \& Sheth (2008) went on to suggest that satellite colors 
should also be much weaker functions of group mass than are the 
colors of centrals, and Skibba (2009) showed that this was indeed 
the case.  Since color is an indicator of $M_*/L$, and satellite 
$L$ is almost independent of group mass, $M_*$ should be similarly 
independent.  This prediction has now been confirmed:  
van den Bosch et al. (2008) find that $M_*$ for satellites changes 
by a factor of only 2 over a range where group mass changes by a 
factor of 100.  For similar reasons, one expects only a small 
increase of mean satellite age with BCG luminosity; this is 
qualitatively consistent with the constant age offset between 
satellites and BCGs in our Figure~\ref{tbcg-tsat}, and the weak 
age-luminosity relation for BCGs in our Figure~\ref{mstf_zoom}.

\section*{Acknowledgments}

M.B. is grateful for support provided by NASA grant LTSA-NNG06GC19G. 
She also thanks Francesco Shankar and Ravi Sheth for helpful conversations.

Funding for the Sloan Digital Sky Survey (SDSS) and SDSS-II Archive has been
provided by the Alfred P. Sloan Foundation, the Participating Institutions, the
National Science Foundation, the U.S. Department of Energy, the National
Aeronautics and Space Administration, the Japanese Monbukagakusho, and the Max
Planck Society, and the Higher Education Funding Council for England. The
SDSS Web site is http://www.sdss.org/.

The SDSS is managed by the Astrophysical Research Consortium (ARC) for the
Participating Institutions. The Participating Institutions are the American
Museum of Natural History, Astrophysical Institute Potsdam, University of Basel,
University of Cambridge, Case Western Reserve University, The University of
Chicago, Drexel University, Fermilab, the Institute for Advanced Study, the
Japan Participation Group, The Johns Hopkins University, the Joint Institute
for Nuclear Astrophysics, the Kavli Institute for Particle Astrophysics and
Cosmology, the Korean Scientist Group, the Chinese Academy of Sciences (LAMOST),
Los Alamos National Laboratory, the Max-Planck-Institute for Astronomy (MPIA),
the Max-Planck-Institute for Astrophysics (MPA), New Mexico State University,
Ohio State University, University of Pittsburgh, University of Portsmouth,
Princeton University, the United States Naval Observatory, and the University
of Washington.

\newpage

\appendix
\section{Effect of the magnitude limit}

\begin{figure*}
 \centering
 \includegraphics[width=0.475\hsize]{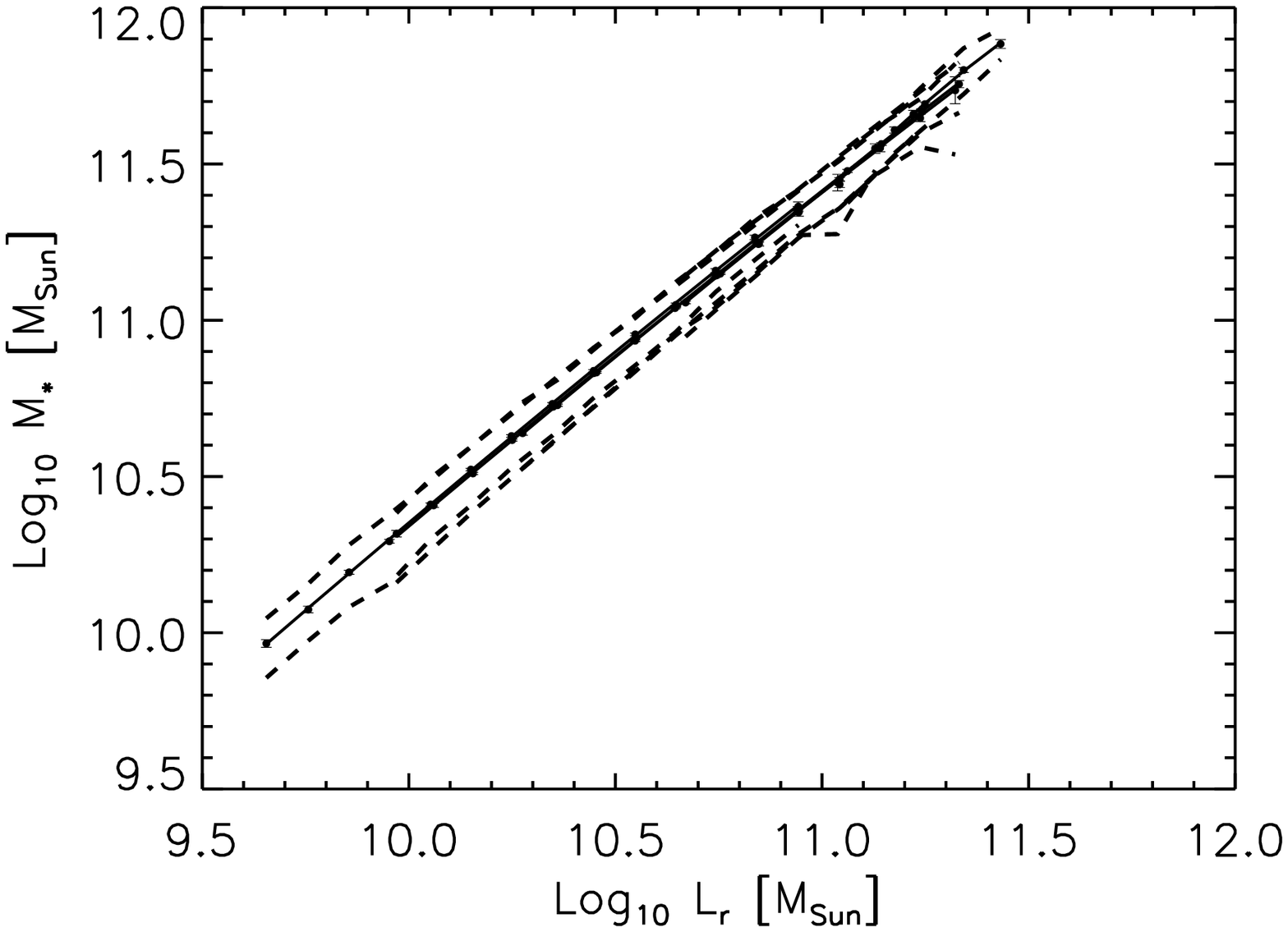}
 \includegraphics[width=0.475\hsize]{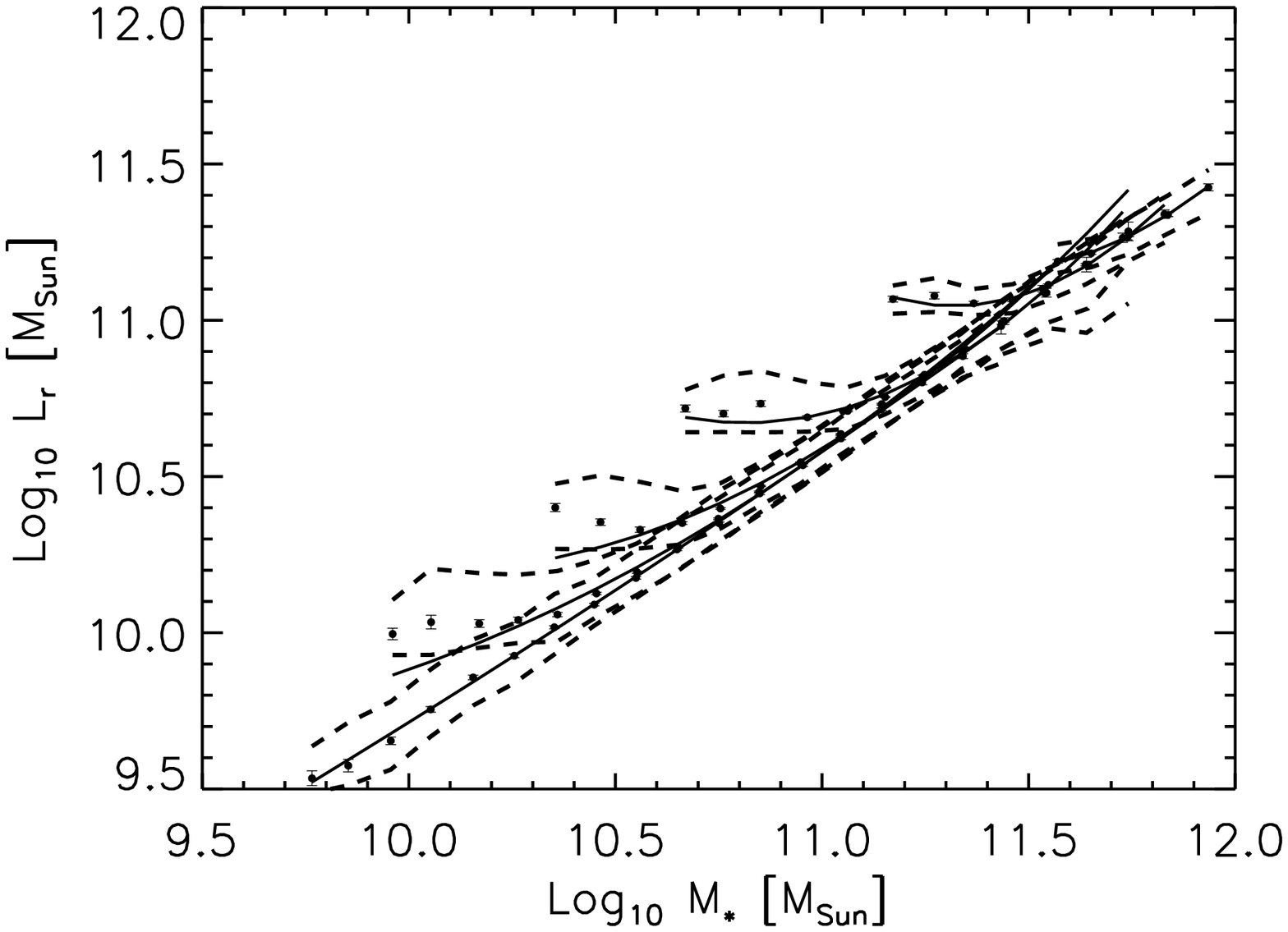}
 \caption{Correlation between stellar mass and luminosity (left) 
          and luminosity and stellar mass (right) in a number of 
          narrow redshift bins.  
          The flattening at (redshift dependent) small $M_*$ 
          in the panel on the right is a selection effect 
          which is due to the magnitude limit of the SDSS.}
 \label{M*L}
\end{figure*}

\begin{figure*}
 \centering
 \vspace{-2cm}
 \includegraphics[width=0.45\hsize]{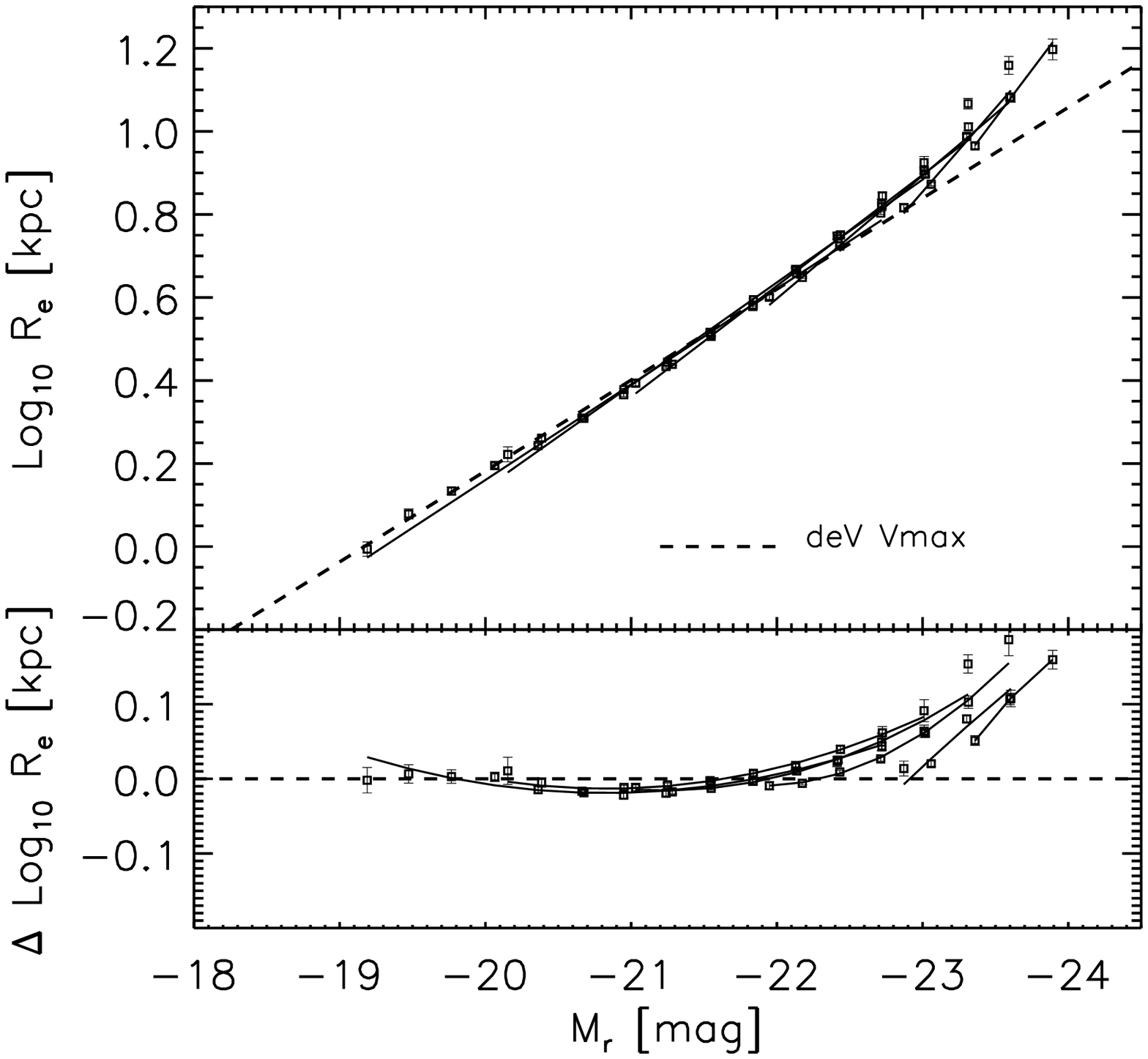}
 \includegraphics[width=0.45\hsize]{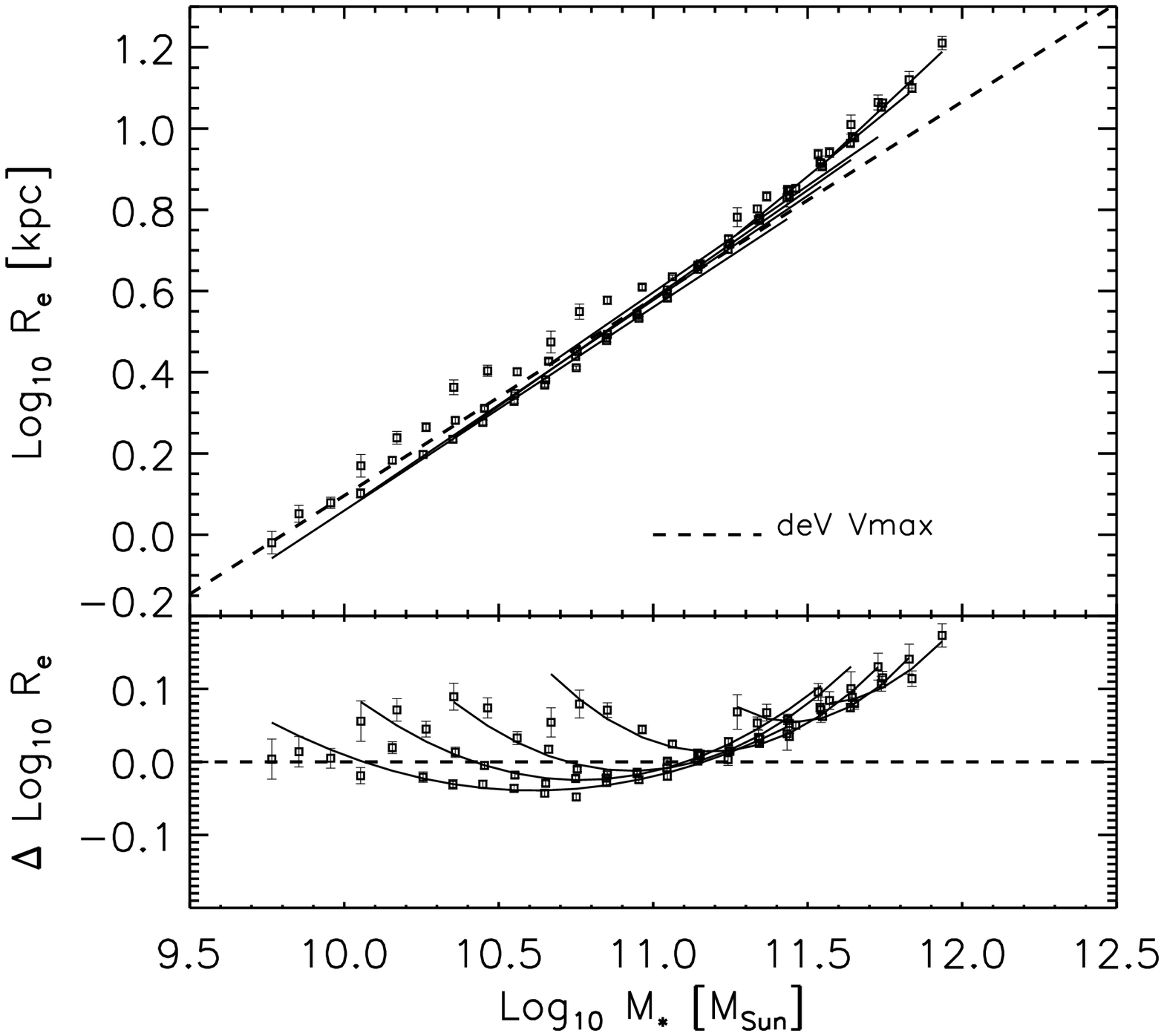} 
 \caption{Correlation between size and luminosity (left) and 
          stellar mass (right) in a number of narrow redshift 
          bins.  The correlation with luminosity curves upwards 
          at high luminosities; at low luminosities, the relation 
          is not curved, and is independent of redshift.  
          The strongly redshift-dependent curvature at low stellar 
          masses in the panel on the right is a selection effect 
          which is due to the magnitude limit of the SDSS.}
 \label{selRMs}
\end{figure*}

\begin{figure*}
 \centering 
 \includegraphics[width=0.475\hsize]{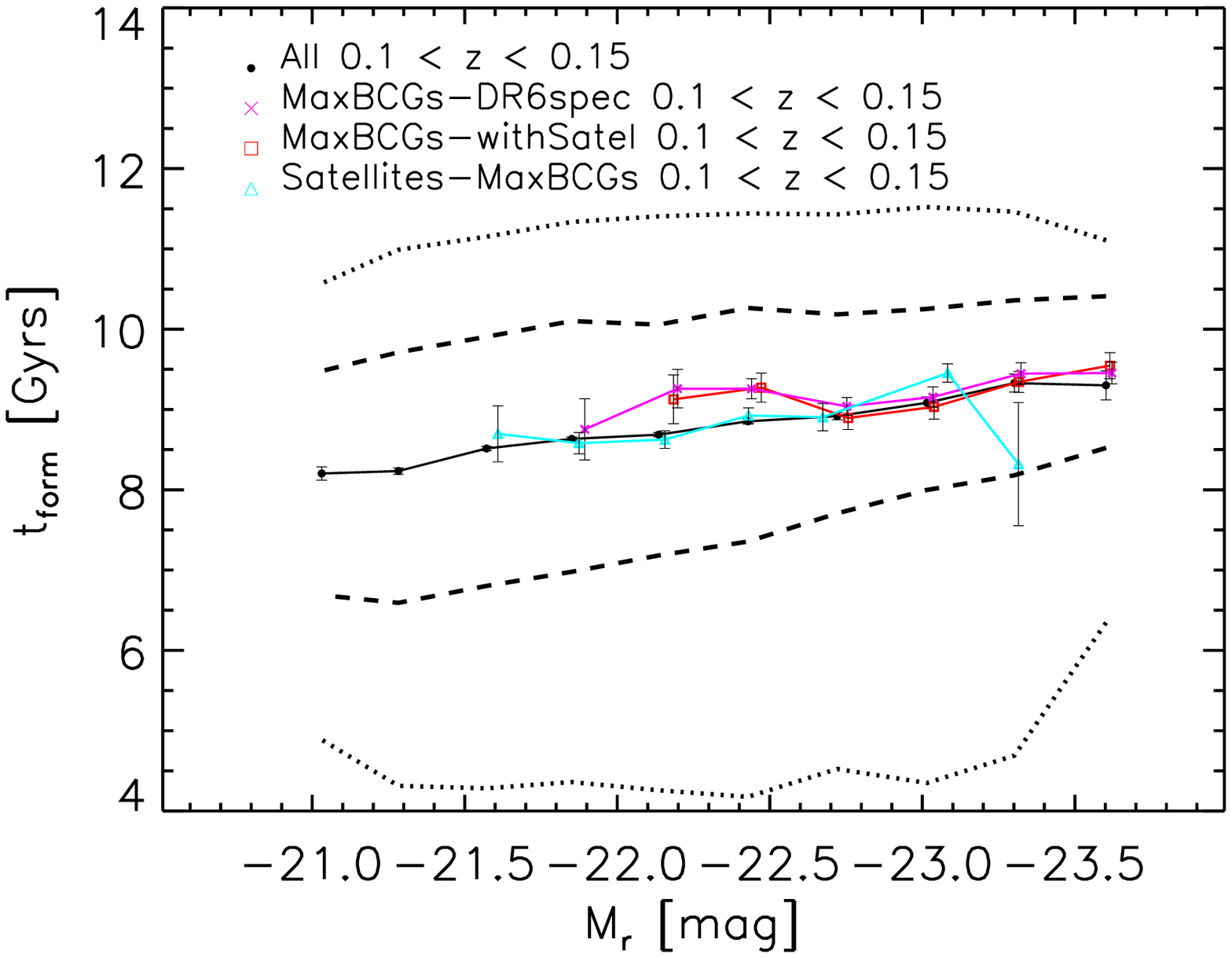}
 \includegraphics[width=0.475\hsize]{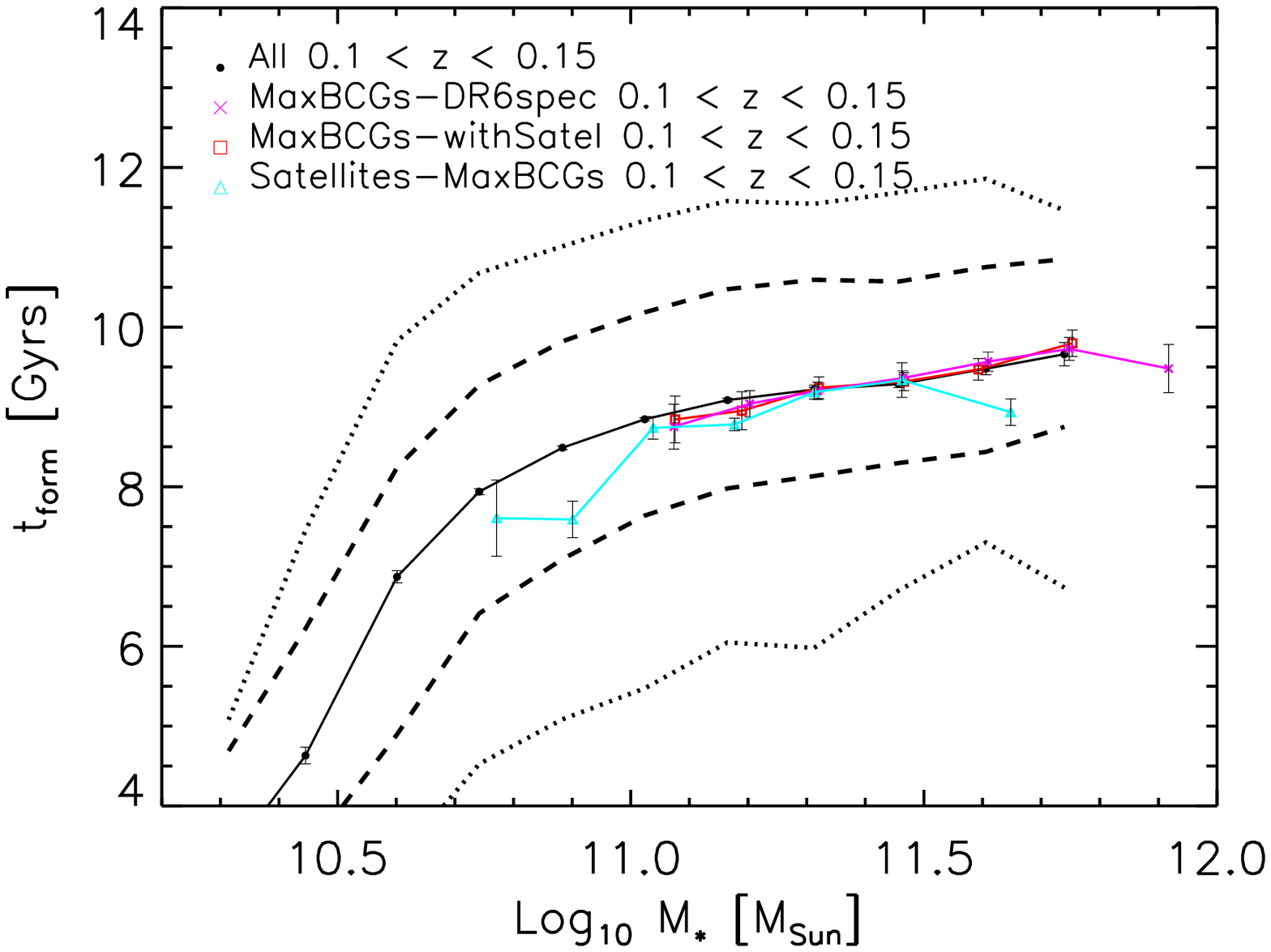}
 \caption{Correlation between (lookback time to) formation and 
          luminosity (left), and stellar mass (right), 
          for BCGs (squares and crosses), 
          satellites (triangles), and the bulk of the population 
          (filled circles with error bars), over the redshift range 
          $0.1<z<0.15$.  The drop at small $M_*$ is a selection effect.}
 \label{mstf}
\end{figure*}

The SDSS is magnitude limited.  As a result, care must be taken 
when interpretting redshift-dependent trends.  In general, accounting 
for the selection effect is only straightforward for correlations 
with luminosity; correlations with stellar mass may be strongly 
affected -- even though $M_*$ and $L$ are tightly correlated.  
Things are even more complicated if one wishes to study age 
related trends, since the errors on the age estimates are 
correlated with those on $M_*$, and these may be substantial.

\subsection{Correlations with $M_*$}
To illustrate, the panel on the left of Figure~\ref{M*L} shows 
the $M_*-L$ relation in a number of redshift bins; there is no 
trend with redshift.  
The standard way of accounting for the magnitude limit is to 
weight each galaxy by the inverse of the volume $V_{\rm max}(L)$, 
over which it could have been seen.  For the $M_*-L$ relation, 
this weighting does not matter, as all galaxies in a given $L$ 
bin have almost the same weight (the weighting matters very much 
for the $L-M_*$ relation!).  The panel on the right shows the $L-M_*$ 
relation in these same bins, when objects have been weighted by 
$V_{\rm max}^{-1}$:  notice how the relation flattens out to 
constant $L$ at small $M_*$.  The mass scale on which this 
bias appears depends on redshift, and is purely a consequence of 
the SDSS magnitude limit -- the $V_{\rm max}$-weighting does not 
solve this problem.

Figure~\ref{selRMs} illustrates that this can have a dramatic effect 
on the $R_e-M_*$ relation if it is measured in a narrow redshift 
bin.  The panel on the left shows the $R_e-L$ relation in the same 
sequence of narrow redshift bins as before.  At small $L$, the $R_e-L$ 
relation is the same in all the redshift bins; at high $L$ where 
there is significant curvature in the relation, there is some 
evidence for evolution.  
In contrast, the $R_e-M_*$ relation appears to evolve dramatically, 
particularly at small $M_*$.  It is easy to see that this is a 
selection effect, and that it produces dramatic effects even 
though $L$ and $M_*$ are tightly correlated.  Consider objects 
in a given narrow redshift bin.  Because of the magnitude limit, 
objects which scatter to lower $L$ for their $M_*$ will be excluded 
from the sample.  The observed sample will contain objects with 
large $L$ for their $M_*$; since $L$ and $M_*$ are strongly 
correlated, this will be more dramatic at small $M_*$; since size 
and $L$ are strongly correlated, the exclusion of small $L$ objects 
biases the sample to large $R$ at small $M_*$.  

Since $M_*/L$ increases with age, the effect of the magnitude limit 
is particularly pernicious for studies which include both $M_*$ and 
the age.  This is shown in Figure~\ref{mstf}.  The left hand panel 
shows that $t_{\rm form}$, the lookback time from the present to when 
the stars formed, increases slightly with luminosity.  
The right hand panel shows the correlation when $L$ is replaced with 
$M_*$.  For $M_*$ smaller than $10^{11}$~$M_{\odot}$ the relation for 
the bulk of the population is changed dramatically.  This is because, to make 
the plot on the right, we have shifted each object in the panel on the 
left by $M_*/L$.  However, because we have restricted to a narrow 
bin in $z$, $M_*/L$ increases with $t_{\rm form}$, so the shift is 
larger for large $t_{\rm form}$.  Now, at small $M_*$, the objects 
with large  $M_*/L$ fell outside the magnitude limit of the survey 
($M_r<-21$), so they are missing from the $t_{\rm form}-M_*$ 
correlation.  Since large $M_*/L$ means large $t_{\rm form}$, the 
correlation between $t_{\rm form}$ and $M_*$ curves sharply downwards 
as a result.  We emphasize that this curvature is a selection effect.   
The satellites and BCGs are far enough from the limiting magnitude 
that they are less affected by this bias.

\begin{figure*}
 \centering 
\includegraphics[width=0.475\hsize]{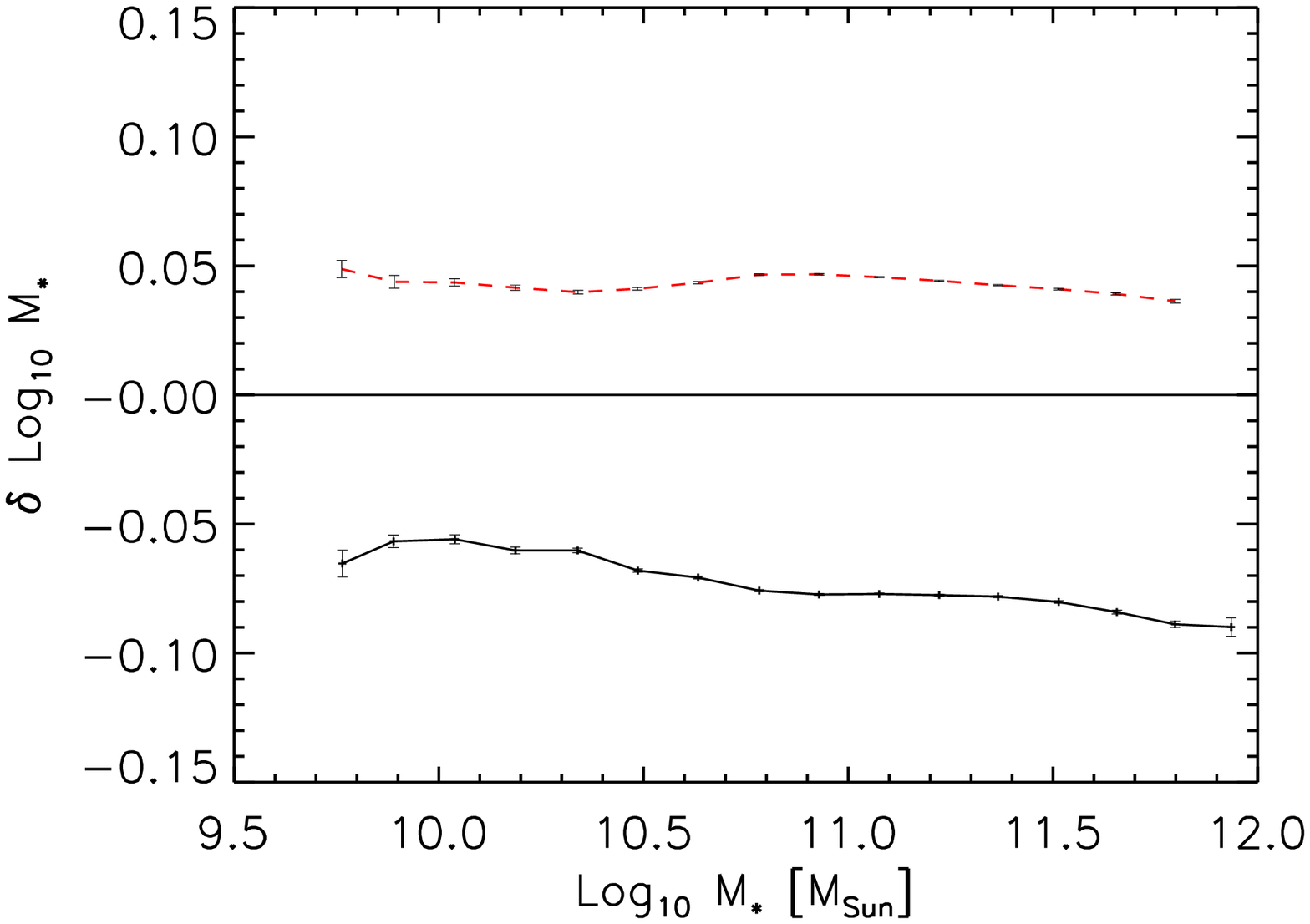}
\includegraphics[width=0.475\hsize]{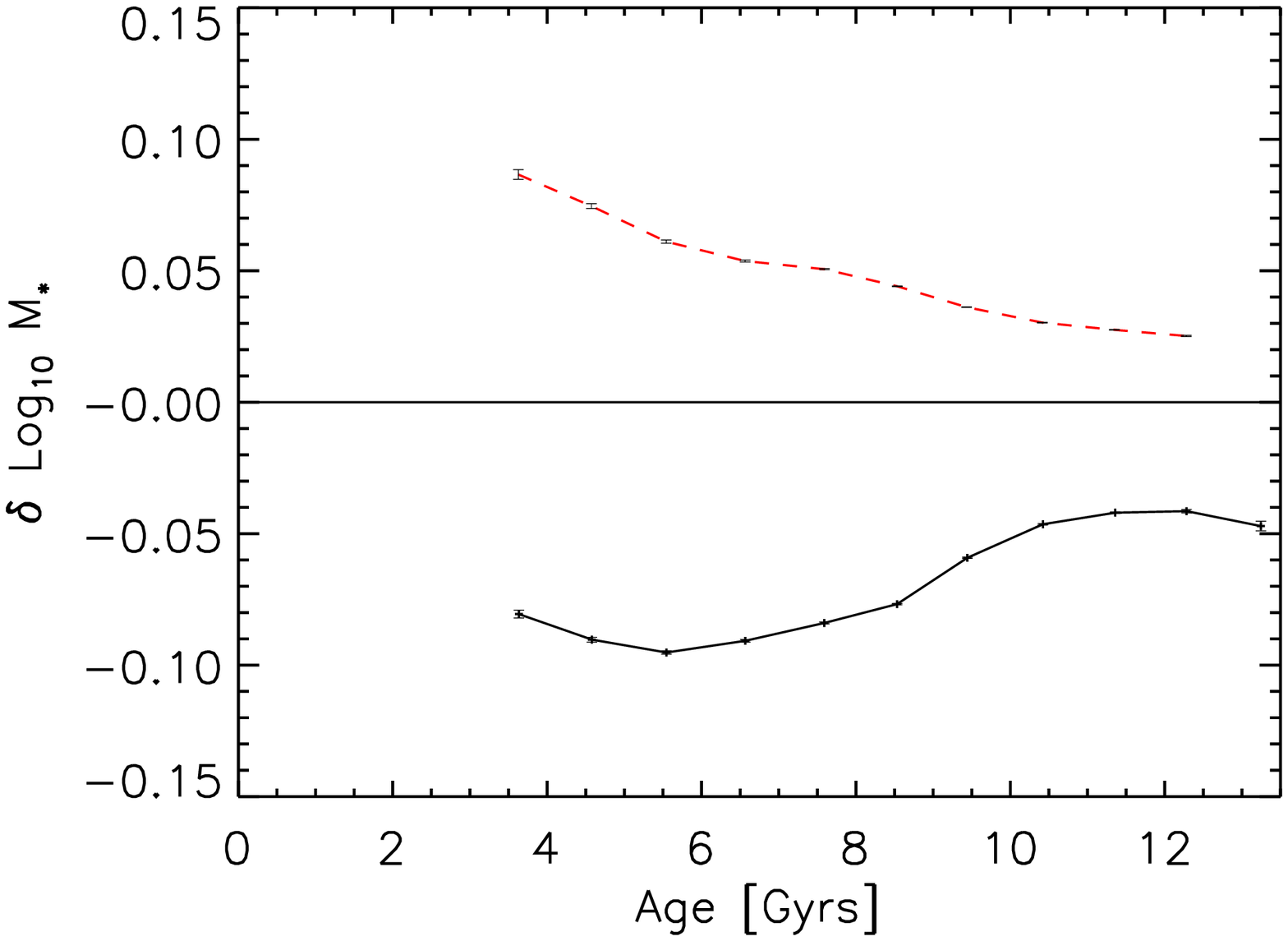}
\includegraphics[width=0.475\hsize]{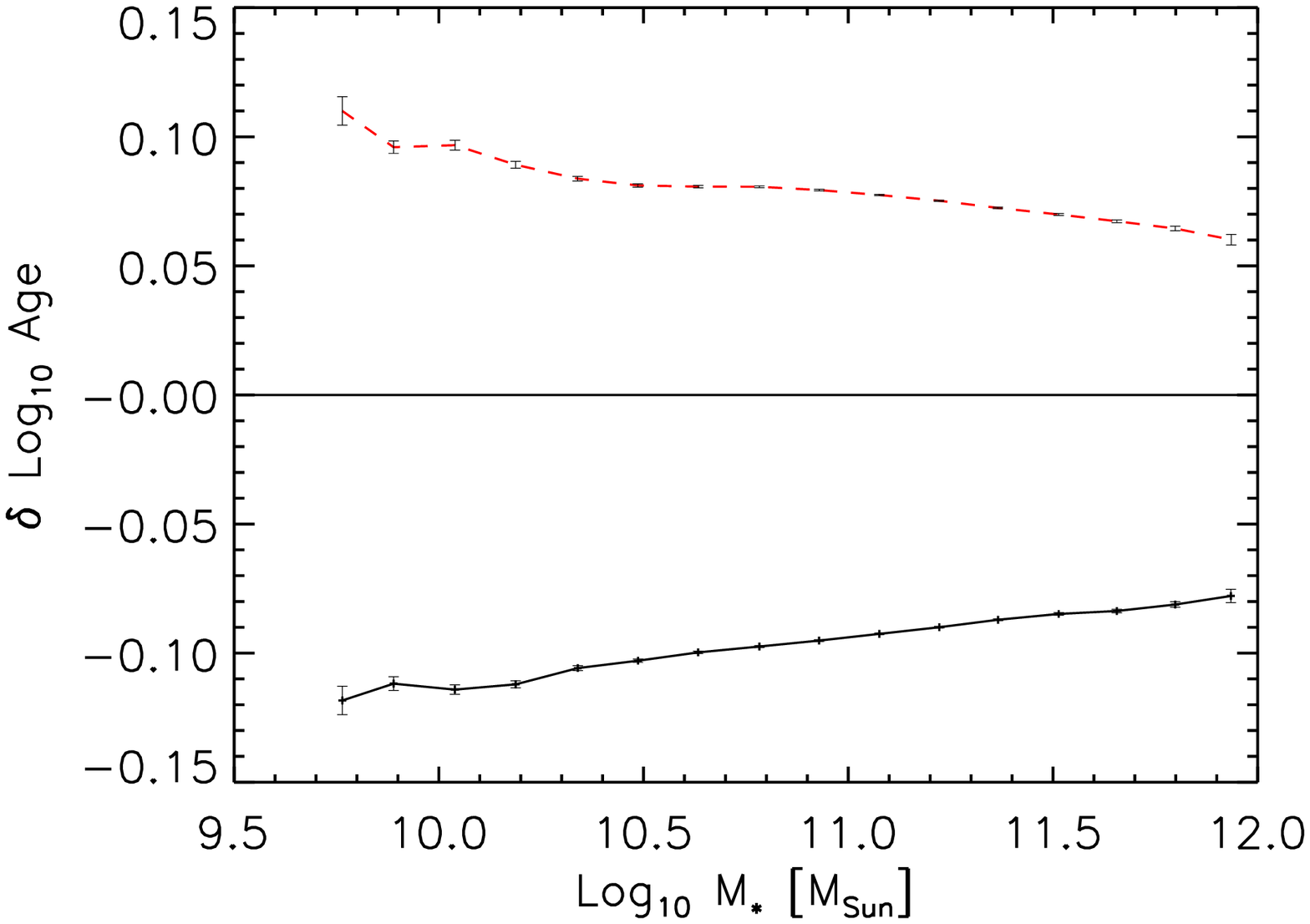}
\includegraphics[width=0.475\hsize]{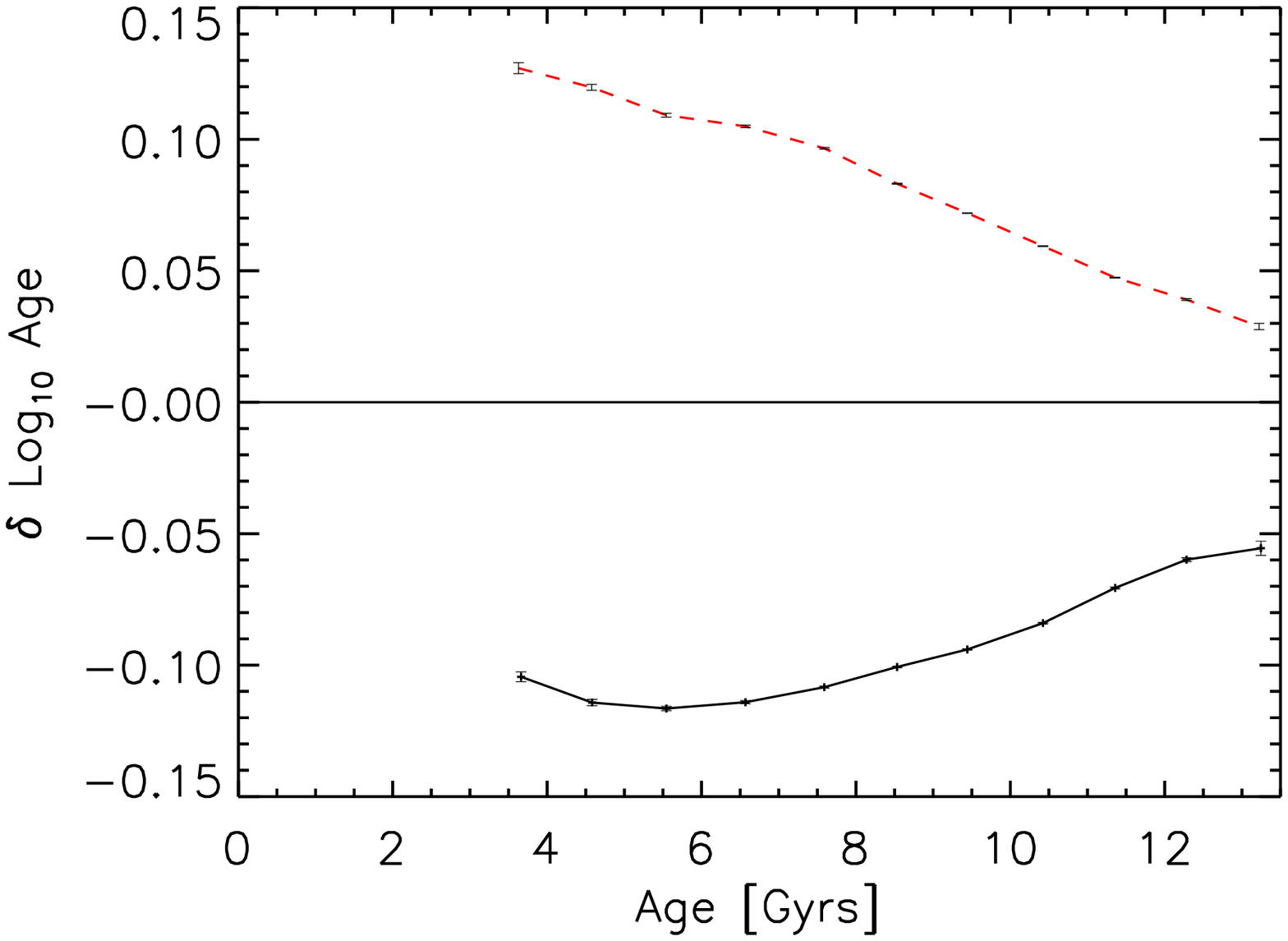}
\includegraphics[width=0.475\hsize]{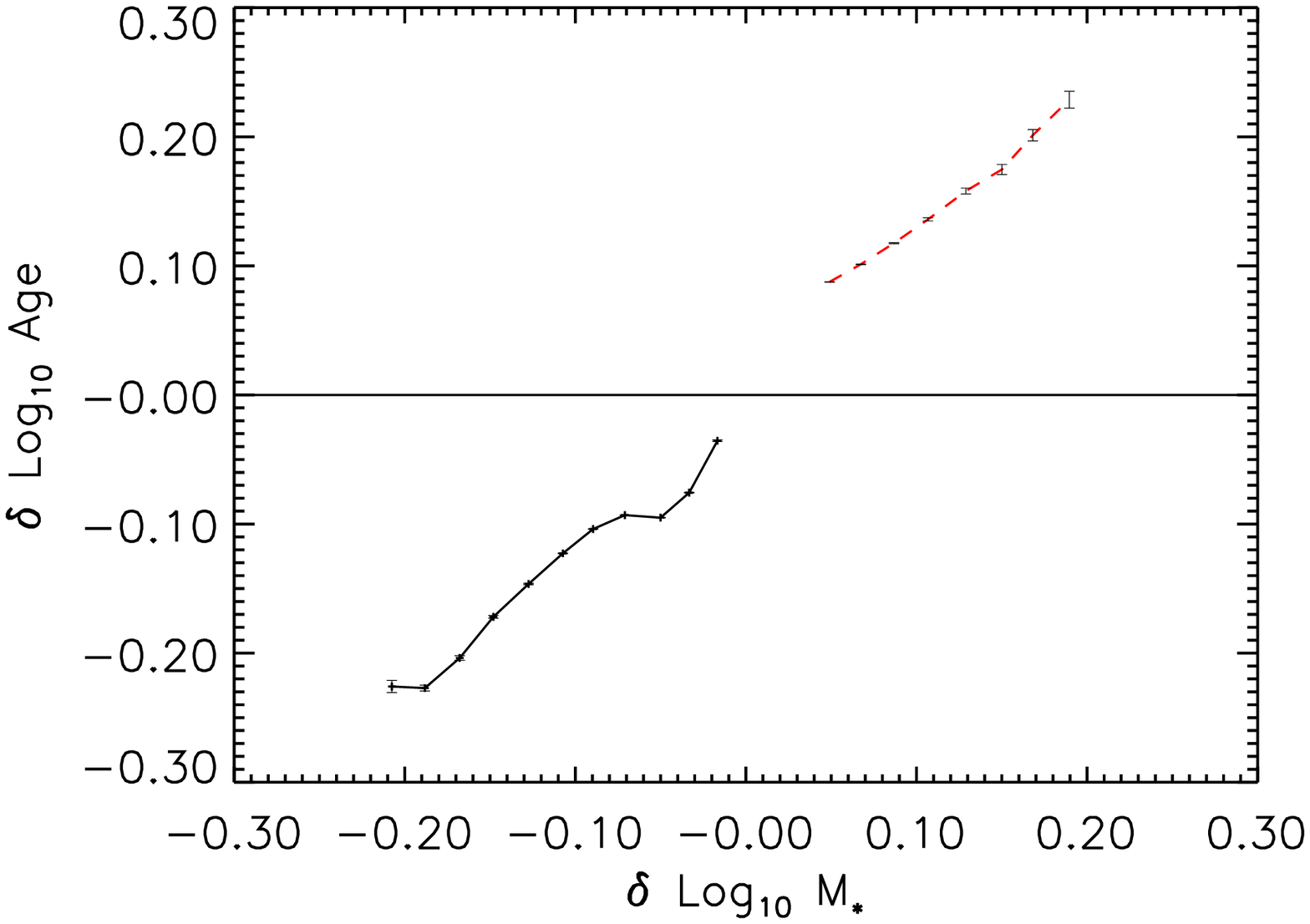}
\includegraphics[width=0.475\hsize]{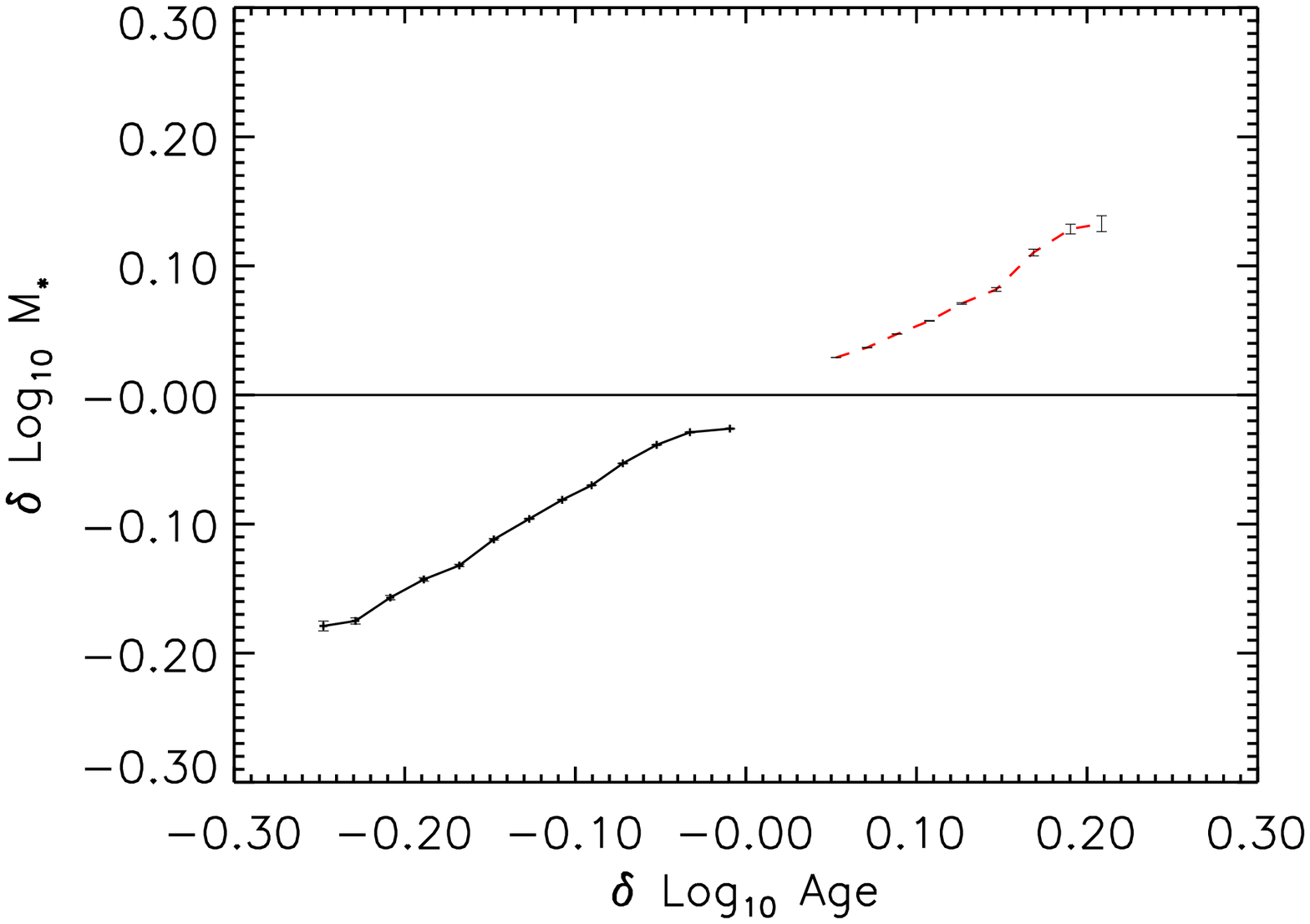}
 \caption{Uncertainties on estimated ages and stellar masses.
      Dashed line shows the difference between the 84th percentile of 
      likelihood distribution of the measured age or stellar mass 
      and the 50th percentile; solid line shows the difference between 
      the 16th and 50th percentile.
      The error on $\log_{10}M_*$ develops a tail which extends 
      to small masses when $M_*$ is large (top left), and it 
      increases dramatically for ages below 9 Gyrs (top right).
      The error on the age increases at small $M_*$ (middle left) 
      and age (middle right).  When the error on $M_*$ is large, 
      so is the error on the age (bottom left), 
      and vice versa (bottom right).}
 \label{errors}
\end{figure*}

\subsection{Correlations with age and $M_*$}\label{ageM*}
The main text studies the $R_e-L$ and $R_e-M_*$ correlations as a 
function of the formation time and the age of the stellar population.  
However, because the age and $M_*$ estimates have significant 
uncertainties, and they are correlated, it is important to use ages 
that are output from the same models which estimate $M_*$.

Although we cannot actually plot the error in the age versus the 
error in $M_*/L$, we expect they will be correlated because, for 
older stellar populations,
 $M_*/L \propto t^{0.75}$ or so, 
where $t$ is the age of the population.  As a result a galaxy that 
is incorrectly assigned a small age will also be assigned a small 
$M_*/L$ ratio.  In addition, if the uncertainty on the age is small 
then the uncertainty on $M_*/L$, and hence $M_*$ will also be small.  
This explains the trends shown in the bottom panels of 
Figure~\ref{errors} (note that these do not show the correlated 
errors themselves -- they show that when one quantity has a large 
error bar, then so does the other).  
Notice that when the estimated age is small, then the uncertainties 
on the age increase dramatically; this increases the uncertainty 
on $M_*$ as well.  

\begin{figure*}
 \centering
 \includegraphics[width=0.475\hsize]{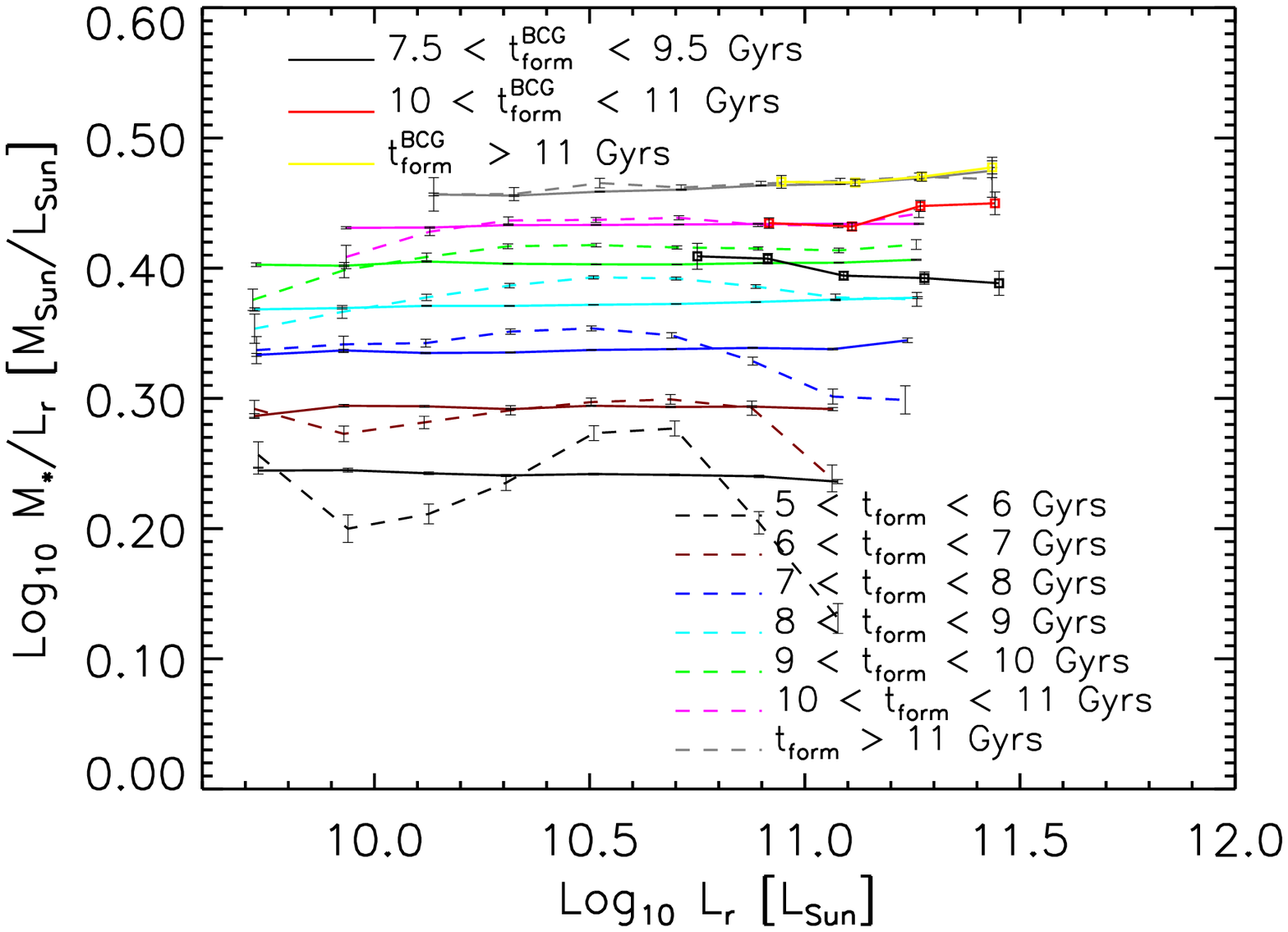}
 \includegraphics[width=0.475\hsize]{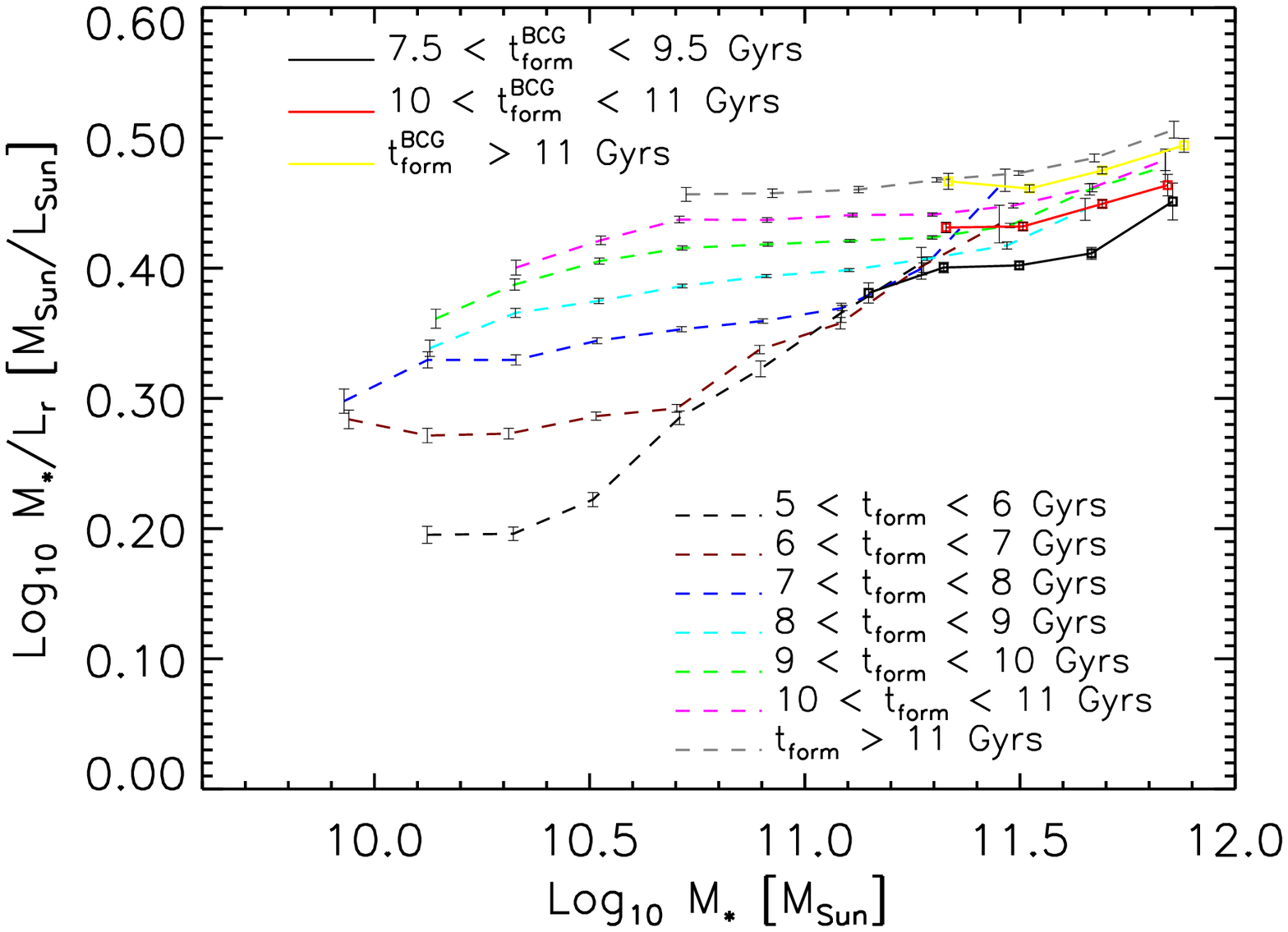}
 \caption{Stellar mass to light ratio as a function of luminosity (left),
          and $M_*$ (right); the slight curvature at the small $M_*$ 
          for each bin in $t_{\rm form}$ is due to the same selection effect 
          as in Figure~\ref{mstf}.  The sudden increase in $M_*/L$ at larger 
          $M_*$ (right panel) and decrease in $M_*/L$ at larger $L$ 
          (left panel) for the younger galaxies is 
          due to correlated errors. Symbols connected by solid and  
          dashed lines show the MaxBCGs-DR6spec sample and bulk of 
          the early-type population, respectively.  Horizontal solid 
          lines in left panel show the expected $M_*/L$ given the age:
          $d\log(M_*/L)/d\log t = 0.75$. }
 \label{M*Ltform}
\end{figure*}

\begin{figure*}
 \centering
 \includegraphics[width=0.5\hsize]{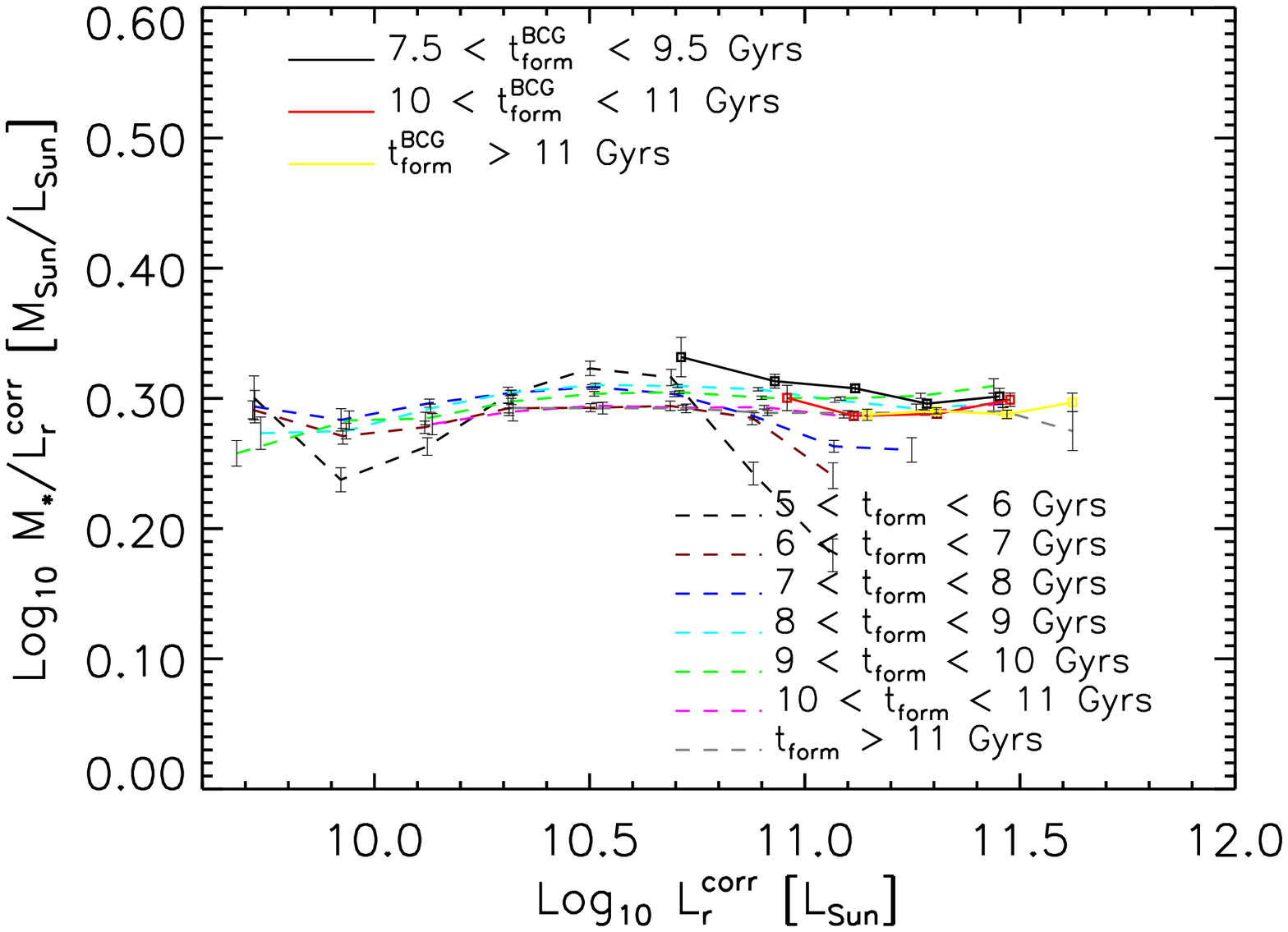}
 \caption{Stellar mass to light ratio as a function of luminosity,
          where the luminosity has been corrected for age effects 
          assuming $d\log(M_*/L)/d\log t = 0.75$. Here, we corrected 
          the luminosity to a population $6.5$~Gyr old, i.e. 
          $\log_{10} L_r^{corr} = \log_{10} L_r + 0.75(\log_{10} t - \log_{10} 6.5)$.
          Lines as in Figure~\ref{M*Ltform}. }
 \label{M*Lcorr}
\end{figure*}

Correlated errors in age and $M_*/L$ complicate analyses of how 
galaxy structure correlates with formation time and age.  To 
illustrate, Figure~\ref{M*Ltform} shows $M_*/L$ as a function of 
$L$ and $M_*$ for a number of bins in formation time.  
The panel on the left is not very surprising -- galaxies which 
formed longer ago have larger $M_*/L$ ratios -- although it 
appears that there may be something amiss in the bin with the 
most recent formation times.  The offset from one bin to another 
is consistent with the expected fading of an old stellar population: 
$M_*/L \propto t^{0.75}$ or so, where $t$ is 
the age of the population.  Figure~\ref{M*Lcorr} shows this 
explicitly:  when the luminosity has been corrected for this age 
effect, then the different formation time bins overlap.  

It is worth noting that using age estimates from a different 
algorithm than the one which provided the $M_*/L$ estimates results 
in qualitatively similar behaviour to that shown in the lefthand 
panel of Figure~\ref{M*Ltform}, except that the offset between the 
different formation time bins is smaller.  This is because, if this 
is done, then the correlated nature of the age and $M_*/L$ errors is 
missing.  As a result, it would be possible for an object to be assigned 
a younger age than its true one, as well as a larger $M_*/L$ ratio than 
its true one, and so objects assigned recent formation times 
would have larger $M_*/L$ ratios on average, and objects assigned 
older formation times have smaller $M_*/L$ ratios, than when the 
errors in age and $M_*/L$ are correlated.  

Unfortunately, the presence of correlated errors produces spurious 
features in the right hand panel of Figure~\ref{M*Ltform}.  In this 
case, errors in $M_*/L$ move objects along lines that slope upwards and 
to the right.  But if an object scatters downwards and to the left 
along such a line (of constant $L$), it is also assigned a younger age, 
and so it contributes to a more recent formation time bin.  If there 
are, in fact, no real galaxies having large $M_*$ but recent formation 
times, then the correlated errors will have produced such a 
population.  Since the uncertainties are largest for the youngest 
galaxies, we believe that it is this effect which causes the 
sharp upturn in the lowest two formation time bins.  

A better procedure, if only the age or only $M_*$ is known, is to 
assume that $\partial \log(M_*/L)/\partial \log t = 0.75$, and to 
use this to correct luminosities for age effects.  Thus, when we 
use $L^{\rm corr}$ as a proxy for $M_*$, we obtain the correct 
spread in ages at fixed $L$ (solid lines in Figure~\ref{M*Ltform}), 
and we find $M_*/L^{corr}\approx $~constant (Figure~\ref{M*Lcorr}).

\section{Comparison with previous work at $z<0.1$}\label{LRlit}
The size-luminosity relation for BCGs has been the subject of 
much recent interest \cite{Lauer07,Bernardi07,vdLinden07,Liu08}.  
Whereas most authors agree that early-type BCGs are very different 
from the bulk of the population, von der Linden et al. (2007) 
find substantially smaller differences.  
This is almost certainly because von der Linden et al. 
use Petrosian-based quantities, and these have been compromised 
by seeing \cite{hb09}.  

\begin{figure*}
 \centering
 \includegraphics[width=0.5\hsize]{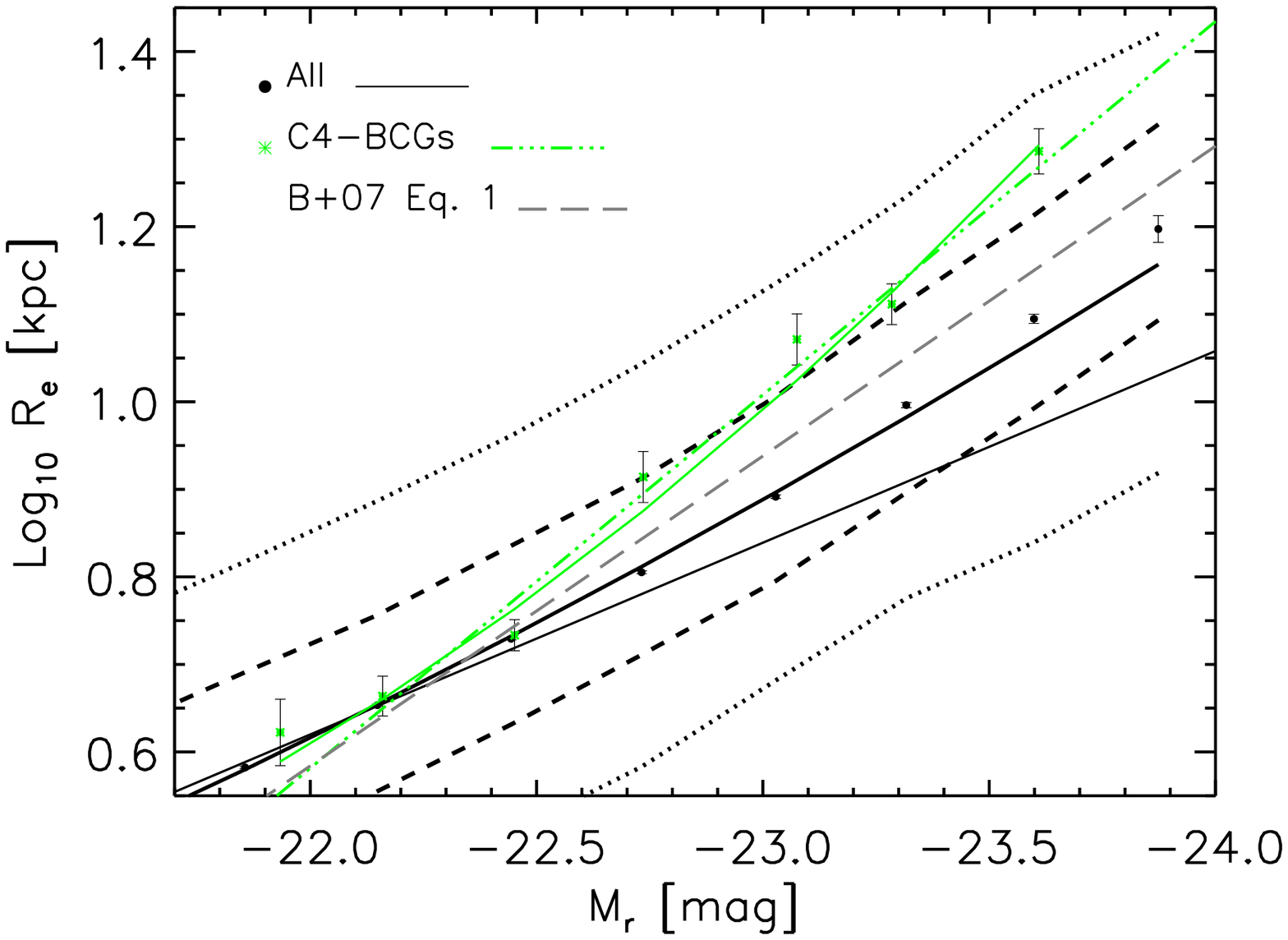}
 \caption{Comparison of our measurement of the size-luminosity 
        relation in the C4 sample with that reported by 
        Bernardi et al. (2007).  
        Our sizes are larger, and the scaling relation steeper,
        because we set $R=\sqrt{ab}$, where $a$ and $b$ are 
        the major and minor axis lengths;  
        Bernardi et al. inadvertantly set $R=b$. Filled circles 
        show the median value and its uncertainty for the bulk of 
        the early-type population; thin and thick black solid lines 
        show the linear and quadratic fits from Table~1 of 
        Hyde \& Bernardi (2009), respectively. Dashed and dotted curves show 
        the regions which enclose 68\% and 95\% of the objects.}
 \label{b07}
\end{figure*}

However, the $R-L$ scaling relation from Bernardi et al. (2007) 
differs slightly from that found in the main text above.  
Figure~\ref{b07} compares the relation for C4 BCGs reported by 
Bernardi et al. (2007) (long dashed line), with our present 
determination (symbols and dashed-triple-dot line) for the 
same BCGs; the Bernardi et al. sizes are slightly smaller, and 
the scaling relation is slightly shallower.  We have traced this 
to the fact that, although both sizes come from 2d fits to the 
surface brightness profile, our size is an effective circular 
size ($R=\sqrt{ab}$), whereas Bernardi et al. inadvertently show 
the minor axis $b$, even though they state that they show 
$\sqrt{ab}$ (a result of correcting the long axis $a$ by two 
powers of $\sqrt{b/a}$ rather than just one).  Correcting for 
this effect makes their relation the same as ours.  For completeness, 
this relation is
\begin{equation}
 \log_{10} \left(\frac{R_e}{\rm kpc}\right) = 0.158 - 0.423\, (M_r+21).
\end{equation}

\begin{figure*}
 \centering
 \includegraphics[width=0.45\hsize]{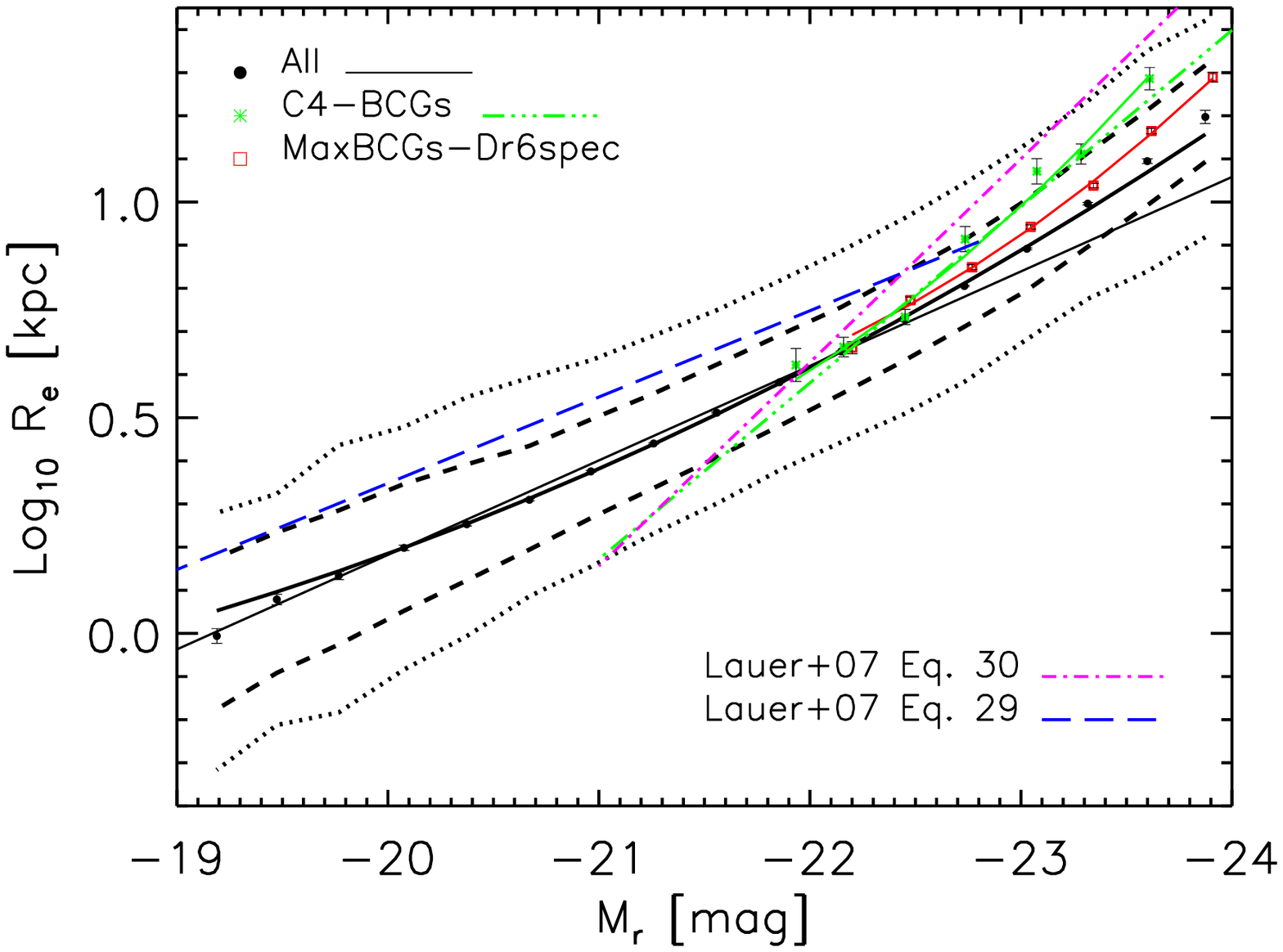}
 \includegraphics[width=0.45\hsize]{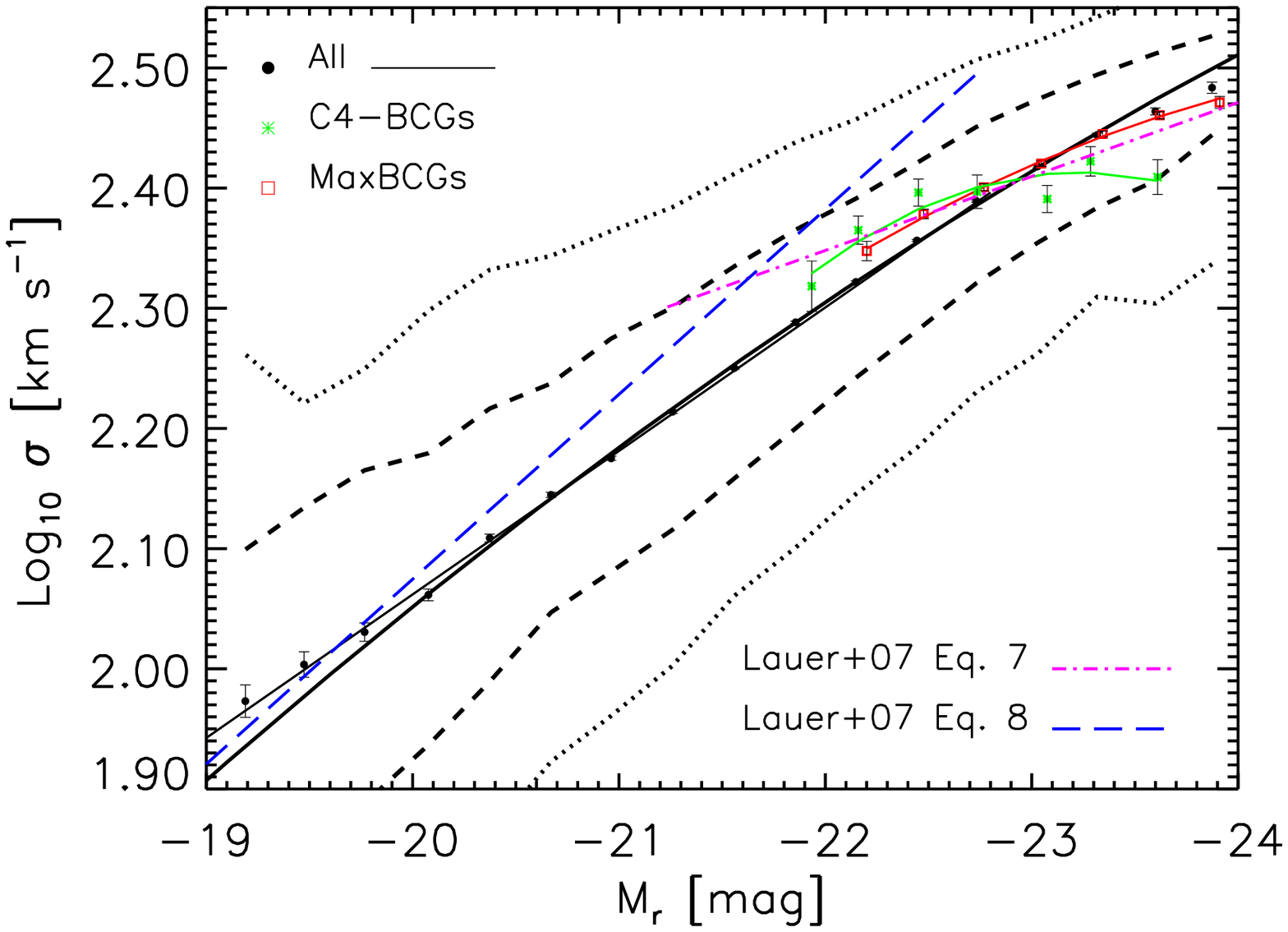}
 \caption{Comparison of the size-luminosity and velocity dispersion-luminosity 
          relations (left and right panels) in our data with that 
          reported by Lauer et al. (2007). Lauer et al. work with
          V-band photometry; we assume V-r = 0.36.}
 \label{L07}
\end{figure*}

However, Figure~\ref{L07} compares the scaling we find for the C4 BCGs 
with that reported by Lauer et al. (2007).  Although both relations are 
clearly more like one another than like the bulk of the population, 
there are significant differences.  Some of this is due to systematic 
differences between the two reductions, and, in light of the results 
in the main text, some may be due to evolution:  the Lauer et al. 
sample is at even lower redshifts than our C4 sample.  
This, of course, does not explain the offset at smaller luminosities.  

\label{lastpage}

\end{document}